\documentstyle[12pt,aaspp4,flushrt,psfig]{article}
\begin{document}

\title{The First Hour of Extra--galactic Data of the Sloan Digital Sky
Survey Spectroscopic Commissioning: The Coma Cluster.}

\author{
Francisco J. Castander\altaffilmark{\ref{Chicago},\ref{Pyrenees},\ref{YaleCalan},\ref{Andes}},
Robert C. Nichol\altaffilmark{\ref{CMU}},
Aronne Merrelli\altaffilmark{\ref{CMU},\ref{Caltech}},
Scott Burles\altaffilmark{\ref{Chicago},\ref{Fermilab}},
Adrian Pope\altaffilmark{\ref{CMU},\ref{JHU}},
Andrew J. Connolly\altaffilmark{\ref{Pittsburgh}},
Alan Uomoto\altaffilmark{\ref{JHU}},
James E. Gunn\altaffilmark{\ref{Princeton}}, 
John E. Anderson\altaffilmark{\ref{Fermilab}},
James Annis\altaffilmark{\ref{Fermilab}},
Neta A. Bahcall\altaffilmark{\ref{Princeton}},
William N. Boroski\altaffilmark{\ref{Fermilab}},
Jon Brinkmann\altaffilmark{\ref{APO}}, 
Larry Carey\altaffilmark{\ref{Washington}},
James H. Crocker\altaffilmark{\ref{JHU}},
Istv\'an Csabai\altaffilmark{\ref{JHU},\ref{Eotvos}},
Mamoru Doi\altaffilmark{\ref{UTokyo}},
Joshua A. Frieman\altaffilmark{\ref{Fermilab},\ref{Chicago}},
Masataka Fukugita\altaffilmark{\ref{CosmicRay},\ref{IAS}},
Scott D. Friedman\altaffilmark{\ref{JHU}},
Eric J. Hilton\altaffilmark{\ref{CMU}},
Robert B. Hindsley\altaffilmark{\ref{NRL}},
\v{Z}eljko Ivezi\'{c}\altaffilmark{\ref{Princeton}},
Steve Kent\altaffilmark{\ref{Fermilab}},
Donald Q. Lamb\altaffilmark{\ref{Chicago}},
R. French Leger\altaffilmark{\ref{Washington}},
Daniel C. Long\altaffilmark{\ref{APO}}, 
Jon Loveday\altaffilmark{\ref{Sussex}},
Robert H. Lupton\altaffilmark{\ref{Princeton}}, 
Harvey MacGillivray\altaffilmark{\ref{Edinburgh}},
Avery Meiksin\altaffilmark{\ref{Edinburgh}},
Jeffrey A. Munn\altaffilmark{\ref{Flagstaff}},
Matt Newcomb\altaffilmark{\ref{CMU}},
Sadanori Okamura\altaffilmark{\ref{UTokyo}},
Russell Owen\altaffilmark{\ref{Washington}},
Jeffrey R. Pier\altaffilmark{\ref{Flagstaff}},
Constance M. Rockosi\altaffilmark{\ref{Chicago}}, 
David J. Schlegel\altaffilmark{\ref{Princeton}},
Donald P. Schneider\altaffilmark{\ref{PSU}},
Walter Seigmund\altaffilmark{\ref{Washington}},
Stephen Smee\altaffilmark{\ref{JHU}},
Yehuda Snir\altaffilmark{\ref{CMU}},
Larry Starkman\altaffilmark{\ref{Washington}},
Chris Stoughton\altaffilmark{\ref{Fermilab}},
Gyula P. Szokoly\altaffilmark{\ref{Potsdam}},
Christopher Stubbs\altaffilmark{\ref{Washington}},
Mark SubbaRao\altaffilmark{\ref{Chicago}},
Alex Szalay\altaffilmark{\ref{JHU}},
Aniruddha R. Thakar\altaffilmark{\ref{JHU}},
Christy Tremonti\altaffilmark{\ref{JHU}},
Patrick Waddell\altaffilmark{\ref{Washington}},
Brian Yanny\altaffilmark{\ref{Fermilab}} and 
Donald G. York\altaffilmark{\ref{Chicago}}
}
\newcounter{address}
\setcounter{address}{1}
\altaffiltext{\theaddress}{The University of Chicago, Astronomy \& Astrophysics
Center, 5640 S. Ellis Ave., Chicago, IL 60637
\label{Chicago}}
\addtocounter{address}{1}
\altaffiltext{\theaddress}{Observatoire Midi-Pyr\'en\'ees, 
14 ave Edouard Belin, Toulouse, F-31400, France
\label{Pyrenees}}
\addtocounter{address}{1}
\altaffiltext{\theaddress}{Current address:Yale University, P.O. Box 208101, New Haven, CT 06520 \& Universidad de Chile, Casilla 36-D, Santiago, Chile
\label{YaleCalan}}
\addtocounter{address}{1}
\altaffiltext{\theaddress}{Andes Fellow
\label{Andes}}
\addtocounter{address}{1}
\altaffiltext{\theaddress}{Department of Physics, Carnegie Mellon University,
     5000 Forbes Ave., Pittsburgh, PA-15232
\label{CMU}}
\addtocounter{address}{1}
\altaffiltext{\theaddress}{Department of Astronomy, California Institute
of Technology, Pasadena, CA 91125
\label{Caltech}}
\addtocounter{address}{1}
\altaffiltext{\theaddress}{Fermi National Accelerator Laboratory, P.O. Box 500,
Batavia, IL 60510
\label{Fermilab}}
\addtocounter{address}{1}
\altaffiltext{\theaddress}{
Department of Physics \& Astronomy, The Johns Hopkins University,
   3701 San Martin Drive, Baltimore, MD 21218, USA
\label{JHU}}
\addtocounter{address}{1}
\altaffiltext{\theaddress}{Department of Physics \& Astronomy,
          University of Pittsburgh,
          Pittsburgh, PA 15260
\label{Pittsburgh}}
\addtocounter{address}{1}
\altaffiltext{\theaddress}{Princeton University Observatory, 
Princeton, NJ 08544
\label{Princeton}}
\addtocounter{address}{1}
\altaffiltext{\theaddress}{Apache Point Observatory, P.O. Box 59,
Sunspot, NM 88349-0059
\label{APO}}
\addtocounter{address}{1}
\altaffiltext{\theaddress}{University of Washington, Department of Astronomy,
Box 351580, Seattle, WA 98195
\label{Washington}}
\addtocounter{address}{1}
\altaffiltext{\theaddress}{Department of Physics of Complex Systems,
E\"otv\"os University,
   P\'azm\'any P\'eter s\'et\'any 1/A, Budapest, H-1117, Hungary
\label{Eotvos}}
\addtocounter{address}{1}
\altaffiltext{\theaddress}{Department of Astronomy \& Research Center 
  for the Early Universe, School of Science, University of Tokyo, Hongo,
  Bunkyo, Tokyo, 113-0033, Japan
\label{UTokyo}}
\addtocounter{address}{1}
\altaffiltext{\theaddress}{Institute for Cosmic Ray Research, University of
Tokyo, Midori, Tanashi, Tokyo 188-8502, Japan
\label{CosmicRay}}
\addtocounter{address}{1}
\altaffiltext{\theaddress}{Institute for Advanced Study, Olden Lane,
Princeton, NJ 08540
\label{IAS}}
\addtocounter{address}{1}
\altaffiltext{\theaddress}{Remote Sensing Division, Naval
  Research Laboratory, 4555 Overlook Ave. SW, Washington, DC 20375
\label{NRL}}
\addtocounter{address}{1}
\altaffiltext{\theaddress}{Astronomy Center, University of Sussex, Falmer , Brighton BN1 9QJ, UK
\label{Sussex}}
\addtocounter{address}{1}
\altaffiltext{\theaddress}{Royal Observatory, Edinburgh, EH9 3HJ, United
  Kingdom
\label{Edinburgh}}
\addtocounter{address}{1}
\altaffiltext{\theaddress}{U.S. Naval Observatory, Flagstaff Station, 
P.O. Box 1149, 
Flagstaff, AZ  86002-1149
\label{Flagstaff}}
\addtocounter{address}{1}
\altaffiltext{\theaddress}{Department of Astronomy \& Astrophysics,
Pennsylvania State University,
University Park, PA 16802
\label{PSU}}
\addtocounter{address}{1}
\altaffiltext{\theaddress}{Astrophysikalisches Institut Potsdam, An
der Sternwarte 16, D-14482 Potsdam, Germany
\label{Potsdam}}

\begin{abstract}
	
On 26 May 1999, one of the Sloan Digital Sky Survey (SDSS) fiber--fed
spectrographs saw astronomical first light. This was followed by the first
spectroscopic commissioning run during the dark period of June 1999. We
present here the first hour of extra--galactic spectroscopy taken during
these early commissioning stages: an observation of the Coma cluster of
galaxies.  Our data samples the Southern part of this cluster, out to a
radius of 1.5 degrees ($1.8\,h^{-1}$ Mpc, approximately to the virial
radius) and thus fully covers the NGC 4839 group.  We outline in this paper
the main characteristics of the SDSS spectroscopic systems and provide
redshifts and spectral classifications for 196 Coma galaxies, of which 45
redshifts are new.  For the 151 galaxies in common with the literature, we
find excellent agreement between our redshift determinations and the
published values, {\it e.g.}, for the largest homogeneous sample of
galaxies in common (63 galaxies observed by Colless \& Dunn 1996) we find a
mean offset of 3\, ${\rm km\,s^{-1}}$ and an RMS scatter of only 24 ${\rm
km\,s^{-1}}$. As part of our analysis, we have investigated four different
spectral classification algorithms: measurements of the spectral line
strengths, a principal component decomposition, a wavelet analysis and the
fitting of spectral synthesis models to the data. We find that these
classification schemes are in broad agreement and can provide physical
insight into the evolutionary histories of our cluster galaxies.  We find
that a significant fraction (25\%) of our observed Coma galaxies show signs
of recent star--formation activity and that the velocity dispersion of
these active galaxies (emission--line and post--starburst galaxies) is 30\%
larger than the absorption--line galaxies.  We also find no active galaxies
within the central (projected) $200\,h^{-1}$ Kpc of the cluster. The
spatial distribution of our Coma active galaxies is consistent with that
found at higher redshift for the CNOC1 cluster survey.  Beyond the core
region, the fraction of bright active galaxies appears to rise slowly out
to the virial radius and are randomly distributed within the cluster with
no apparent correlation with the potential merger or post-merger of the NGC
4839 group.  We briefly discuss possible origins of this recent galaxy
star-formation.

\end{abstract}

\section{Introduction}

The Coma cluster is the richest cluster of galaxies in our local universe
and has thus attracted considerable attention over the last century (see
reviews by Biviano 1998 \& West 1998).  In the optical, for example,
Goldwin, Metcalfe \& Peach (1983; hereafter GMP83) have published an
extensive photometric study of the cluster providing accurate positions,
colors, magnitudes and ellipticities for 6724 bright galaxies over 2.63
square degrees centered on Coma. Several authors have explored the fainter
dwarf galaxy population of Coma (see Bernstein et al. 1995; Kashikawa et
al. 1998; Adami et al. 1998 \& 2000). The dynamics of the cluster have also
been well studied. \markcite{KG82}Kent \& Gunn (1982) assembled
approximately 300 optical redshifts from the literature to determine the
cluster mass distribution. This initial work was extended by Colless \&
Dunn (1996; CD96), who collected 556 redshifts (based on new and literature
redshifts), and Geller, Diaferio \& Kurtz (1999) who have extended the
dynamical study of Coma to large radii (10 degrees) and larger numbers
(1693 redshifts) thus measuring the density profile of Coma well beyond the
virial radius of the cluster. Hughes (1989) measured the total mass of Coma
using early X-ray observations of the cluster. More recently, ROSAT
observations of Coma have provided unprecedented detail of the intracluster
gas morphology (e.g., \cite{bri92}; \cite{whi93}). ASCA observations of the
cluster have provided important information on the temperature structure of
the X--ray emitting gas (Honda et al. 1996) which have been recently
complemented by XMM-Newton observations (Briel et al 2000; Arnaud et al
2000; Neumann et al 2000)

The Coma cluster has long been regarded as the archetypal relaxed massive
cluster of galaxies. However, recent studies of the cluster have shown
otherwise. The current view of Coma is that the cluster is the product of
one recent, and one ongoing, cluster--group merger.  A group centered on
NGC 4839, in the southwest region of the cluster, is falling into (CD96),
or has just passed through (\cite{bur94}), the main body of the cluster and
may have triggered new star--formation in galaxies in Coma (Caldwell \&
Rose 1997). The velocity dispersion of this southwest group is
approximately $\frac{1}{3}$ of that of the main cluster.  Meanwhile, the
core of Coma has two dominant galaxies, NGC 4874 and NGC 4889, which seem
to be the relic central galaxies of previous groups that have merged into
the current cluster. The X--ray and dynamical data reveal that both of
these dominant galaxies do not appear to sit at the bottom of the cluster
potential (CD96; \cite{whi93}). The lack of a cooling flow and the
existence of an extended radio halo support this merging history of the
cluster.

The Coma cluster was thus an ideal first target for the Sloan Digital Sky
Survey (SDSS; \cite{yor00}) spectroscopic commissioning program because of
its location (the North Galactic Pole), the pre--existence of wide--field
galaxy photometry (e.g. GMP83), and the high--density of known redshifts
for comparison and testing. Moreover, there remains interesting scientific
questions that can be addressed using the unique SDSS spectroscopic
hardware {\it e.g.} the influence of cluster merger events on the
star--formation rates of galaxies. The quality and quantity of SDSS
spectral data will allow us to study such problems in great detail and in
this paper, we start by outlining and comparing the different analysis
techniques for classifying SDSS galaxy spectra as well as quantifying their
star--formation rates.

Robust, automated, spectral classification methods are necessary to help us
understand the distribution and evolution of the galaxies physical
properties that define and characterize their spectral energy
distributions.  The problem of galaxy spectra classification has been
copiously treated in the astronomical literature. In general, methods are
based on the measurement of spectral continuum and line features (e.g., the
Lick/IDS system: Faber et al 1985; Burstein, Faber \& Gonz\'alez 1986)
which are then used to classify and derive the galaxies physical
properties. Stellar population synthesis models can, for example, be
compared to these measurements, or the entire spectrum using template
fitting, to provide a physical understanding of the galaxy
properties. Recently, other techniques have been investigated. Amongst
them, the principal component analysis has attracted a lot of interest
(e.g, Connolly et al 1995; Bromley et al 1998; Folkes et al 1999; Ronen et
al 1999). The technique is based on decomposing the spectra into a basis
that highlights the galaxy differences. Such decomposition can therefore be
used to classify the spectra. Different implementations change the way the
basis is constructed or how the resulting coefficients of the decomposition
are used to generate a classification method. Some authors, for instance,
use artificial neural networks to build a classification scheme from the
principal component coefficients (e.g., Folkes, Lahav \& Maddox
1996). Wavelets also provide an orthogonal basis in which the spectra can
be decomposed and therefore can in principle be used as a classification
method. Along these lines, Pando \& Fang (1996) and Theuns \& Zaroubi
(2000) have used wavelets to study quasar spectra.

In this paper, we present the first hour of extra--galactic data taken by
the SDSS spectroscopic system. Only one of the ten plates originally
designed in the Coma region has been observed producing nearly 200
Coma redshifts and thus illustrating the capabilities of this new
instrumentation.  These spectra will allow us to investigate several
classification schemes and test their applicability to the future SDSS
dataset. The paper is structured as follows.  In \S 2, we briefly
highlight the main characteristics of the SDSS spectroscopic system. In \S
3, we describe the selection of galaxy targets in the Coma cluster
region. The observations of the Coma cluster plates are presented in \S
4. In \S 5, we describe the data analysis. We discuss the redshift
measurements in \S 6 and classify the galaxy members in \S 7. Finally, we
discuss our data and present our conclusions in \S 8.

\section{The SDSS Spectroscopic System}

The SDSS will use a dedicated 2.5m telescope with a 3 degree
field--of--view and two fiber--fed double spectrographs to measure the
redshifts of approximately a million galaxies and one hundred thousand
quasars over the next 5 years. The SDSS will also image the same area of
the sky to an approximate depth of $\sim$23 mag in five optical bands
($u'$, $g'$, $r'$, $i'$ and $z'$; \cite{fuk96}; \cite{fan00}). For more
details on the SDSS, the reader is referred to \cite{yor00} for a brief
overview of the survey hardware, software and strategy$\footnote{In
addition, the reader can refer to the online SDSS Project Book at {\tt
http://www.astro.princeton.edu/PBOOK/welcome.html}}$.  The full details of
the SDSS spectroscopic system will be presented in Uomoto et al. (2001; in
preparation) and Frieman et al. (2001; in preparation).  For completeness,
we provide here an outline of the SDSS spectroscopic system.

The SDSS spectrographs are designed to cover the wavelength range of 3900
$\rightarrow$ 9100 {\AA}. This is achieved using a dichroic beamsplitter
centered at $\simeq6000$ {\AA} to separate the incoming light onto a red
and blue camera in both spectrographs. The SDSS spectrographs achieve a
spectral resolution of 1800 across its entire spectral range.  Each
spectrograph accepts 320 fibers, each of which subtends a diameter of
3$^{''}$ on the sky. The fibers are plugged into aluminum plug--plates
sitting in the focal plane of the SDSS telescope.  Each plate contains 640
science holes (one for each science fiber), light--trap holes (drilled to
avoid the reflection of light from bright stars off the plug--plate back
into the telescope optics), guide star holes (11 in total which are fed to
the guide camera) and quality assurance holes.

During normal operations, plates scheduled for observation are manually
plugged during the day. Each plug--plate is mounted in an individual
cartridge that possesses a full set of 640 science fibers and 11 coherent
fiber bundles that are plugged into the guide star holes and used to guide
the telescope. Once plugged, the cartridges are automatically mapped with a
video system that measures the plug--plate location of each fiber ({\it
i.e.} sky coordinates) as they are successively illuminated by a laser
diode at the slithead.  After mapping the cartridges are stored ready for
mounting on the SDSS telescope during the night.

\section{Target Selection}
\label{target}

In normal survey operations, the SDSS will select targets for spectroscopic
observation using the SDSS photometric survey.  However, in the case of the
Coma cluster, the SDSS imaging photometric camera had not yet observed this
region of the sky and thus target selection was performed using an external
catalog.

We have used the SuperCOSMOS scans of two photographic plates (J and O),
centered on the Coma cluster (see http://www-wfau.roe.ac.uk/sss/ for
details) kindly provided to us by Harvey MacGillivray.  The passband of the
J plate is the combination of the Kodak IIIaJ emulsion and a GG395 filter
(same as the UK Schmidt $b_j$ system; see Nichol, Collins \& Lumsden 2000),
while the passband of the O plate is the combination of the IIIaF emulsion
and a OG590 filter. The raw object lists of these two SuperCOSMOS scans
contain approximately 150,000 objects per plate and therefore must be
trimmed for our purposes.

We first merged the two photographic plates catalogs keeping only objects
that matched to within 1.5 arcseconds.  This merged catalog was separated
into a list of stars and galaxies using the SuperCOSMOS source
classifier. We produced two lists; one of stellar sources, which were
objects classified as stars on both plates, and one of galaxies, which were
objects classified as galaxies on at least one of the plates. We used these
criteria since star-galaxy separation becomes increasingly difficult at the
fainter magnitudes. By including sources which were classified as a galaxy
in one plate, but as a star in the other, we hoped to recover some of the
faint galaxy population that would have otherwise been
missed. Unfortunately, we can expect higher stellar contamination
(especially from blended stars) in our galaxy list.

We photometrically calibrated the SuperCOSMOS data using the GMP83
catalog. We matched our galaxy list to the GMP83 data and then derived a
transformation to convert the SuperCOSMOS instrumental magnitudes into the
GMP83 magnitude system which was calibrated using $B$-band photoelectric
magnitudes for their blue band and $B-R$ colors for their red band (see
GMP83 for details). Our photometry is thus measured in the blue and red
"photographic bands" (emulsion + filter described above) but transformed to
the GMP83 system with a zero-point and color terms.  Hereafter, we will
refer to these magnitudes as, $b$ and $r$, following the notation in GMP83.
Finally, a small astrometric correction was applied by correlating our star
catalog with the ACT star reference catalog.\footnote{The ACT catalog is
produced by the US Naval Observatory, see {\tt
http://aries.usno.navy.mil/ad/act/act.html}}.

\begin{figure}
\centerline{\psfig{file=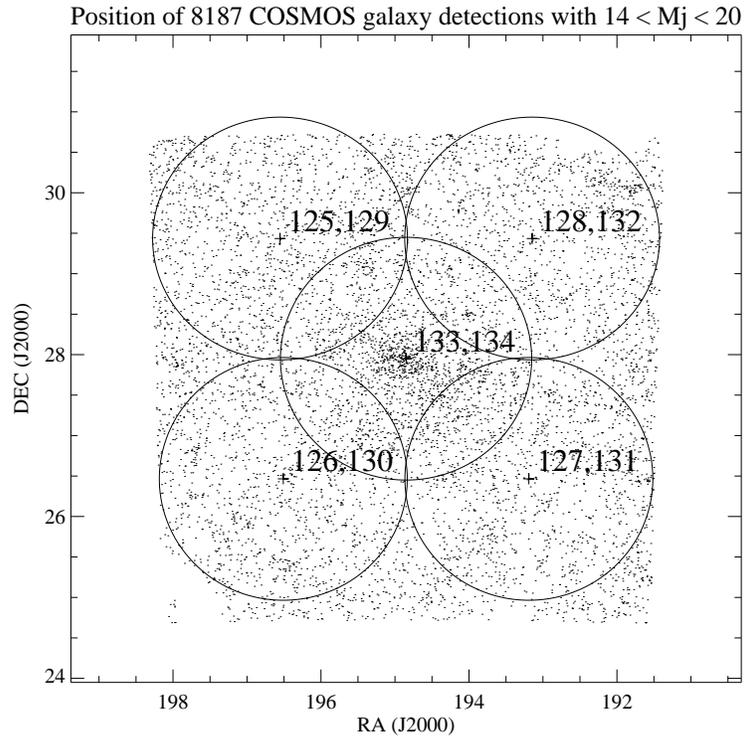,height=10.0cm}}
\caption{The distribution of SuperCOSMOS galaxies in the Coma cluster
region.  The circles superimposed on the data show the position of our
10 designed plug-plates. We designed two plates at each position: one
containing the brighter sources, the other, fainter sources. The
numbers are the SDSS plate identifiers. Due to engineering and weather
constraints during the commissioning period in June 1999 only plate
133 was observed.
\label{fig1}
}
\end{figure}

\begin{figure}
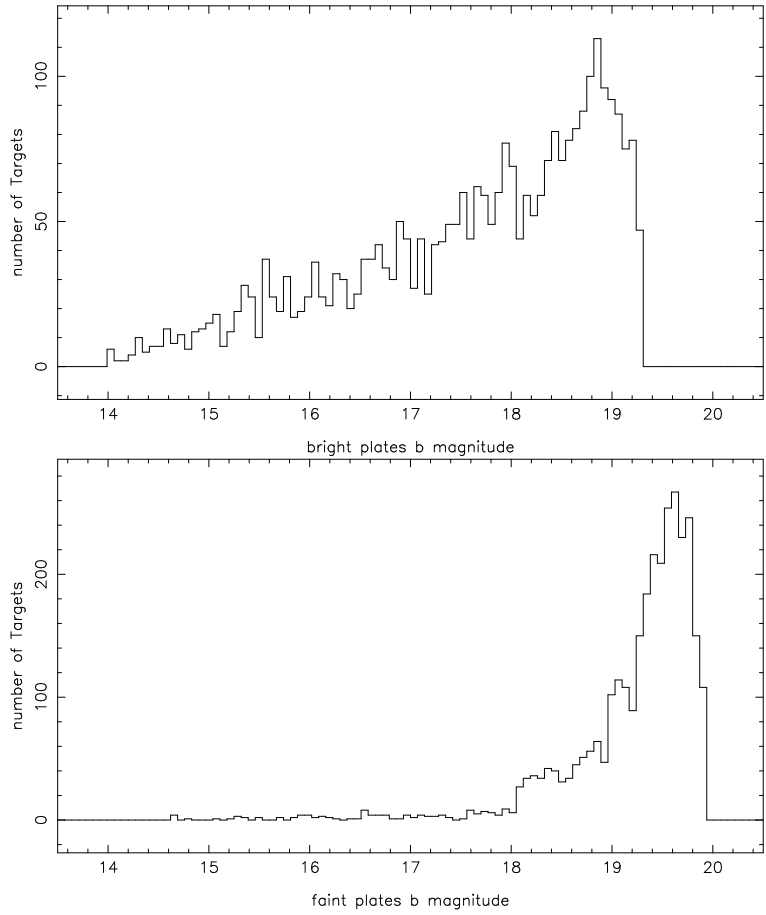

\centerline{\psfig{file=castander.fig2a.ps,height=6.0cm,angle=270}}
\centerline{\psfig{file=castander.fig2b.ps,height=6.0cm,angle=270}}
\caption{Magnitude distribution of selected sources in the five {\it
bright} plates (top) and the five {\it faint} plates (bottom).
\label{fig2}
}
\end{figure}

The resulting list of galaxies contained nearly 33,000 objects. This was
further trimmed to 8187 targets by imposing a magnitude cut of 14 $\le b
\le$ 20. From these targets, a total of 10 plates were designed, with two
plates at each of the five plate centers shown in Figure~\ref{fig1}. Each
pair of plates was allocated unique targets, with one plate containing the
brighter galaxies and the other containing the fainter galaxies (see
figure~\ref{fig2}). This was done for two reasons; first, to accommodate
the minimum spacing restriction between SDSS fibers (55 arcseconds) and
second, to allow for different exposure times for these plates.  Finally,
each plate was allocated a set of 10 bright stars ( $14 < b < 15$ ) for the
guide camera fibers. Since these were selected from the same SuperCOSMOS
data there was little concern about their relative astrometric reference
frame.

\section{Observations}
\label{observe}

One of the SDSS spectrographs saw first astronomical light on 26 May 1999
in bright time. After this milestone, the first spectroscopic commissioning
run took place in the dark period of June 1999.  At the time of the Coma
SDSS spectroscopic observations, several parts of the nominal SDSS
spectroscopic system were either not available or were being commissioned
for the first time. For example, only one of the two SDSS spectrographs was
mounted on the back of the telescope.  Therefore, only 320 science fibers
were available per plate instead of the 640 which is now the standard
number.  In addition, the plate--mapper, which automatically matches the
plugged fibers to the target catalog, was not fully automated and we were
forced to map the plate by hand.  The telescope was not properly collimated
at the time of the observations and the hardware needed to obtain
spectroscopic calibrations was not installed. Finally, condensation on the
front of the red camera dewar produced a series of bright ``doughnuts'' in
the center of the red CCD images rendering most of the red spectrograph
data in this commissioning run almost impossible to use.

On 6 June, the SDSS observed plate 133 (the bright plate at the center of
Coma). We took two 900 seconds exposures at mean airmasses of 1.53 and 1.66
without guiding. On 8 June, plate 133 was re--observed taking three 1200
second exposures. These were the first astronomical exposures in which the
SDSS spectroscopic system guided on the sky. The seeing was between 1.2 and
1.5 arcseconds and the mean airmasses were 1.07, 1.10 and 1.15 for the
three exposures. Being near the zenith, the atmospheric differential
refraction effects were small. In addition to plate 133, we also observed
plates 135 and 136 on 8 June 1999, which were designed using SDSS
photometry in areas of the sky unrelated to the Coma cluster and are thus
more representative of the field galaxy population.

Plate 133 is the only Coma plate that has been observed during the SDSS
spectroscopic commissioning phase. Figure~\ref{fig3} shows the distribution
on the sky of the identified cluster members in this plate. Coma will be
re--observed as part of the main SDSS spectroscopic and photometric survey
in the coming years.  These data will be far superior to the data discussed
herein.

\section{Data Reduction}
\label{datared}

In normal production mode the SDSS will reduce the spectroscopic data
through a specifically designed pipeline (Frieman et al 2001; in
preparation). However, due to the uniqueness of these observations and the
non--standard observing set--up, the Coma observations were processed using
a modified version of an earlier copy of the spectroscopic pipeline. The
reduction stages are, nevertheless, standard in multi--fiber
spectroscopy. The blue and red camera data were reduced separately given
the aforementioned problem with the red camera.

First, we subtracted the bias signal using the overscan columns at the edge
of the CCD and co--added the individual 2--dimensional images, rejecting
cosmic rays in the process. We then traced the fibers and optimally
extracted the 1-dimensional spectra using the Horne (1986) algorithm. In
the red this procedure had to be individually supervised and sometimes
changed for troublesome fibers. No flat-fielding of the data was possible
given the lack of uniformly illuminated exposures either for the whole CCD
(2-dimensional flat) or through the fibers (1-dimensional flat).  We
wavelength calibrated the spectra using the sky emission lines. In the
blue, we used the Hg, [OI] and NaD lines, while in the red, we used the
numerous OH lines as well.  We fitted a third order polynomial to the
wavelength dispersion, and in the blue we obtained residuals of $<0.1$
{\AA} for most fibers.  For a few fibers, we did witness residuals of 0.2
{\AA} (which is $\simeq15$ ${\rm km\,s^{-1}}$). However, these errors are
computed at the position of the sky lines and are therefore, probably
under--representative of the wavelength calibration error for the whole
spectrum (see later). We obtained a similarly accurate wavelength solution
in the red.  On average, the blue spectra spanned the wavelength range 3770
to 6100 {\AA} with a median dispersion of 1.14 {\AA}/pix, while in the red,
we obtained a wavelength coverage of 5770 to 9120 {\AA} with a median
dispersion of 1.64 {\AA}/pix.  These are close to the original design
specification.

To sky--subtract the spectra, we combined the dedicated sky fibers into one
``super--sky'' spectrum which was rebinned to the resolution of the
individual spectra and scaled to match the flux of the [OI] $\lambda$5577
night sky line of each science fiber to compensate for any fiber throughput
differences. Then each re-scaled ``super--sky'' was subtracted from the
corresponding science fiber. As we did not possess calibration frames, and
given the poorer quality of the red side, we did not attempt to merge the
red and blue spectra. For the rest of the paper, except on the galaxy
classification in Section~\ref{classifications} when we use H$\alpha$, we
only use the blue camera data.

To correct for the intrinsic response of the instrument in the blue, we
have used bright, early-type stars observed as part of the plate 133.  This
was achieved by selecting all stars of a spectral type earlier (hotter)
than K0 which also possessed a signal-to-noise ratio per pixel of greater
than 100 at 5000 {\AA}. In total, this criterion provided 15 stars, the
earliest type being a A9-F0 star, the latest a K0. We divided these stars
by the best fit stellar template in the Jacoby et al. (1984) stellar
atlas. We smoothed, normalized, and combined the resulting spectra to
obtain the response function of the instrument. We then divided all the
spectra by this response function.  We note this procedure does not provide
an absolute spectrophotometric calibration; it only corrects for the
response function of the instrument.

\section{Redshift Determinations}

We first determined the redshifts of our spectra via visual inspection and
fitted Gaussians to all obvious absorption/emission lines.  The main
absorption lines used were CaII H and K $\lambda\lambda$3968, 3934, CaI
$\lambda$4227, the Balmer series lines, MgI $\lambda\lambda\lambda$5167,
5173, 5184 and NaD $\lambda\lambda$5790, 5796. In emission we used [OII]
$\lambda\lambda$3726, 3729, the Balmer series lines and [OIII]
$\lambda\lambda$4959, 5007.  Our visual inspections of the spectra revealed
196 galaxies in the Coma cluster, 49 field galaxies, 5 quasars, 47 stars
and 12 spectra that could not be classified. The remaining 11 fibers were
sky fibers.

We also used the cross-correlation technique to obtain a more accurate
galaxy redshift. We utilized the RVSAO package (\cite{KM98}) within the
IRAF environment. Given that the majority of synthetic spectra have
resolution coarser than our data, we decided to construct our own templates
for the cross-correlation using the stellar objects observed on Plate 133;
this ensures that we have the same resolution for both the template and
object spectra. We chose to use as templates the fifteen stars discussed
above (for correcting the response function of the instrument) as well as
two additional later-type stars (K1 and K4) that also passed the
aforementioned signal-to-noise criterion.  We also added emission lines to
these stellar templates creating a set of 34 templates {\it i.e.}, 17
absorption line stars and 17 absorption and emission stellar templates. All
templates were shifted to their rest--frame.

To determine the true error on our wavelength solutions and thus our
redshifts, we first re--calculated the wavelength solution for each of our
stellar templates using at least 9 lines, and normally 14, per spectrum.
The availability of more lines, and the better sampling of the wavelength
range, allowed us to improved the wavelength solution compared to that
derived from just the sky lines alone. We then cross-correlated these
re--calibrated stellar templates against higher--quality stellar spectra
taken by the SDSS in later commissioning observations in Spring 2000. From
these tests, we estimated that our dispersion solutions had an error of
$\simeq30$ ${\rm km\,s^{-1}}$ which we added in quadrature to the error
resulting from the cross-correlation technique.

In Figure~\ref{fig4}, we present the redshift distribution of the 245
galaxies for which we obtained a redshift on plate 133. The Coma cluster
can be clearly seen as the broad peak around $z=0.0232$. Figure~\ref{fig5}
is an expanded view of this region together with a 3-$\sigma$ clipped
Gaussian fit to the data. Cluster membership is easily assigned for our
galaxy sample since there is only one galaxy in the 3 to 4$\sigma$ region
of the distribution. This galaxy would not normally be assigned to the
cluster based on our data alone, however, CD96 measured a higher number of
galaxy redshifts obtaining a larger velocity dispersion which would put
this galaxy inside our 3-$\sigma$ cut. Therefore, for the rest of our
analysis, we included this galaxy as a cluster member, bringing the total
number of Coma member on plate 133 to 196.

Table~\ref{tbl1} lists the main galaxy data on our Coma galaxies.  Column 1
gives the extraction ID number (running from 1 to 320) while Column 2 is
the GMP83 number when appropriate (three of our galaxies do not have a
corresponding GMP83 detection). Columns 3 and 4 give the right ascension
and declination of the fiber center (not necessarily the galaxy
centroid). In column 5 we present the cross-correlation redshift corrected
to the heliocentric reference frame. The redshift error, in Column 6,
includes the error in the wavelength calibration combined in quadrature
with the error resulting from the cross-correlation technique.  The usual
$R$ cross--correlation coefficient is listed in column 7. In most cases, we
simply chose the cross--correlation redshift with the largest $R$ value out
of the 34 possible template cross--correlation redshifts. There were a few
cases where the best $R$ was in clear disagreement with our visual redshift
measurements and in these cases we re-inspected the spectrum and quote, as
our final redshift, the template with the highest $R$ coefficient
consistent with our visual redshift. These rare cases were typically due to
non-corrected cosmic rays or sky residuals that produced a confident
cross-correlation with one of the emission--line templates. Finally, in
Columns 8 and 9, we provide the magnitudes for the galaxy taken from the
SuperCOSMOS catalog and calibrated to the GMP83 system.

We have compared our redshift measurements to the data available in the
literature. We have matched our catalog with the redshift compilation of
CD96 as well as data from the NED$\footnote{NED is the NASA/IPAC
Extragalactic Database, operated by the Jet Propulsion Laboratory, Caltech,
under contract with NASA}$ database. We found 151 galaxies that match
previous known redshifts. The remaining 45 galaxies, approximately a
quarter of our Coma redshifts, represent new measurements.  In
Figure~\ref{fig6}, we show the comparison between our redshift measurements
and those in the literature. The solid histogram represents the 151
matches, where 148 come from the compilation of CD96 (including their own
measurements and values in the literature), and the other three are taken
from NED. The solid histogram is a Gaussian fit to the entire
distribution. The fit is poor because of the extended tails.  We can
improve the fit considerably if we reject 3-$\sigma$ outliers (15 galaxies;
see Figure \ref{fig6}). We attribute these outliers to the heterogeneity of
the data where different data sets from the literature have different
calibration errors. Also, these discrepancies could be in part attributed
to the internal dynamics of the observed galaxies because of the different
fiber and/or slit placements. We can improve the fit even further by
restricting ourselves to the largest homogeneous dataset from the
literature {\it i.e.}, the 63 new redshifts measurements of CD96. We then
find excellent agreement between our measurements (the dotted histogram and
Gaussian fit) with a mean offset of 3 ${\rm km\,s^{-1}}$ and an RMS scatter
of 24 ${\rm km\,s^{-1}}$. Overall, the agreement of our redshift
determinations with the literature is remarkably good. The RMS scatter
between datasets indicates that our estimate of the errors in the
wavelength calibration is accurate.

\section{Spectral Classifications}
\label{classifications}

In this section, we consider the spectral classification of our Coma
galaxies. We have investigated four different algorithms, all of which
could be implemented for the main SDSS galaxy survey. The first algorithm
is based on visual inspections of the spectra but quantified using
measurements of the equivalent widths\footnote{All equivalent widths given
in this paper are rest-frame equivalent widths $W_o=W_{observed}/(1+z)$ and
follow the following convention unless otherwise stated: Absorption line
equivalent widths are positive and emission line equivalent widths are
negative.} of the lines seen in the spectrum.  The next two algorithms,
Wavelets and Principal Component Analysis (PCA), were used to objectively
classify the spectra using the visual inspections to define relevant
thresholds in wavelet and eigenspace.  The final classification scheme
attempts to define a physical classification scheme based on synthetic
models of galaxies.

\subsection{Line Strength Classifications}
\label{lines}

We started by visually classifying the spectra into five classes which
could then be compared to the other algorithms. We first divide the spectra
into two broad classes: emission and absorption line galaxies.  Absorption
line spectra were then sub--divided into normal absorption line systems and
objects with strong Balmer lines or post--starburst galaxies. The spectra
showing emission lines were sub--divided into three categories depending on
the strength of the emission lines.

For our visual and line--strength classification scheme, we were able to
use both the blue side of the spectrum and the presence of H$\alpha$ in the
red end of the spectrum. Unfortunately the other classification algorithms
could not use the red side of the spectrum due to the condensation problems
discussed above. Since H$\alpha$ is a powerful indicator of
star--formation, we believe that one of the classification schemes should
utilize these data even if the others could not. Moreover, the other
algorithms used our line strength classifications to quantify
star--formation in Coma and therefore, it was justified to make the line
strength classification as strong as possible by using all data available.

Using our equivalent width measurements, we divided the spectra in the five
following types: 1) Absorption line galaxies ({\bf AB}) which have spectra
without H$\alpha$ in emission and with the sum of the equivalent widths of
H$\delta$, H$\gamma$ and H$\beta$ smaller than 15 {\AA}. 2) Post--starburst
galaxies ({\bf PS}) which have no H$\alpha$ in emission, $W_o({\rm
H}\alpha)>0$, and $W_o({\rm H}\delta+{\rm H}\gamma+{\rm H}\beta) > 15$
{\AA}. 3) Absorption line dominated galaxies but with H$\alpha$ in emission
and modest H$\beta$ emission ({\bf AB+EM}): $W_o(H{\alpha})<0$ and
$W_o(H{\beta})>-5$ {\AA}. 4) Emission and absorption line galaxies ({\bf
EM+AB}) which we define using $W_o({\rm H}\alpha)<0$ and $-5<W_o({\rm
H}\beta)<-15$ {\AA}. 5) Emission line dominated spectra ({\bf EM}) which
have $W_o({\rm H}\alpha)<0$ and $W_o({\rm H}\beta)<-15$ {\AA}.  We present
these classifications in Table~\ref{tbl2} where Column 1 is the same
extraction number as given Table~\ref{tbl1}, Column 2 is the classification
based on the line strength criteria given above, Column 3 is the PCA
classification and Column 4 is the wavelet classification (see below). The
results of the synthetic model fitting and the parameters of the PCA
classification, both of which are discussed below, are listed in Columns 5
through 10 and Columns 11 through 16, respectively.

In Figure~\ref{fig3}, we show the distribution on the sky of the 196
galaxies given in Tables~\ref{tbl1} and ~\ref{tbl2}.  The half circle shape
observed on the data is due to the availability of only one spectrograph.
The presence of the NGC 4839 group is obvious in the southwest region of
the cluster. The plotting symbols reflect the 5 different galaxy types
defined above.

\subsection{Principal Component Analysis}
\label{PCA}

To quantify the visual classification discussed above, we have implemented
a Principal Component Analysis (PCA) of our Coma spectra (Mittaz et al
1990; Connolly et al 1995). The basis of this method consists of describing
a multi--dimensional distribution of variables with a minimum number of
dimensions.  In our case, we are searching for the minimum number of
eigenspectra needed to describe the whole spectral dataset.  We can
thereafter classify each spectrum as a function of these eigenspectra. We
have used here a version of the official SDSS spectroscopic reduction
software which already implements the PCA analysis of Connolly et
al. (1995).

\subsubsection{Data Preparation}

As mentioned before, only the blue spectra were used for the PCA analysis.
To prepare the spectra, we first interpolated over the [OI] $\lambda$5577
and NaD $\lambda\lambda$5890, 5896 doublet sky emission lines since many of
the spectra contained significant residuals of these strong sky lines.
Next, each spectrum was blueshifted to the galaxy's restframe and rebinned
to the wavelength range $3750 \rightarrow5900$ {\AA} at a dispersion of
1.125 {\AA} per pixel. Then, we subtract the continuum from the spectra
using a fourth order polynomial. We have also carried out the PCA analysis
without continuum subtraction, but we obtain better discriminating power to
reproduce the line strength classification employing the continuum
subtracted sample.  Finally, we smoothed each spectrum with a Gaussian with
a width set by the spectrograph resolution of $R=1800$ which produced a set
of 673 pixel spectra, binned over a common wavelength range, which could
then be used as input to the PCA algorithm.

\subsubsection{Eigenspectra Derivation}

Using these prepared spectra, we performed the PCA methodology presented in
Connolly et al (1995) and Connolly \& Szalay (1999).  We present a very
brief summary below, but for a more detailed description see Connolly et al
(1995). We regard each spectrum as a vector, $f_{\lambda}$, where $\lambda$
is an index running on wavelength. We represent the whole spectra set by a
matrix $f_{\lambda,i}$, where the index $i$ runs through the individual
spectra. We apply a uniform normalization to each spectrum, so that the
weighted sum of the squared flux is unity. Thus, our normalized spectra are
computed as

\begin{equation}
F_{\lambda,i} = \frac{f_{\lambda,i}}
{\sqrt{\sum_{\lambda}^{} f_{\lambda,i}^2 \, W_{\lambda} }}
\end{equation}
 
In all cases a uniform weight was applied, so that $W_{\lambda} = 1$ for
all wavelengths $\lambda$.  We want to construct an orthogonal basis for
the space spanned by the spectra, which we call eigenspectra. We derive
these eigenspectra by diagonalizing the correlation matrix $C_{ij}$, given
by

\begin{equation}
C_{ij} = \sum_{\lambda}^{} F_{\lambda,i} F_{\lambda,j} \, W_{\lambda}
\end{equation}

By diagonalizing the correlation matrix, we find the matrix $R$ that
produces the diagonal matrix $\Lambda$ whose components are the
eigenvalues, $\gamma_i$, of the orthogonal eigenspectra, $e_{\lambda,i}$,

\begin{eqnarray}
R^{\dagger} C R & = & \Lambda \\
R^{\dagger}_{i,j} (\sum_{\lambda}^{} F_{\lambda,i} F_{\lambda,j}) R_{i,j} \
	& = & \Lambda_{i,j} \nonumber \\
R^{\dagger}_{i,j} F_{\lambda,j} F_{\lambda,j} R_{i,j} & = & \
e_{\lambda,i} e_{\lambda,j}^{\dagger} \;=\; \gamma_i\, \delta_{i,j}\nonumber\\
\Rightarrow R^{\dagger}_{i,j} F_{\lambda,j} & = & e_{\lambda,i} 
\end{eqnarray}

In Figure~\ref{fig7}, we show the distribution of the first 20
eigenvalues. The two curves are the percentage contribution of the N$^{th}$
eigenvalue (decreasing curve on a log scale) and the percentage
contribution of all the eigenvalues less than or equal to N (the increasing
curve on a linear scale). As one can see the relative contribution of each
eigenspectrum falls off rapidly; it is possible to account for $\sim90\%$
of the variance in our sample using a reasonably small number of
eigenspectra ($\le 10$). The remaining $\sim10\%$ of the variance is
probably mostly noise in the data, but may also partly be due to complex,
real variations in the galaxy spectra.  Following the methodology of
Connolly et al. (1995), we limit our classification to the first three
components, as these eigenspectra contain virtually all the usable
information, from a classification standpoint, as we will illustrate
below. In our analysis, these three components account for 87\% of the
variance in our sample.  By contrast, the first 2-3 components for
eigensystems derived from synthetic spectra (e.g. Connolly \& Szalay 1999,
Ronen et al. 1999), generally account for 98-99\% of the variance, while
the preliminary eigensystem derived for the 2dF Galaxy Redshift Survey
(Folkes et al 1999) contained about 66\% of the total variance within the
first three components. Our galaxy sample lies between these two extremes
in terms of complexity and signal to noise, so this result is qualitatively
consistent with these other works.

These first three eigenspectra are displayed in Figure~\ref{fig8}. The
first eigenspectrum represents the mean spectrum in our sample, so as
expected it is a typical elliptical galaxy spectrum. It has spectral
features typical of an old stellar population {\it e.g.}, the strong
absorption lines of Ca II H \& K, G band and magnesium. The second
eigenspectrum contains strong Balmer absorption and emission lines, and
strong [OIII] emission.  This component will clearly correlate with current
star formation.  The third eigenspectrum again contains strong Balmer
absorption, but now the strong H$\beta$ and O[III] lines are shown in
absorption.  The eigenspectra for eigenvalues beyond the top three
components do not show major spectral features, so we believe these are
mostly shaped by the noise in the data. Also, the eigencoefficients for
these higher order components do not show any significant trends with our
visual spectral types above.

\subsubsection{Galaxy Classification}
The eigensystem can be used to define eigencoefficients for each galaxy
spectrum, which are simply the dot products of the galaxy spectrum with
each eigenspectra. Since we are using uniform weights, the relationship is
straightforward. For example, the $i^{th}$ eigencoefficient of the $j^{th}$
galaxy spectrum is simply:

\begin{equation}
c_{i,j} =  \sum_{\lambda}^{} e_{\lambda,i} f_{\lambda,j} 
\end{equation}

The resulting set of coefficients are normalized such that sum of their
squares is one, that is, $\sum_{}^{} c_{i}^2 = 1$, where the summation
index i runs through all eigencoefficients.  In Figure~\ref{fig9}, we show
the distribution of the first, second and third
eigencoefficients. Immediately one can see that the different spectral
types cluster into specific areas of eigenspace.  The relationship between
the three eigencoefficients is best related by using a projection scheme as
in Connolly et al. (1995), where three coefficients are converted into
spherical coordinates.  By assuming that all the useful information is
contained within the first three components, we throw away the remaining
coefficients and then treat the three top coefficients as a vector in a
three dimensional space and convert them into a spherical coordinate
system. In this view, the spectra can be described very simply by the two
spherical angles $\theta$ and $\phi$ (see Figure~\ref{fig10}). The relative
lengths of the vectors are not important for classification, since they
only represent how well each spectrum is represented by the three
components.

Figure~\ref{fig10} shows the distribution of projected angles. The dashed
lines show where we have made cuts based on the aforementioned
classifications except that we have merged the {\bf EM+AB} and {\bf EM}
types into one class. In figure~\ref{fig11}, we show the composite spectra
from each of our four remaining classes (these composites are simply
averages of the spectra).  The cuts we have made in Figure~\ref{fig10} are
arbitrary and were chosen to reproduce the line strength classification
scheme.  The spectra have a continuum of possible shapes, which follow as a
function of these parameters. Figure~\ref{fig10} shows that the angle
$\theta$ correlates with the strength of the Balmer features in absorption
and therefore can be used to differentiate post--starburst galaxies. The
angle $\phi$ correlates with the emission line strength and the K--star
spectral features thus being a good discriminator of the overall star
formation activity.  The dependence on the K--star spectral features can be
seen in Figure~\ref{fig11} where we investigate a further division of the
{\bf AB} class into two sub--groups {\bf ABa} and {\bf ABb}, along a line
parallel to the {\bf PS} boundary. The {\bf ABb} class have stronger Balmer
absorption but weaker MgI and G band features than the {\bf ABa} type.

We present the PCA classifications in Column 3 of Table \ref{tbl2}. For
completeness, we also provide the coefficients of the PCA used in this
paper: Columns 11 to 13 contain the top three eigencoefficients, Columns 14
\& 15 are the $\theta$ \& $\phi$ angle coefficients, and the square of the
radius for each spectrum is given in Column 16.  We do not include any
information on the eigencoefficients in the case where the continuum is not
subtracted as the results are qualitatively the same although
quantitatively they differ.

\subsection{Wavelet Classification}

In addition to the PCA analysis on the Coma spectra, we have also
investigated the use of Wavelets, since they provide a multi-resolution
approach of classifying spectra that allows us to study -- in a single
analysis -- both the small--scale emission/absorption features together
with low-frequency continuum information.  The goal of our wavelet analysis
was to automate the line strength classification given above using as few a
number of wavelet coefficients as possible thus providing an overall
compression of information as well as producing an objective classification
scheme that could be replicated elsewhere. For the analysis discussed
herein, we have used the IDL Wavelet Workbench\footnote{Based on Wavelab at
Stanford University by Donoho, Johnstone, Buckheit, Chen \& Scargle. See
ftp.rsinc.com/pub/user\_contrib/wwb}.  We do not provide here a detailed
explanation of the theory behind Wavelet since there now exists a large
volume of literature on this subject (see, for example, Press et al. 1992).

To prepare the data for our wavelet analysis, analogously to the PCA
analysis, we first shifted all spectra to the galaxy's rest--frame and
define a common (rest--frame) wavelength range ($3890 \rightarrow5901$
{\AA}) which only uses data from the blue side of the SDSS spectrograph.
We then computed the average pixel value for the whole spectrum and
subtracted this off each pixel thus re--normalizing the spectrum to have a
mean of zero. This procedure removed the large discontinuities at the ends
of the spectrum.  The final step was to ``zero--pad'' the spectrum on
either side of the data to create a pixel array with an integer power of
two (2048 in this case). We then applied cosine--bell smoothing to the
edges of the original spectrum to gradually taper the original data to zero
thus avoid sharp edge--effects at the ends of the spectrum which would
result in ``ringing'' or the Gibb's phenomena. For our spectra, we smoothed
5\% of the original data at either end of the spectrum.

We note here that as in the PCA analysis above (Section~\ref{PCA}), we did
not use the red side of the spectrum because of the problems discussed in
Section~\ref{observe} and thus we have no information about H$\alpha$ in
emission. However, for the wavelets, we did retain the shape of the
continuum which is different from the PCA analysis. The multi-resolution
nature of the wavelets does not require continuum-subtracted
data. Nonetheless, we also performed the same wavelet analysis with
continuum subtracted spectra obtaining different parameters values but a
similar classification using the same strategy as described below.
  
We then performed a wavelet transform on these spectra which returned a
series of wavelet coefficients at ten different resolution levels; the
number of resolution levels is set by the size of the padded data array
since each level has a factor of two lower resolution than the previous
layer. An example of such a wavelet transform for one of the SDSS Coma
spectra is shown in Figure~\ref{fig12} along with the different resolution
levels and the corresponding amplitude of the wavelet coefficients at those
resolution levels.

The first decision faced in the implementation of our wavelet analysis was
the choice of the Mother Wavelet {\it e.g.}, a Daubechies, Symmlet or
Coiflet wavelet function.  Moreover, we also needed to choose the level of
smoothness for these wavelets {\it e.g.}, Daubechies2 (which is just a Haar
Wavelet), Daubechies4, Daubechies8 {\it etc.}, with the higher numbered
wavelets being smoother versions of the Mother Wavelet. To aid in these
decisions, we performed a series of empirical tests on the data to
determine which wavelet provided the maximal compression of the information
for the smallest number of coefficients used in the reconstruction of the
data. In these tests, we re--constructed the spectra using the largest $N$
coefficients (in absolute value), regardless of which level they were taken
from, as well as implementing the hard thresholding scheme {\it i.e.}, all
coefficients below the $N$ largest--valued coefficients were set to zero in
the re-construction.

In Figure~\ref{fig13}, we show the measured mean $\chi^2$ for all 196 Coma
spectra as a function of $N$, the number of wavelet coefficients used in the
reconstruction of the spectra. For our thresholding scheme, the Daubechies4
wavelet provided the most information compression, or largest fractional
change in $\chi^2$, for a given $N$ coefficients. As expected, the $\chi^2$
decreases exponentially as a function of $N$, and for small values of $N$, the
Daubechies4 clearly provides significant gain over the Symmlet wavelets.
Therefore, we have used this Mother Wavelet (Daubechies4) in our analysis of
these SDSS Coma spectra. This Mother Wavelet is rather sharp compared to the
other wavelets which is an advantage for quantifying sharp discontinuities in
these spectra {\it e.g.}, the $4000$ {\AA} Balmer break discontinuity.

We note here that we are simply used the relative change in $\chi^2$ to
differentiate between the Mother Wavelets available and the absolute
goodness--of--fit from the $\chi^2$ statistic is not being used because these
spectral data are slightly correlated (due to the sky--subtraction and the
resolution element being approximately 3 times the dispersion of these
spectra). For reference however, we note that each spectrum had, on average,
2000 degrees--of--freedom and thus a reduced $\chi^2$ equal to one, which
suggests a reasonable fit between the model and observed data, is achieved
after adding $\simeq30$ wavelet coefficients (see Figure~\ref{fig13}). The
fact we can explain these spectra with a relatively small number of
coefficients agrees with our visual inspections and is reasonable given that
we are studying a rather homogeneous subsample of galaxy spectra {\it i.e.} a
majority of the spectra are dominated by older spectral populations. We
caution the reader to not over--interpret the absolute $\chi^2$ results since
we have not accounted for the correlations in the data.
 
In Figure~\ref{fig14}, we plot the amplitude of the second and third
largest wavelet coefficients for all 196 Coma spectra presented in Table
\ref{tbl1}.  We have ignored the largest wavelet coefficient since it
simply measures the amplitude of the whole spectrum and thus contains no
information about the general shape the spectrum (this is because a wavelet
is defined to sum to zero). Figure~\ref{fig14} demonstrates that we can
extract worthwhile information about these spectra from just two of wavelet
coefficients; galaxies with recent star--formation, as indicated by the
presence of emission lines or post--starburst characteristics (A star
spectrum), preferentially possess large (negative) second and third wavelet
coefficients, while absorption, non--star--forming, galaxies populate the
remainder of the phase space. The explanation for Figure~\ref{fig14} is
that these two wavelet coefficients are measuring the ``color'' of the
continuum and are providing a similar discrimination as one may obtain from
broadband multi--color photometry. This is exemplified by the fact that all
the post--starburst galaxies in Figure~\ref{fig14} reside in the low--left
arm since they all have ``blue'' spectra.  In Figure~\ref{fig15}, we show
one of the post--starburst galaxies in Coma along with the re--constructed
spectrum based on the three largest wavelet coefficients which illustrates
that we have just measured the tilt of the spectrum.  In this case, we have
not exploited any high--frequency emission or absorption features, but have
still managed to broadly segregate the galaxies based on their recent
star--formation histories.

To extend our analysis further, we have exploited the multi--resolution
nature of wavelets by designing an algorithm to classify galaxies as
emission line, post--starburst and absorption line galaxies using as few a
number of wavelet coefficients as possible.  Here, we have merged the
different sub--divisions of the emission lines as discussed in
Section~\ref{lines} and focus on the {\bf AB}, {\bf PS} and {\bf EM}
classes.  To help define our algorithm, we co-added all the obvious
post--starburst ({\bf PS}) Coma galaxies (based on our visual
classification scheme) and identified the strongest features in that
co-added spectrum. In Figure~\ref{fig16}, we show this summed spectrum and,
as expected, the Hydrogen Balmer absorption series is strong because of the
young A stars in the galaxy. Also visible are the Calcium H \& K absorption
features (with the $4000\AA$ break Balmer discontinuity) which is
indicative of an older stellar population (K star spectrum). As discussed
by Dressler \& Gunn (1992) and Zabludoff et al. (1996) these
post--starburst galaxies are also known as {\rm E+A} or {\rm K+A} galaxies
{\it i.e.}, Elliptical plus A star or K plus A star spectra.

The algorithm we have adopted is as follows. For each spectrum, we first
scanned for the presence of any emission lines. This was empirically found
to be equivalent to having one of the eight largest coefficients be a
negative value located in the first five resolution levels {\it e.g.},
levels with more that 32 wavelet coefficients per level.  We further
restricted the search to look at wavelengths $>4000$ {\AA} thus avoiding
problems with the Balmer discontinuity.

To identify post--starburst galaxies, we then re-scanned the first twelve
largest coefficients for positive wavelet coefficients near to the
wavelengths of H$\delta$, H$\gamma$ and H$7$ (see Figure~\ref{fig12} for an
example of this algorithm).  These features were used since they are the
strongest of the Balmer absorption lines as shown in Figure~\ref{fig16}. We
did not use H$\beta$ because of possible contamination from emission
especially if H$\alpha$ is present about which we have no information in
the current application of the method.  If three such coefficients were
detected, the galaxy was defined to be a post--starburst galaxy.  We
present these automated classifications in Column 4 of Table \ref{tbl2}.

\subsection{Synthetic Models}

As a final test, we have fit synthetic models of galaxy stellar populations
to the Coma spectra. The advantage of this approach is that the spectra can
be directly related to physical phenomena and parameters. The disadvantage
is that the parameterization of the model, {\it e.g.}, initial mass
function (IMF) or the star formation history of the stellar population
studied, can differ substantially from the real star formation processes
thus introducing large uncertainties based on the particular model
used. Moreover, it is difficult to compare this scheme to the
aforementioned three empirical classification schemes. Despite these
caveats, we present this work here which will be primarily used in a
forthcoming companion paper.

Our method for comparing the stellar models with our Coma spectra is
to minimize the following $\chi^2$ function,

\begin{equation}
\chi^2(a_j) = \sum_{i=1}^n  \left(\frac{(F_{i\;observed}^{\lambda} -
F_{i\;model}^{\lambda}(a_j))} {\sigma_i(F_{i\;observed}^{\lambda})}\right)^2
\end{equation}

\noindent where $a_j$ are the free parameters of the synthetic spectrum
being fitted, $i$ is an index running through pixels,
$F_{i\;observed}^{\lambda}$ and $F_{i\;model}^{\lambda}(a_j)$ represent
each pixel of the spectral energy distributions observed and modelled
respectively and $\sigma_i(F_{i\;observed}^{\lambda})$ the error of the
observed spectrum in pixel $i$.

In practice, the synthetic spectral energy distributions (SEDs) have a
lower spectral resolution than our observed Coma spectra. Thus, we smooth
both energy distributions to the same resolution; this means that the
errors of the observed spectrum are therefore no longer uncorrelated. We
also mask the spectral regions around strong sky emission lines to avoid
the residual spikes coming from sky subtraction.

We used two sets of synthetic spectral energy distributions.  First, we
employed the evolutionary synthesis model PEGASE (\cite{FRV97}) to generate
the model spectra. We minimized the $\chi^2$ using three free parameters:
an overall normalization, the age of the stellar population and the star
formation prescription. We assumed solar metallicity and did not fit for
the internal extinction. The star formation prescription was parameterized
using two different laws: 1) we assumed that the star formation rate (SFR)
depends on the galaxy gas fraction as $\Phi(t)=\nu\,f_g(t)$ and we
minimized with respect to the astration rate $\nu$. In fact, we minimized
this astration rate parameter using 8 discrete values. Following Fioc \&
Rocca-Volmerange, we used the Rana \& Basu (1992) IMF and values of the
astration parameter that reproduce the colors of the different spectral
type galaxies observed locally. We also added a one Gigayear starburst
model.  2) we assumed an exponential SFR law
$\Phi(t)=\tau^{-1}\,exp(-t/\tau)$ and minimized with respect to the time
scale $\tau$.  In this case, we used a standard Salpeter IMF with lower and
upper mass limits of 0.1 and 120 $M_{\odot}$, respectively.

The age and star formation prescription of the best fitting model spectrum
are presented in Table~\ref{tbl2}. For the first SFR prescription we
provide the age (Column 5), the number of the best fitting model (Column 6)
and the reduced $\chi^2$ (Column 7) of this best fit. The numbering of the
models presented in Column 6 is the following: number 1 corresponds to a
one Gigayear starburst model; numbers from 2 to 8 correspond to decreasing
values of the astration rate parameter $\nu$, where large $\nu$ values
resemble star formation histories typical of early-type galaxies and small
$\nu$ values, typical of late-type galaxies. The degrees of freedom of the
fit depend on the actual masking of each spectrum but for the majority of
them, there are 2005-2020 d.o.f. We note here that our figure of merit, the
$\chi^2$ we compute, does not strictly behave as a real $\chi^2$
distribution. It is certainly close to one, but not exactly given that our
errors have not been determined with high accuracy and there has been some
smoothing which we have neglected not taking into account the correlated
nature of the errors.  For the second SFR prescription, we present the age
(Column 8), the time scale $\tau$ (Column 9) and the reduced $\chi^2$
(Column 10).

\section{Discussion and Conclusions}

\subsection{SDSS Spectroscopic System}

We present in this paper the first extra--galactic spectroscopic data taken
by the SDSS during the initial spectroscopic commissioning run in June
1999. Although the data was primarily obtained to commission all the SDSS
spectroscopic subsystems (guider, plug--plates, spectrographs, control
software), and are therefore far from the nominal quality expected for the
SDSS when it is fully operational, these Coma spectra are of reasonable
quality and provide a new insight into the galactic content of the Coma
cluster, especially at large distances from the cluster center. This is a
combination of the high throughput of the SDSS spectrographs, the large
field of view of the spectroscopic observations ($3^{\circ}$ diameter), the
extensive wavelength coverage and resolution, and the ability to obtain
many hundreds of spectra simultaneously. This single plug--plate
observation of Coma illustrates the overall efficiency of the SDSS
spectroscopic system and demonstrates the potential for further cluster and
field galaxy observations.

As of October 2000, the entire SDSS spectroscopic system has been tested
and commissioned. Both spectrographs are operational and are routinely
collecting data. The target selection procedure, from determining target
objects through to drilling plug-plates and shipping them to APO, has been
exercised several times and the emphasis has now shifted to improving the
overall efficiency of the observations. At the timing of writing, the SDSS
has obtained $\simeq50,000$ spectra (approximately two thirds galaxies and
one third quasars) and these data are being used, amongst other things, to
define our integration time, completeness limits as well as debug the SDSS
spectroscopic software.  These data also represent one of the largest
samples of galaxies and quasars presently in existence and are being used
for several scientific programs.

We have obtained 320 spectra (half the number of a nominal SDSS plug-plate)
in the central region of the Coma cluster. The availability of one
spectrograph only allowed us to observed the South half of plate 133 (see
Figure~\ref{fig3}). Other engineering tasks and bad weather conditions
prevented us from obtaining further Coma spectra and future data taken in
this region will be acquired as part of the regular SDSS survey (both
imaging and spectroscopy). Nevertheless, in one hour of observations with
only half of the spectroscopic system commissioned, and without optimizing
the target selection, we obtained 196 spectra of Coma galaxies. This number
represents approximately $\frac{1}{3}$ of the number of redshifts presently
available in the literature within the same 1.5 deg radius from the cluster
center (as estimated from the NED database in August 2000). Amongst the 320
fibers available, 11 were placed on blank sky (to allow sky-subtraction
from the rest of the fibers), while the rest of the science fibers yielded
196 Coma galaxies spectra, 49 field galaxies, 5 quasars (which were
previously known and targeted on purpose), 47 stars and 12 unidentified
spectra.  Four of the unidentified spectra were due to ``spill-over'' from
bright adjacent fibers$\footnote{ Fibers are only separated on the CCD
camera by approximately 6 pixels. The FWHM of a fiber on the CCD is
approximately 2 pixels. Therefore, the light from bright target fibers can
{\it contaminate} adjacent fibers.}$.  We note here that our target
selection attempted to maximize the number of available galaxy targets by
including all candidate galaxies {\it i.e.}, galaxies that were only
classified as galaxies on one of the two photographic plates scanned by
SuperCOSMOS.  In hindsight, this produced considerable stellar
contamination even at bright magnitude, but does ensure a higher level of
completeness.

Of the 196 Coma redshifts obtained, 45 represent new identifications.
Because of the SDSS large, circular field--of--view, these new redshifts
are located at large cluster radii in areas scarcely sampled by other
spectroscopic surveys of Coma, {\it i.e.}  the difference between a survey
instrument and a more targeted observation focused on a particular science
goal. Given this uniformity, and the quality of our data, we are thus able
to obtain a better understanding of the spatial distribution of galaxy
classifications within Coma.

For the 151 redshifts found in common with the literature, we found a
negligible mean offset between our redshift determinations and those
already published.  In particular, when we compared our redshifts to the
largest homogeneous sample of redshift available in the literature, the 63
galaxies measured by CD96, we find only a 3 ${\rm km\,s^{-1}}$ mean offset
with an RMS scatter of only 24 ${\rm km\,s^{-1}}$ (this is less than 2 SDSS
spectrograph pixels).  This is remarkable given the non--optimal observing
conditions, the lack of proper calibration frames, the difference in the
spectral resolutions of the two instruments used, as well as the
non--standard wavelength calibration used. Moreover, we have looked for
correlations between the observed difference between CD96 and our redshift
as a function of CCD position, plate position and matching distance and
find no significant dependences with these observational parameters.  In
summary, it is likely that we have over--estimated the error due to our
wavelength calibration given the low RMS difference with CD96; the SDSS
spectroscopic systems appear to be remarkably stable and linear over a one
hour observation.

\subsection{Dynamics and Spectral Classifications}

In Figure~\ref{fig5}, we show the redshift distribution of our 196 Coma
members which is qualitatively similar to that seen for the 465 redshifts
in CD96 (their Figure~5).  This is not surprising since we have 148
galaxies in common with their analysis.  However, we do sample out to
larger radius than CD96 and the fraction of galaxies measured in the NGC
4839 group (in the SW portion of the cluster), with respect to the total,
is larger in our sample than in CD96. A single Gaussian fit to our data
gives a mean velocity of $6950\pm66\: {\rm km\,s}^{-1}$ and a velocity
dispersion of $\sigma_v=916\pm50\: {\rm km\,s}^{-1}$, where we have
corrected for the fact that we measure the velocity dispersion with a
3-$\sigma$ clip criterion. These values are in agreement with previous
measurements and, as expected, are an average of the mean velocities, and
mean velocity dispersions, of the main cluster core and the NGC 4839 group
(see CD96). As already noted by CD96, a single Gaussian fit to the data is
an over-simplification with a clear excess of galaxies, compared to the
single Gaussian, between $\simeq 6950 \rightarrow 7600\: {\rm km\,s}^{-1}$
due to the NGC 4839 group. The central velocity of these excess galaxies is
approximately coincident with the mean redshift quoted by CD96 for the NGC
4839 group, $7339\: {\rm km\,s}^{-1}$. We have also investigated the
dynamical differences between the absorption line galaxies and the active
(emission--line and post--starburst galaxies) galaxies. We find that the
{\bf AB} galaxies given by the line strength classification have a mean
velocity of $6923\pm70\: {\rm km\,s}^{-1}$ and a velocity dispersion of
$\sigma_{\rm v}=846\pm54\: {\rm km\,s}^{-1}$, while the rest of the
galaxies have $<cz>=7027\pm160\: {\rm km\,s}^{-1}$ and $\sigma_{\rm
v}=1160\pm134\: {\rm km\,s}^{-1}$. The velocity dispersion of the active
galaxies is then 30\% higher than that of the absorption line
galaxies. This is similar to the difference found by CD96 for Coma and
Carlberg et al (1997) when comparing the velocity dispersions measured from
the red and blue galaxies in the CNOC1 cluster survey.  We defer a more
detailed examination of the dynamics of the whole Coma region until the
future when we have both high quality photometric and spectroscopic data
from the main SDSS survey in hand.

In this paper we present several different spectral classifications for our
196 Coma galaxies. The homogeneity and quality of the data allowed us to
experiment with these different classification schemes to understand their
applicability to the main SDSS survey. The first classification scheme was
based on the presence or absence of emission lines. In this case, the
H$\alpha$ line served as the main classifier. For typical IMF and
metallicities, H$\alpha$ cannot be detected in emission $\sim 25$ Myr after
the end of active star formation. Therefore, this criterion efficiently
separates galaxies with current star formation from those without.  The
strength of the H$\beta$ line in emission, and the strength of the
H$\delta$, H$\gamma$ and H$\beta$ in absorption, are used as additional
diagnostics to improve the classification (see Section \ref{lines}). Using
these specific lines, we were able to sub--divide the galaxies into 5
spectral types which we discuss below.

Soon after the termination of a star formation episode, even if it only
includes a few percent of the stellar population, the blue part of the
optical spectrum of a galaxy is dominated by A stars which are
characterized by the Balmer absorption features. If there is no further
star formation activity the strength of these lines diminishes gradually
with time. The Balmer absorption lines strength thus provides a rough
estimate of the age of the stellar population, although the lines are also
somewhat sensitive to metallicity. We use this criterion to separate
absorption line systems into two sub--classes.  Galaxies without recent
star formation, based on the absence of H$\alpha$ in emission and the
absence of strong Balmer absorption, were classed as {\bf AB}. Galaxies in
which their last episode of star formation ended recently, or
post--starburst galaxies ({\bf PS}), were characterized by their strong
Balmer absorption lines as well as the absence of H$\alpha$ in emission.
The threshold between {\bf AB} and {\bf PS} was set at an equivalent width
larger than 15~{\AA} for the sum of H$\delta$, H$\gamma$ and H$\beta$
equivalent widths. This corresponds to roughly 400-700 million years after
the truncation of the active star--formation for solar metallicity.  We
note here that our criteria are based on equivalent widths; thus this
threshold also depends upon the underlying stellar population (or the
previous star formation history) {\it i.e.}, how many stars are
contributing to the continuum.  Emission line galaxies were divided into
three sub--classes ({\bf AB+EM}, {\bf EM+AB} and {\bf EM}) depending upon
the strength of the H$\beta$ line. The $W_o$(H$\beta$) cuts are difficult
to translate into star formation rates given the different continuum
strengths and the fact that we have not attempted to correct the intrinsic
extinction. Nevertheless, the equivalent width of H$\beta$ represents a
measure of the strength of the star formation process with respect to the
underlying stellar population.

This classification scheme has the advantage of being simple and easy to
measure in optical spectra with sufficient wavelength coverage, and is
meaningful as it classifies spectra according to their star formation
activity relative to the overall stellar population. Undoubtedly, the use
of H$\alpha$, instead of H$\beta$, to subdivide the emission line spectra
would have been preferred, as it is a better indicator of star formation
and is less affected by extinction. However, the nature of our data made us
settle for H$\beta$. We can use this classification as the test base to
which compare our other schemes.

Other classification schemes based on spectral features have also been used
in the literature. In particular, the MORPHS group (Dressler et al 1999;
Poggianti et al 1999) and the CNOC group (Balogh et al 1999) have recently
put forward similar classification schemes based on the [OII] and H$\delta$
lines, in which they provide a more physical interpretation of their types
based on evolutionary spectral synthesis models. They use the [OII] and
H$\delta$ lines as indicators of current star formation and the time
elapsed after the last episode of star formation, respectively. The large
wavelength coverage of our spectra allow us to use more spectral features,
H$\alpha$, H$\beta$, H$\gamma$ and H$\delta$, for the same purpose. Here,
we have taken the phenomenological approach of providing a simple
classification scheme based on observational quantities which are directly
related to the star formation processes taking place within galaxies. We
defer to a future paper a more detailed interpretation of these classes in
terms of the evolutionary phases of galaxy stellar populations.

There are 44 galaxies (22\% of our sample) that show evidence of current
star formation to some degree. There are five extra galaxies, classified as
post starbursts, which ceased their star formation process in the last
$\sim500$ Myr. If we include these five galaxies into the active galaxy
population, then 25\% of our galaxy sample show some sign of recent star
formation activity.  In Figure~\ref{fig3}, we show the spatial distribution
of all 196 Coma galaxies; the plotting symbol type reflects our line
strength classification scheme discussed above and in Section \ref{lines}.
Clearly, there is no obvious correlation in the position of the different
types of galaxies except that the active star--forming galaxies avoid the
cluster center.

In addition to the line strength classification scheme, we have explored
two objective schemes based on PCA and wavelets. For our PCA analysis,
Figure~\ref{fig7} shows that the first three eigenspectra for our dataset
contains 87\% of the variance while most of the remaining variance is
likely due to the noise in our spectra. Each spectrum can therefore be
expressed as a linear combination of these three eigenspectra which in turn
can be presented in a spherical coordinate system with only two important
parameters (see Figure~\ref{fig10}). We have defined regions in this
two--dimensional space according to our line strength classification
scheme.

In Table~\ref{tbl3}, we compare the line strength and PCA classifications
to check whether the two--dimensional space defined by the eigencoefficient
projection angles can be understand in term of the absorption and emission
lines of the spectra. The agreement between both classifications is
obviously good but nevertheless remarkable as there was no a priori
guarantee that such a good separation should be achieved given that the PCA
classification is based on the overall continuum subtracted SED while the
line strength classification only on the Balmer lines. Figure~\ref{fig10}
demonstrates that simple arbitrary cuts can efficiently separate galaxies
according to their spectral line strengths.  The only cut which does not
clearly reproduce the line strength classification above is the division
between {\bf AB} and {\bf AE} galaxies. In fact this is mostly due to fact
that we have used H$\alpha$ to split these two classes in the line strength
classification but have only used the blue part of the spectrum in the PCA
analysis. Some of these galaxies show H$\alpha$ in emission while the rest
of the Balmer series is seen in absorption. PCA thus classifies the galaxy
as an {\bf AB} galaxy while in the line strength classification it is an
{\bf AB+EM} galaxy based on the presence of faint H$\alpha$ emission.  In
addition, two galaxies classified as {\bf AB+EM} by their line strengths
are classified as {\bf PS} by the PCA analysis (fibers 91 and 140). Both
galaxies show strong broad Balmer absorption features which are filled with
narrower emission. In these cases, it is difficult to separately measure
the absorption and emission line components equivalent widths. We estimate
that galaxy 91 has $W_o(H{\delta}+H{\gamma}+H{\beta}) \sim 10$~{\AA} in
absorption and galaxy 140 has $W_o(H{\delta}+H{\gamma}+H{\beta}) \sim 15$
{\AA}. The latter would have been classified as a {\bf PS} galaxy by the
line strength criteria if it did not have the emission. These two galaxies
have certainly experienced star formation activity that has been
substantially decreased in the last Gigayear thus giving rise to the strong
Balmer absorption lines. However, there is still some residual star
formation activity going on and thus the weak emission on top of the
absorption.

We have used a version of the PCA based on the implementation presented in
Connolly et al (1995) and Connolly and Szalay (1999).  Our method uses the
data themselves to generate the eigenbasis in which the observed spectra
are projected. We then use another classification scheme, the line strength
classification, to subdivide the eigencoefficent parameter space according
to it.  The line strength classification can be translated strikingly well
into separated regions on the two-parameter space given by the angles
$\theta$ and $\phi$. Other classification schemes employing PCA have also
been used in the literature. For example, Bromley et al (1998) use a PCA to
classify the Las Campanas Redshift Survey. Their implementation is very
similar to ours. They use the data themselves to generate the eigenbasis as
we do. They weigh their spectra by a smoothed version of the mean spectrum,
while we continuum subtract each spectrum with a low order
polynomial. Their sample is composed of field galaxies instead of galaxy
cluster members and therefore their first eigenspectrum (their Figure~1) is
more emission line dominated. Unlike us, they only use their first two
eigencoefficients to classify their spectra, neglecting the third
eigencoefficient that appears to be able to differentiate post-starburst
galaxies. They subdivide their two eigenspectra parameter space into six
``clans'' or spectral types, without using any other external
classification scheme for that subdivision. Their mean spectra for their
clans show (their Figure 3), nevertheless, a good correlation with star
formation. Post-starburst galaxies are, however, not represented. Folkes et
al (1999) classify a preliminary sample of the 2dF galaxy survey using a
PCA as well. Their sample is made up of field galaxies. Their PCA
implementation is similar to us except that they subtract a mean spectrum
from their spectra before applying the PCA. Their eigenspectra (their
Figure 5) reflect both facts with their mean spectrum being similar to our
first eigenspectrum (modulo our continuum subtraction), their first
eigenspectrum being similar to our second and their second similar to our
third. They use another external classification to subdivide their first
two eigencoefficient parameter space to generate their classification. They
argue that their third eigencoefficients do not add much additional
information for their classification scheme. The ESO-Sculptor Survey has
also been classified with a PCA method by Galaz \& de Lapparent
(1998). They employ the Connolly et al (1995) implementation, that is, the
same method we use. Again, their eigenspectra are more emission line
dominated than ours as they study a field galaxy population. They decompose
their spectra into the same projection angles we do. However, they classify
their spectra comparing them to the Kennicutt (1992) spectra projected onto
the same eigenbasis generated by the data themselves. They find that their
spectra can be interpreted as a one parameter family in their
$\theta$-$\phi$ space. Their sample is of comparable size and
signal-to-noise to ours but of coarser resolution. On the other hand, both
Bromley et al (1998) and Folkes et al (1999) samples are larger and contain
noisier spectra than ours. Their classification (as well as Galaz \& de
Lapparent's) rely on the spectra being a one parameter family going along
stellar activity strength. Our spectra, however, being a cluster sample,
has more absorption line galaxies and more post-starburst galaxies that
require another parameter to be differentiated from the star forming
sequence. We argue that our classification requires two parameters: one,
describing the stellar formation activity; the other, the strength of the
Balmer absorption features.

There have been other approaches to construct classification schemes using
PCA. Connolly et al (1995), Connolly \& Szalay (1999), Sodr\'e \& Cuevas
(1997) and Ronen et al (1999) use spectral synthesis models (or galaxy
spectra already classified) to construct their PCA eigenbases. They, then,
interpret their classifications in terms of their input, either their
already-classified spectra or the synthesis model parameters.  Given that
spectral synthesis models currently available do not have the spectral
resolution of our observed spectra, we have chosen to use our own spectra
to build our eigenbasis and apply the line strength classification to
subdivide the eigencoefficient parameter space. Therefore, not losing the
spectral resolution of our data.

We have also used wavelets to define a simple classification for these
spectra which uses both the absorption and emission lines in conjunction
with the continuum shape. As with the PCA analysis, we only utilize the
blue part of the spectrum and thus have no information on H$\alpha$. We
also do not try and differentiate between the {\bf EM}, {\bf AB+EM} and
{\bf EM+AB} line strength types since, at present, we cannot measure the
equivalent width of lines directly from the wavelet analysis (although in
principle this could be achieved). In Table~\ref{tbl4}, we compare the
wavelet classification to the line strength classification. As above, there
is good agreement between the two where the line strength sub--classes of
{\bf AB+EM} and {\bf EM+AB} classifications are mostly shared between the
{\bf EM} and {\bf AB} wavelet classes (19 {\bf AB+EM} galaxies were placed
in the wavelet {\bf AB} class as expected since there is no information
about H$\alpha$). The only surprises in Table~\ref{tbl4} are the four
galaxies (fibers 84, 138, 201 \& 244) classified as {\bf AB} in wavelet
space but {\bf EM+AB} by the line strength criteria. A visual inspection of
these spectra shows that the signal--to--noise of these data is, on
average, low. However, all four have moderately weak emission lines that
probably missed the top 8 wavelet coefficients because of their size. All
four do have strong Calcium H \& K indicative of an old stellar
population. We also note that the wavelet algorithm has more {\bf PS}
galaxies than the line strength classification. This is primarily due to
the hard threshold of $W_o(H{\delta}+H{\gamma}+H{\beta})>15$ {\AA} imposed
in the line strength classification. These extra wavelet {\bf PS} galaxies
(three classified as {\bf AB} and 7 as {\bf AB+EM} galaxies by the line
strength method) do have Balmer absorption lines, but either they have
H$\alpha$ in emission and/or the Balmer absorption is weak (although still
stronger than galaxies dominated by old stellar populations). This apparent
discrepancy in the classification schemes can be easily understood in terms
of the thresholds used in assigning the {\bf PS} classification, {\it i.e.}
if the 15 {\AA} threshold used in the line classification scheme had been
lowered, then many of these galaxies would have also been classified as
{\bf PS} by the line strength scheme. The wavelet scheme has simply picked
extra {\it older} post--starburst galaxies with weaker Balmer absorption
lines.

In Table~\ref{tbl5}, we show the cross-comparison between the PCA and wavelet
classifications.  Again, the agreement is very good. The only exceptions are
the same cases discussed above given that the PCA classification cuts were
applied to reproduce the line strength classification. It is worth mentioning
here that the extra wavelet {\bf PS} classified galaxies discussed above
reside in the ``bridge of galaxies'' that join the {\bf AB} and {\bf PS}
regions shown in Figure \ref{fig10}. This again explains why we find more
wavelet classified {\bf PS} galaxies compared to the other schemes since we
are simply sampling more of the galaxy population between these two classes.

We have also fitted evolutionary spectral synthesis models to our Coma spectra
(only the blue side of the spectrum) to understand the composition and history
of the stellar populations in these galaxies. We note that these fits are
performed for a particular model, with a chosen star formation prescription,
and should be interpreted as such. The star formation history of a galaxy is
more complex that the simple parameterized models used (e.g., Abraham et al
1999) and the best fit gives only an indication of the properties of the
combined stellar population and helps to compress the spectral information
into a handful of meaningful parameters.

We have chosen to fit the age of the stellar population and a parameter
describing the star formation prescription using two families of models (see
section~7.4 for further details). Both parameterizations give similar results
(see table~\ref{tbl2}). It is nevertheless clear from table~\ref{tbl2} that in
some cases the star formation prescription utilized is unable to fit properly
the observed spectrum and a different prescription with combinations of
different bursts is necessary. We have chosen, at this stage of simple
comparison with other classification schemes, not to fit for the intrinsic
absorption or the galaxy metallicity, assuming fixed solar metallicity and no
intrinsic absorption. Indeed, there are known degeneracies between
age--metallicity and age--intrinsic absorption when fitting synthesized SED to
observed spectra (e.g., Worthey 1994; Thompson, Weymann \& Storrie-Lombardi
2000). In general, a higher metallicity and the effect of neglecting internal
extinction will be compensated by an older computed age.

Figure~\ref{fig17} presents the age and time scale, $\tau$, of the best fit
synthesized SED using an exponential SFR law. The symbols represent the
different line strength classifications while the size of the symbol
represented the observed magnitudes of the galaxies.  Most of the Coma
galaxies cluster along a strip in the age--$\tau$ space. The large dynamic
range shown for the $\tau$ parameter gives the impression that galaxies of
different line strength classes populate similar regions in this parameter
space. However, the general trends of age-$\tau$ values with respect to the
spectral line strength classification are as expected: {\bf AB} galaxies
populate regions of old ages ($\ge3-4$ Gyr) and short time--scales of star
formation which produce old stellar population while the {\bf AB+EM}, {\bf
EM+AB} and {\bf EM} galaxies occupy regions next to the {\bf AB} galaxies
but gradually moving to younger ages and longer $\tau$ as the emission
lines strength increases. Finally, the {\bf PS} galaxies populate the
region of short time scale star formation, which correspond to short and
intensive star formation histories, and short ages after the strong
formation episode. To help visualize these separate regions of the
age-$\tau$ space, we show in Figure~\ref{fig18} the 1, 2 and 3$\sigma$
contours of the age-$\tau$ regions occupied by three representative
galaxies drawn from our three main spectral types {\it i.e.}, an {\bf AB}
type galaxy (fiber 143), a {\bf PS} type galaxy (fiber 95) and a {\bf
EM+AB} type galaxy (fiber 117). It is clear from this plot that the {\bf
AB} and {\bf EM+AB} galaxy inhabit different regions of the age-$\tau$
space (at the 3$\sigma$ level) while there is no intersection for the {\bf
AB} and {\bf PS} galaxies either. Only the {\bf EM+AB} and {\bf PS}
galaxies share a similar parameter space (at the 3$\sigma$
level). Nevertheless, it is clear from Figures~\ref{fig18} and \ref{fig17}
that the fits of the exponential decaying star formation models to the
observed spectra are degenerate along a line in the log $tau$--log age
parameter space of slope $\sim$ 3/2. In summary, regardless of this
degeneracy, our spectral classifications have a broad relationship to
expected physical models of star--formation and therefore, they can be used
to study star--formation and galaxy evolution with the Coma cluster.

\subsection{Galaxy Star--Formation in Coma}

Our analysis of the Coma cluster has revealed that a substantial fraction
(25\%) of galaxies in Coma (out to the virial radius) exhibit recent
star--forming activity. Such analyses are at the heart of studies of galaxy
evolution in clusters and, in more general terms, the ``Butcher--Oemler''
effect (Butcher \& Oemler 1984).  Our large fraction of recent star--forming
galaxies is more than initially expected for such a low redshift cluster but
is in reasonable agreement with other spectroscopic studies of galaxy
star--formation in clusters. For example, Caldwell \& Rose (1997) found that
$\sim15\%$ of galaxies in five nearby clusters showed signs of recent
star--formation while the work of Ellingson et al. (2000) shows that
$\sim20\%$ of galaxies within the virial radius ($\sim r_{200}$) of CNOC1
clusters ($0.18<z<0.5$) show signs of recent star--formation. However, we do
not address here the redshift dependence of the BO effect since that will
require a large, homogeneous sample of clusters spanning a significant range
in redshift (See Andreon \& Ettori 1999; Margoninier \& Carvalho 2000). We
will address this issue using the main SDSS sample.

\begin{figure}
\centerline{\psfig{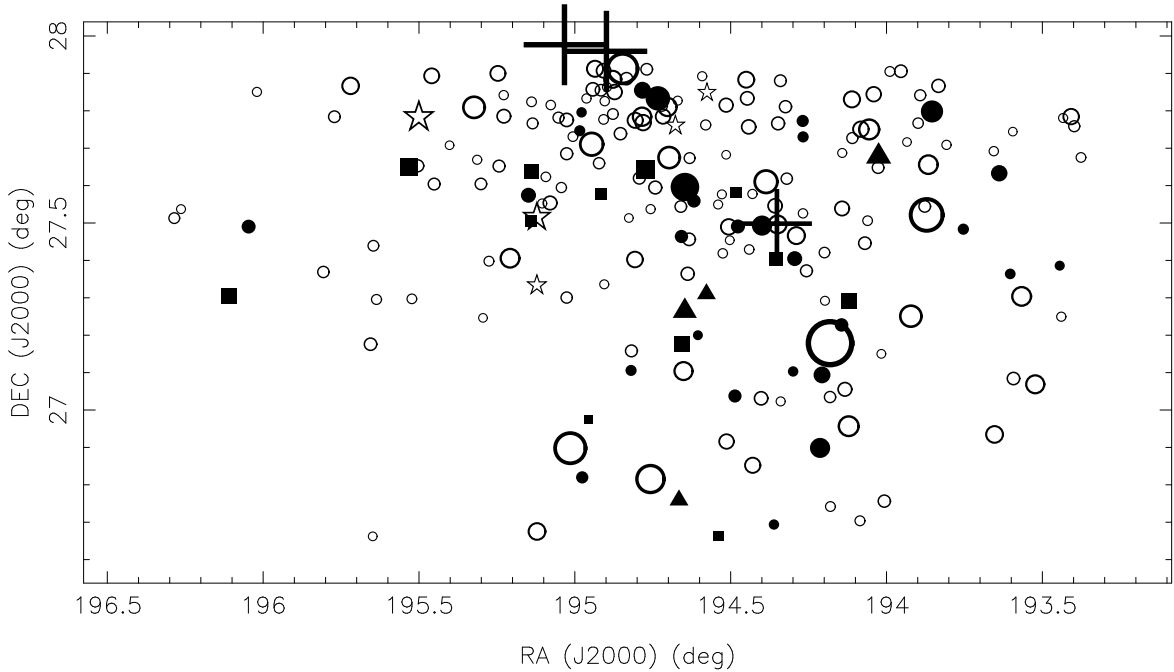}}
\caption{Distribution on the sky of the observed Coma galaxies. All galaxy
spectra were obtained with the central bright Coma plate (number 133). Note
the semicircle shape sampling due to the availability of only one
spectrograph. The different symbols represent different galaxy types: {\bf
AB} open circles; {\bf AB+EM} solid circles; {\bf EM+AB} solid squares;
{\bf EM} solid triangles; {\bf PS} stars (see text). The position of
NGC4874, NGC4889 and NGC4839 are also shown as plus signs. The size of the
symbols denotes the $b$ magnitude of the object, from the brightest
(largest symbol) $b=14.0$, to the faintest (smallest) $b=18.0$, excluding
the plus signs.
\label{fig3}
}
\end{figure}

\begin{figure}
\centerline{\psfig{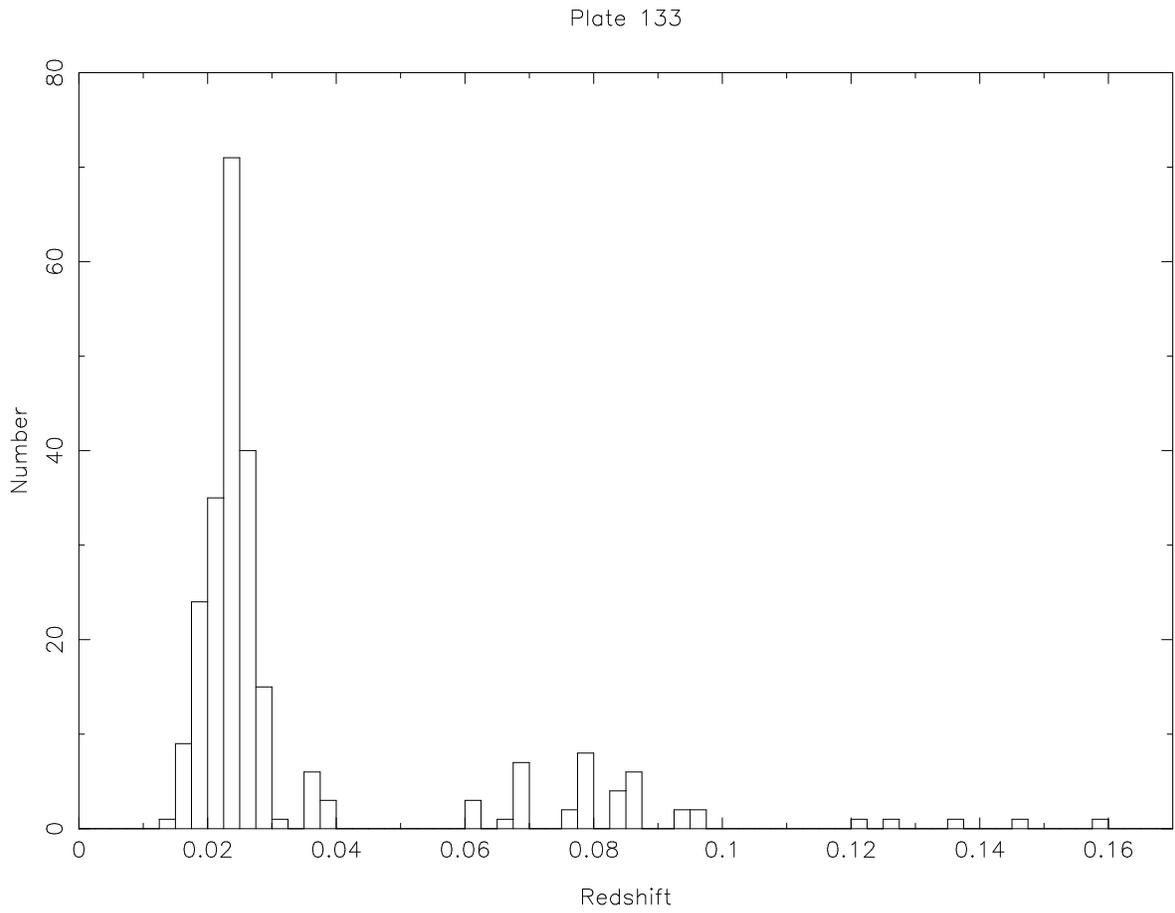}}
\caption{The redshift distribution of the all identified galaxies in plate 133.
\label{fig4}
}
\end{figure}

\begin{figure}
\centerline{\psfig{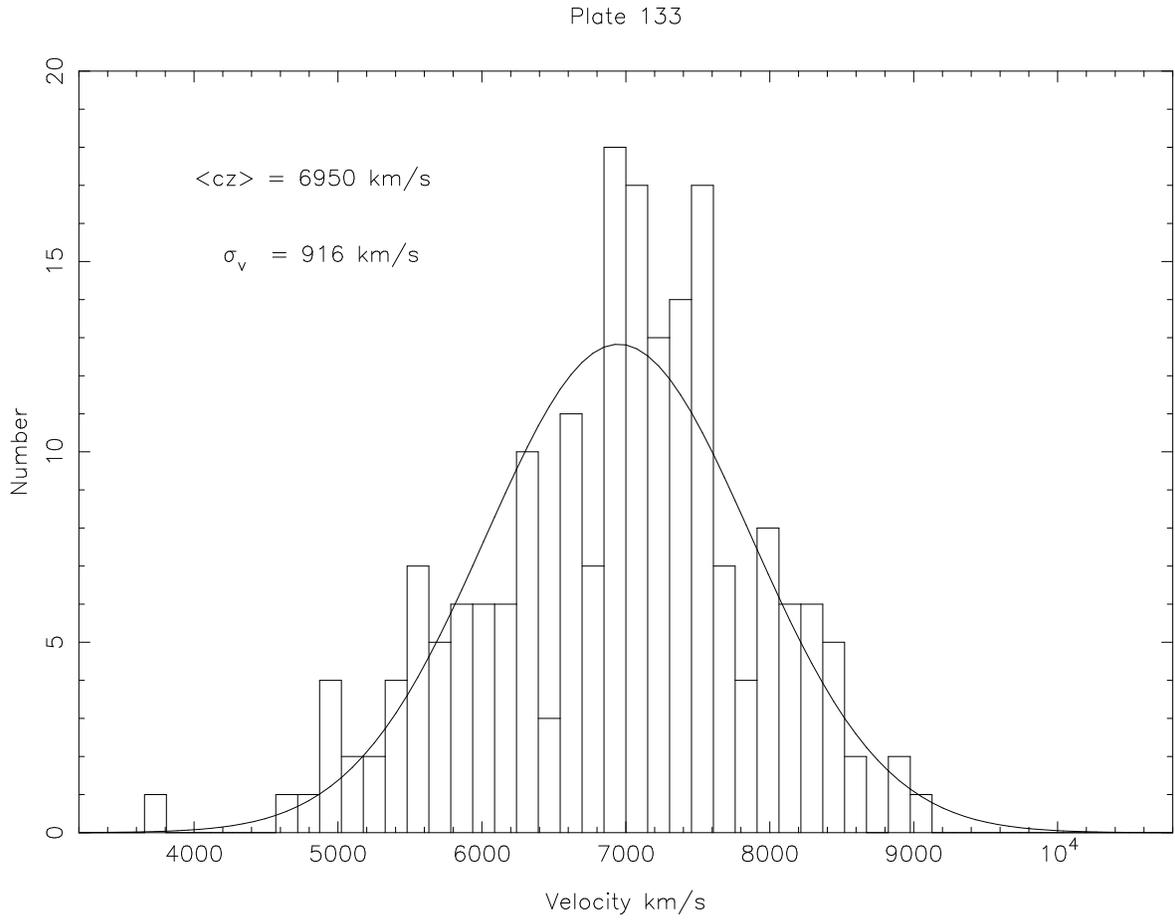}}
\caption{The redshift distribution of the Coma galaxies in plate 133.
\label{fig5}
}
\end{figure}

\begin{figure}
\centerline{\psfig{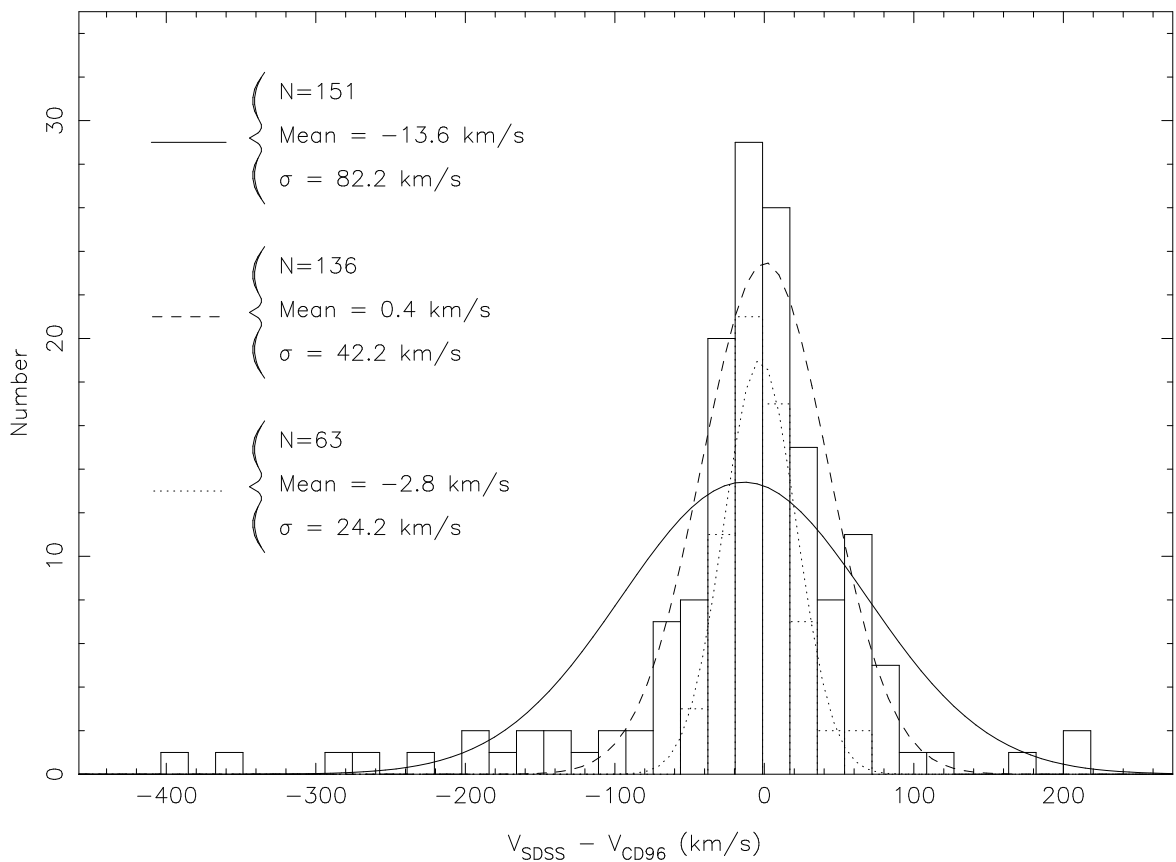}}
\caption{Distribution of the difference in $cz$ between the our redshift
measurements (Table~\ref{tbl1}) and those in the literature.  The
solid histogram represents all 151 matches. The solid line is a Gaussian
fit to the whole histogram. The dash line is a Gaussian fit to the
histogram with 3-$\sigma$ clipping rejection. The dotted histogram counts
only the galaxies that were measured by CD96 and not taken from the
literature by these authors. The dotted histogram is a Gaussian fit to it.
(See text for further details).
\label{fig6}
} 
\end{figure}

\begin{figure}
\centerline{\psfig{file=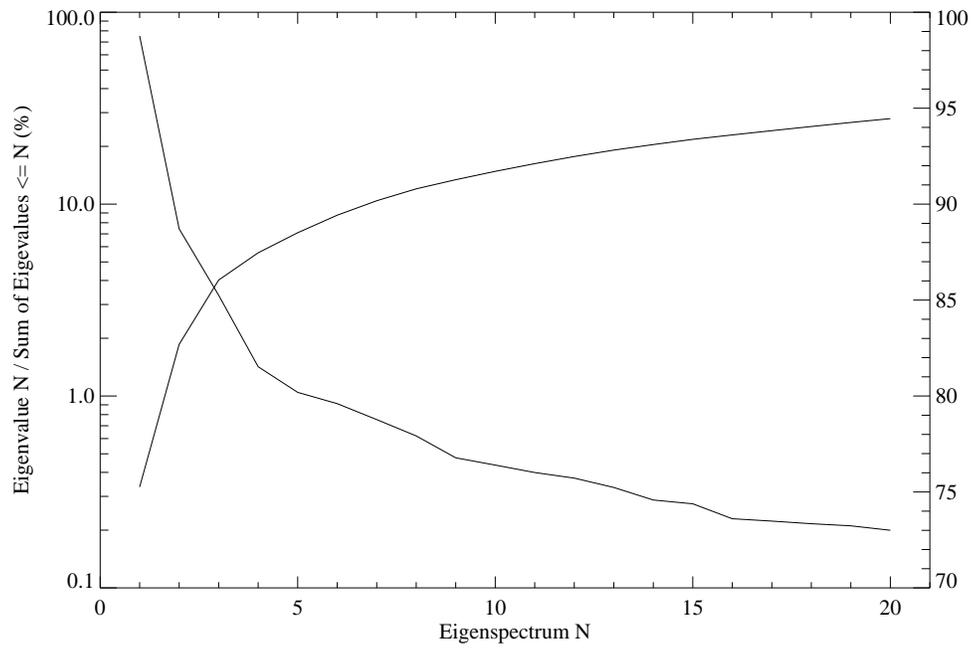,width=13.0cm}}
\caption{Distribution of the first 20 eigenvalues. The 
decreasing function is the percentage contribution of the Nth 
eigenvalue, and uses the left logarithmic axis. The increasing 
function is the cumulative percentage contribution by the first 
N eigenvalues, and uses the right linear axis.
\label{fig7}
}
\end{figure}

\begin{figure}
\centerline{\psfig{file=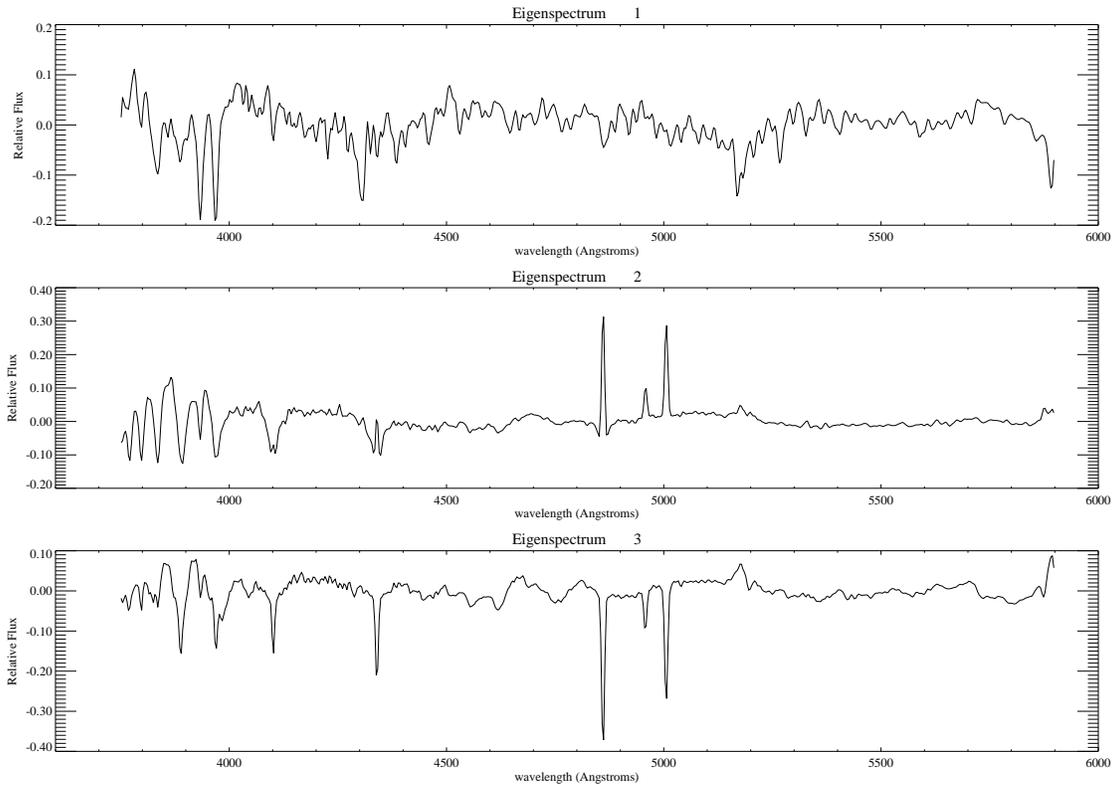,width=15.0cm}}
\caption{The first three Eigenspectra (see text)
\label{fig8}
}
\end{figure}

\begin{figure}
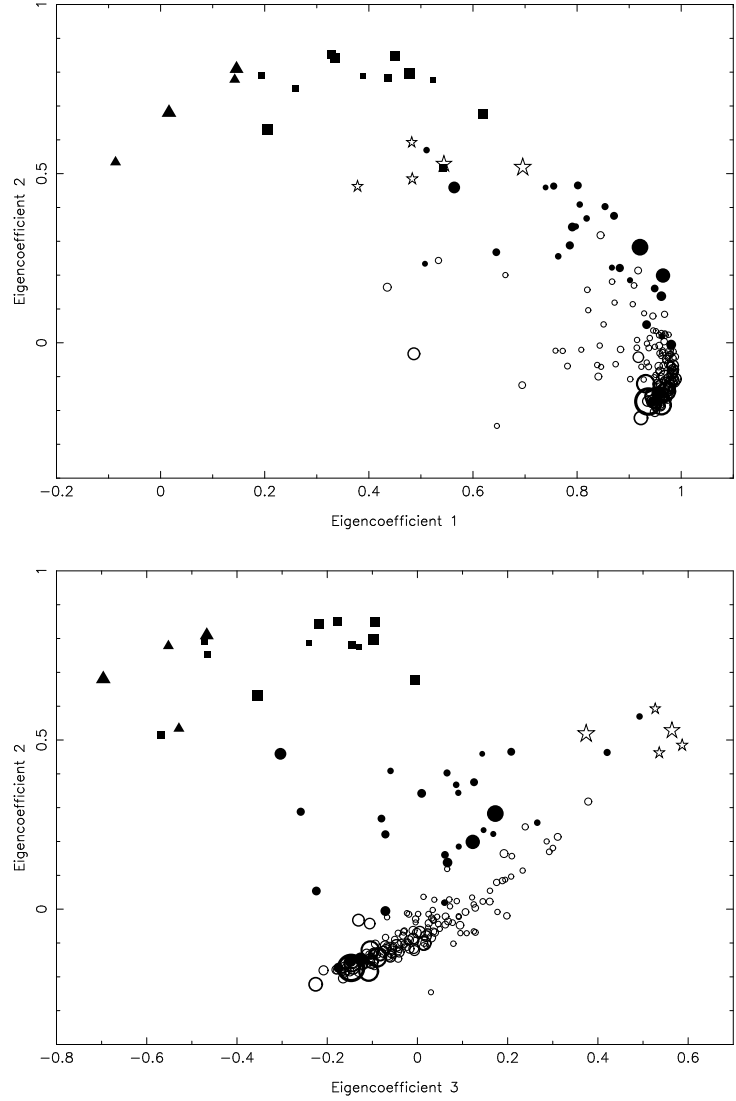

\centerline{\psfig{file=castander.fig9a.ps,height=7.0cm,angle=270}}
\vspace{0.5cm}
\centerline{\psfig{file=castander.fig9b.ps,height=7.0cm,angle=270}}
\caption{The first three eigencoefficients.
The symbols are our line strength classification scheme;
{\bf AB} are open circles, {\bf AB+EM} are
solid circles, {\bf EM+AB} are solid squares, {\bf EM} are 
solid triangles and {\bf PS} are
stars. The size of the symbol is proportional to the magnitude of the
galaxy as in Figure 3.
\label{fig9}
}
\end{figure}

\begin{figure}
\centerline{\psfig{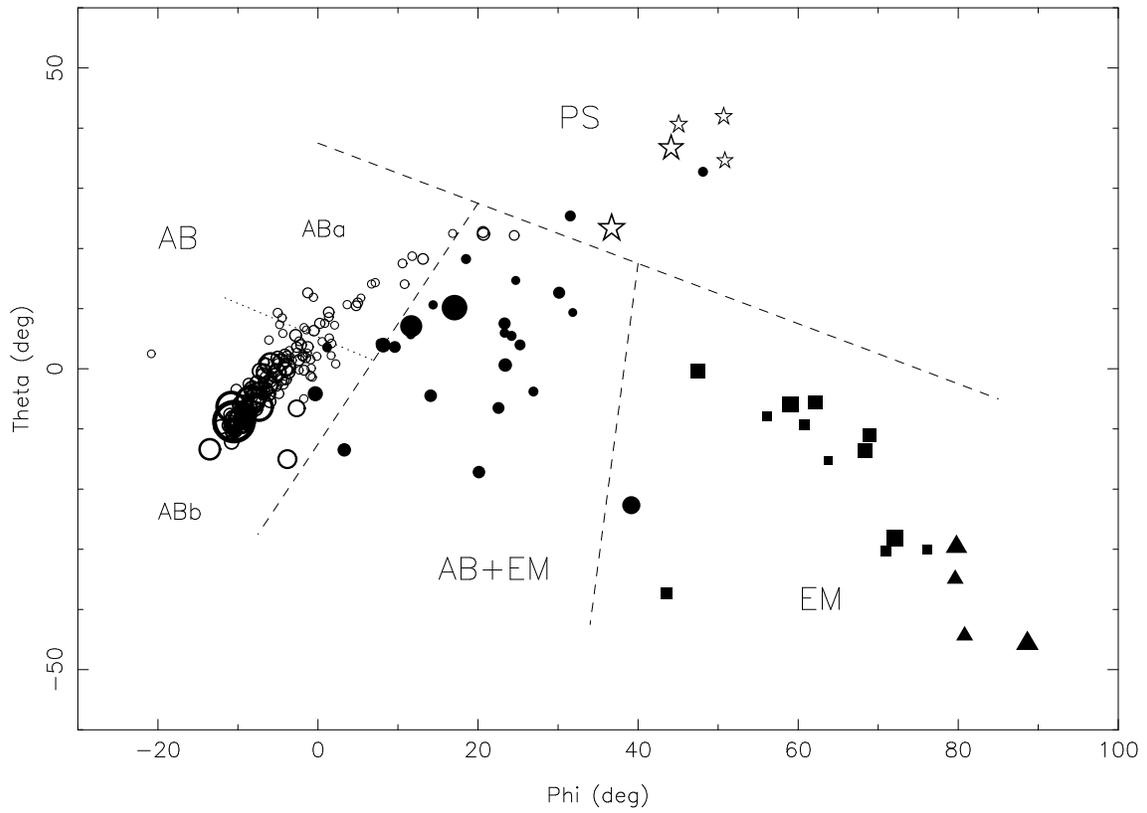}}
\caption{Plot of the spherical projection of the first three 
eigencoefficients. The plot symbols are the same as in Figure \ref{fig3} and \ref{fig9}. 
We show the cuts in eigenspace we have used to objectively
classify our spectra. For this scheme, we
have merged the {\bf EM} and {\bf EM+AB} types into the one type ({\bf EM}).
\label{fig10}
}
\end{figure}

\begin{figure}
\centerline{\psfig{file=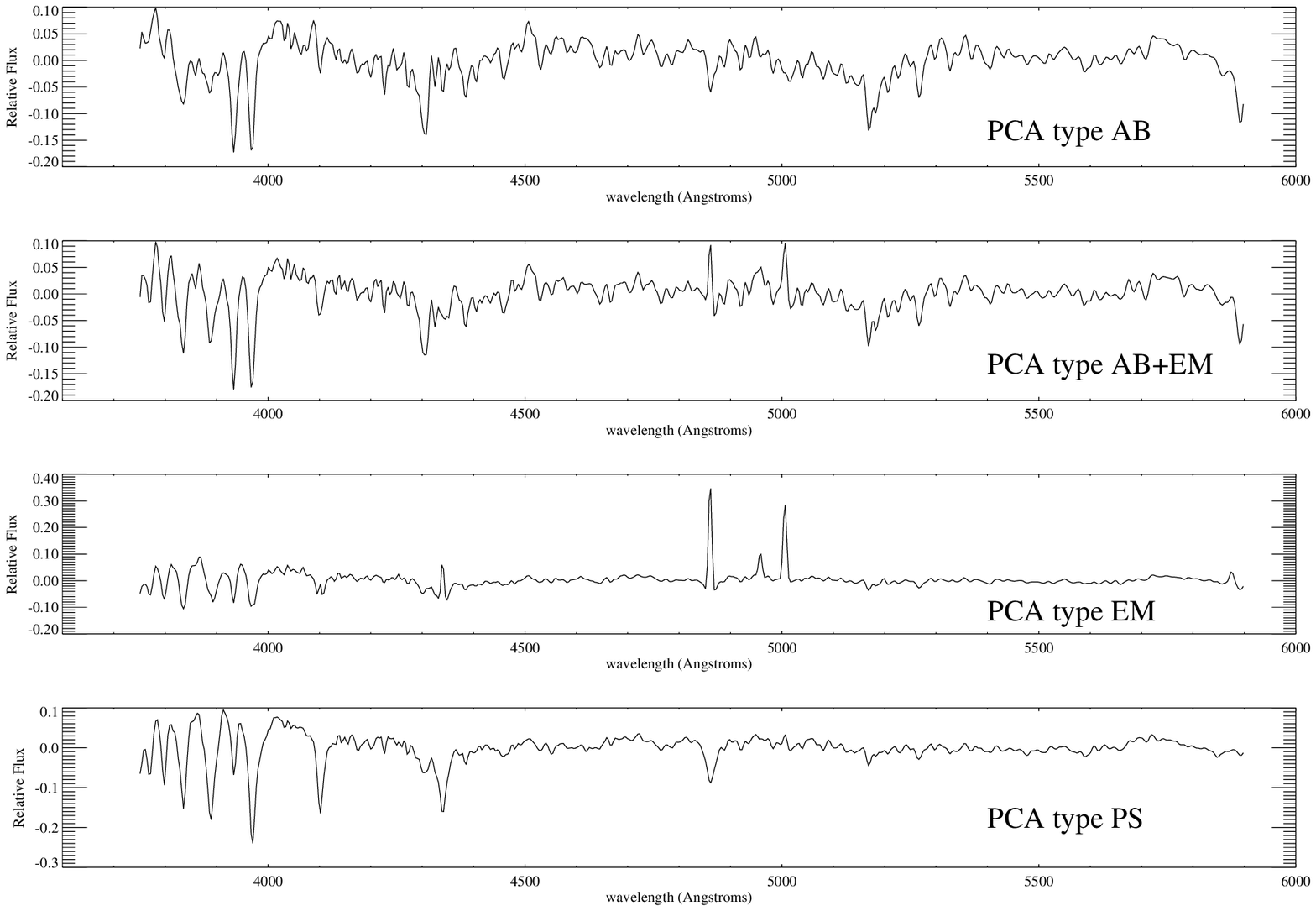,height=10.0cm}}
\centerline{\psfig{file=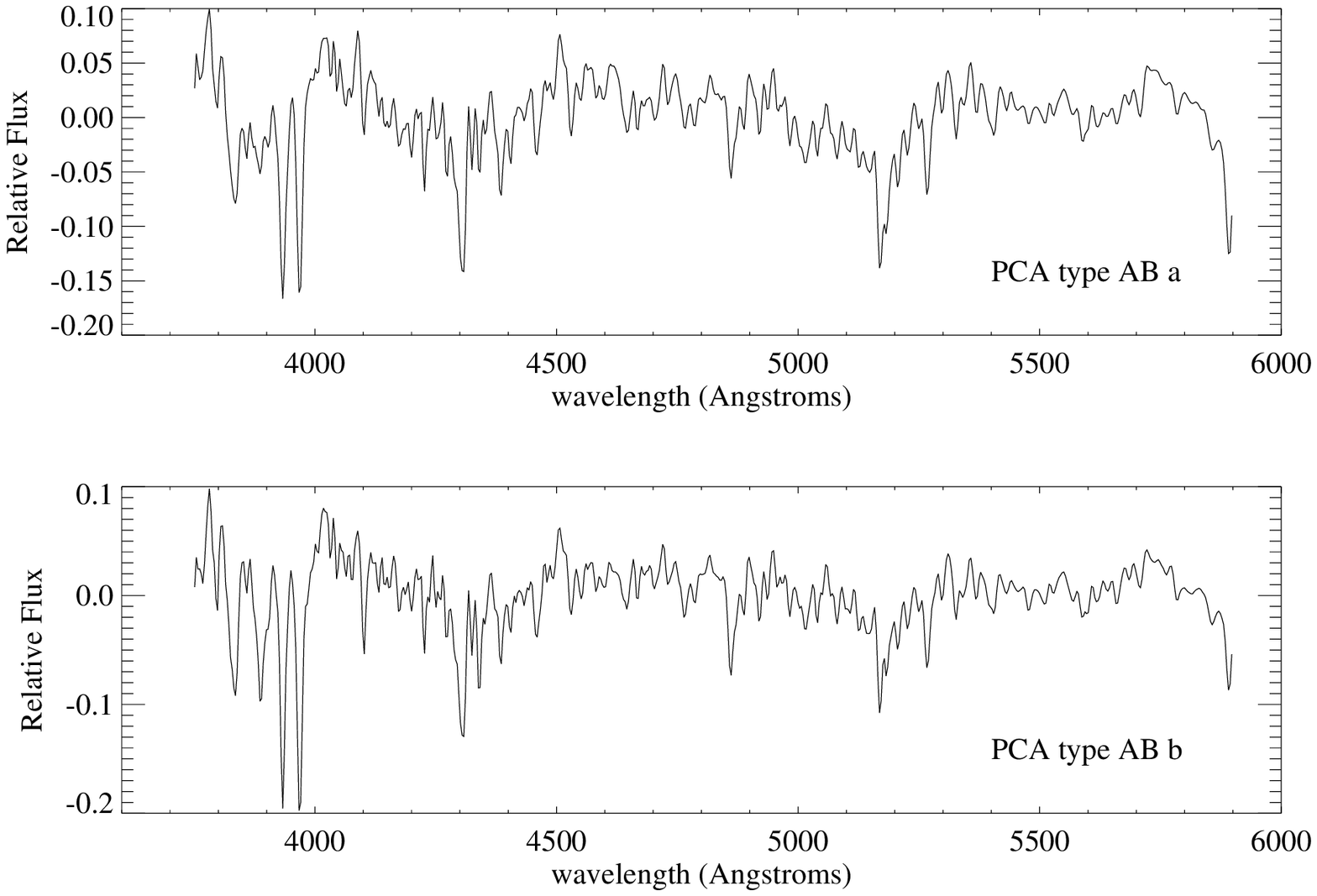,height=10.0cm}}
\caption{Composite spectra of the four PCA spectral types, 
and the two sub A types. The spectra for all the members of 
each group have been averaged with a uniform weight.
\label{fig11}
}
\end{figure}

\begin{figure}
\centerline{\psfig{file=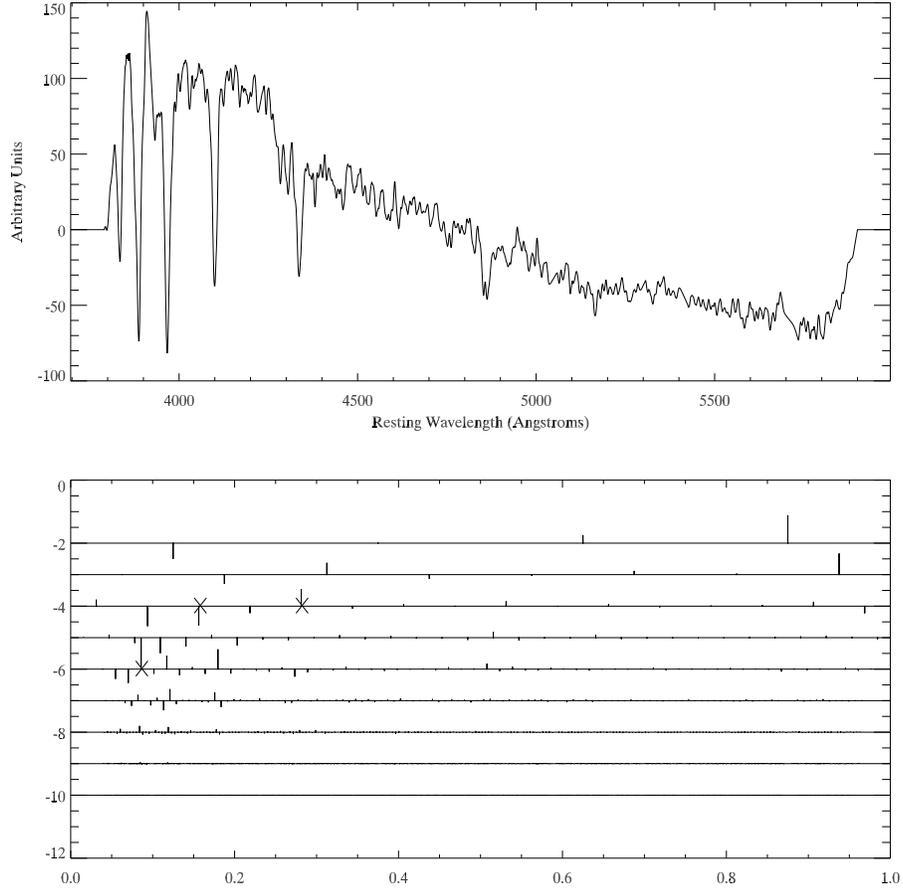,height=5.in}}
\caption{Top: The spectrum of one of the Coma post--starburst galaxies.  We
have subtracted off the mean pixel value as well as padded the ends with
zeros to be 2048 pixels. Cosine--bell smoothing was applied to the ends to
smoothly taper them to zero.  Bottom: The different wavelet resolution
levels start at the top with the lowest resolution and moving to higher
resolutions towards the bottom of the figure. We show the relative height
and sign of the wavelet coefficients. The crosses make the three wavelet
coefficients used to classify this spectrum as a post--starburst (see
text).
\label{fig12}
}
\end{figure}

\begin{figure}
\centerline{\psfig{file=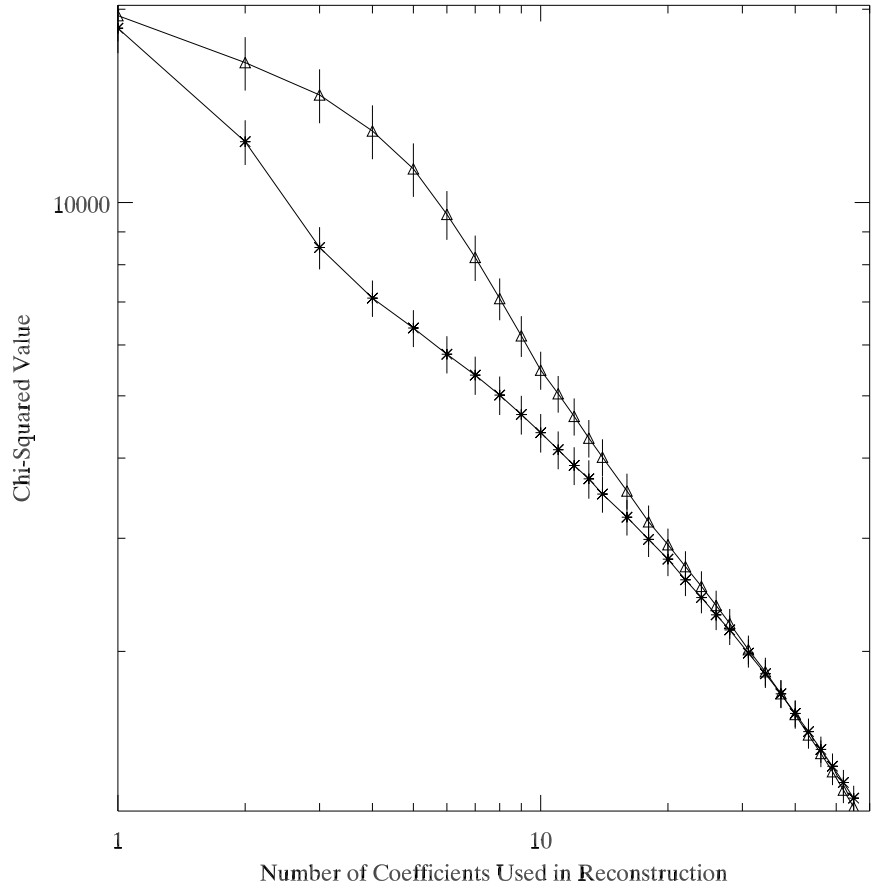,height=5.in}}
\caption{The mean chi-squared for all 196 Coma spectra for both the
Daubeuchies4 (stars) and Symmlet8 (triangles) wavelets as a function of the
number of wavelet coefficients used in the reconstruction of the
spectra. The error is just the standard deviation observed between the Coma
spectra.  To avoid overcrowding, we have not plotted the similar curves for
the Symmlet6 and Daubechies6 wavelets since they are worse than those
presented here {\it i.e.}, they have a higher $\chi^2$.
\label{fig13}
}
\end{figure}

\begin{figure}
\centerline{\psfig{file=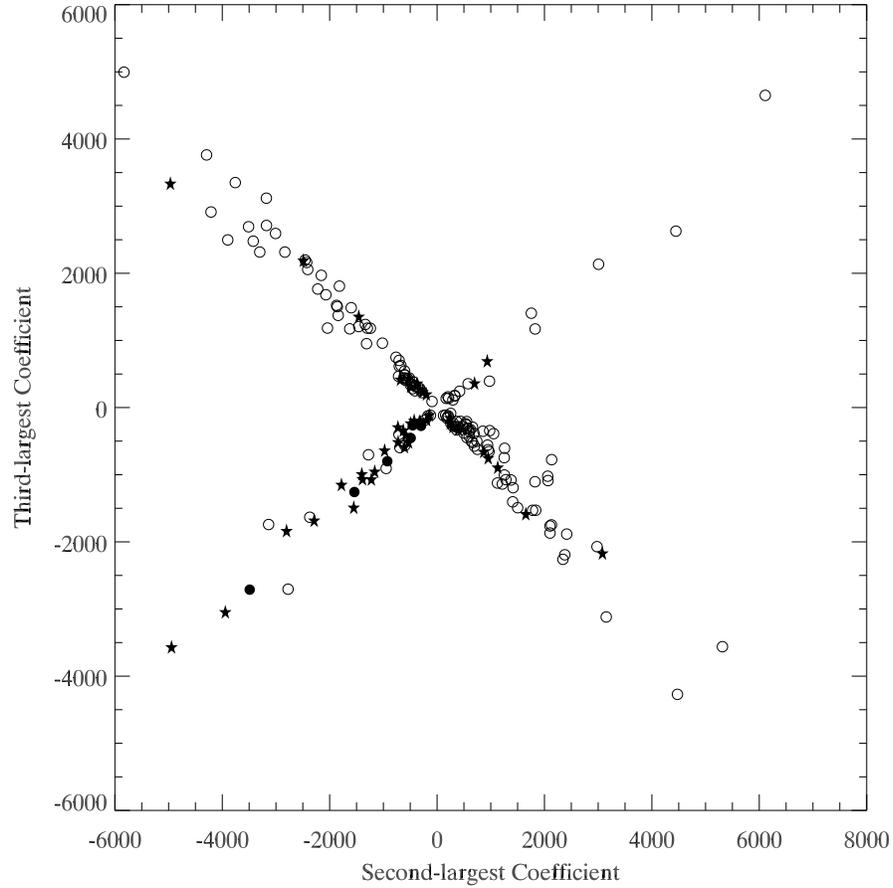,height=5in}}
\caption{The second versus the third largest wavelet coefficient for all
196 Coma spectra. Open circles are absorption galaxies, solid triangles are
emission line galaxies and solid circles are post--starburst (based on the
initial visual classification).  The ``star--pattern''seen in the
distribution of these points is simply the produced of our definition {\it
i.e.}, the second coefficient must be larger (in absolute value) than the
third coefficient.
\label{fig14}
}
\end{figure}

\begin{figure}
\centerline{\psfig{file=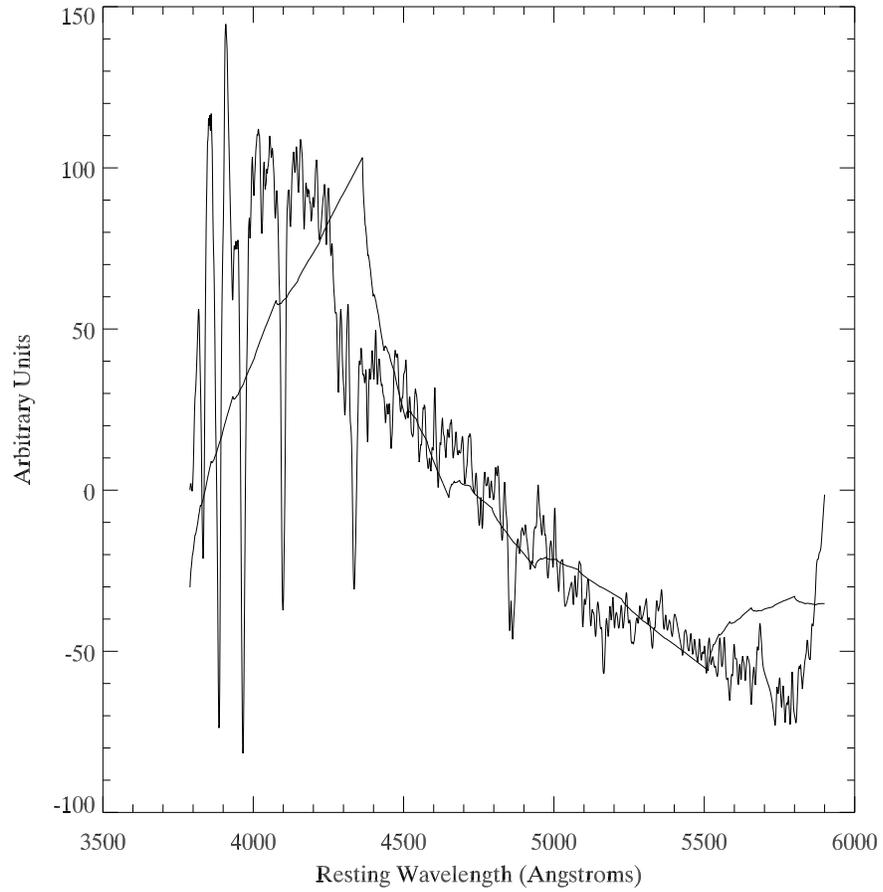,height=5in}}
\caption{One of the strong post--starburst Coma galaxy spectra (same as
presented in Figure~\ref{fig12}) and the wavelet reconstruction based on just
the three largest wavelet coefficients (see Figure~\ref{fig14}).
\label{fig15}
}
\end{figure}

\begin{figure}
\centerline{\psfig{file=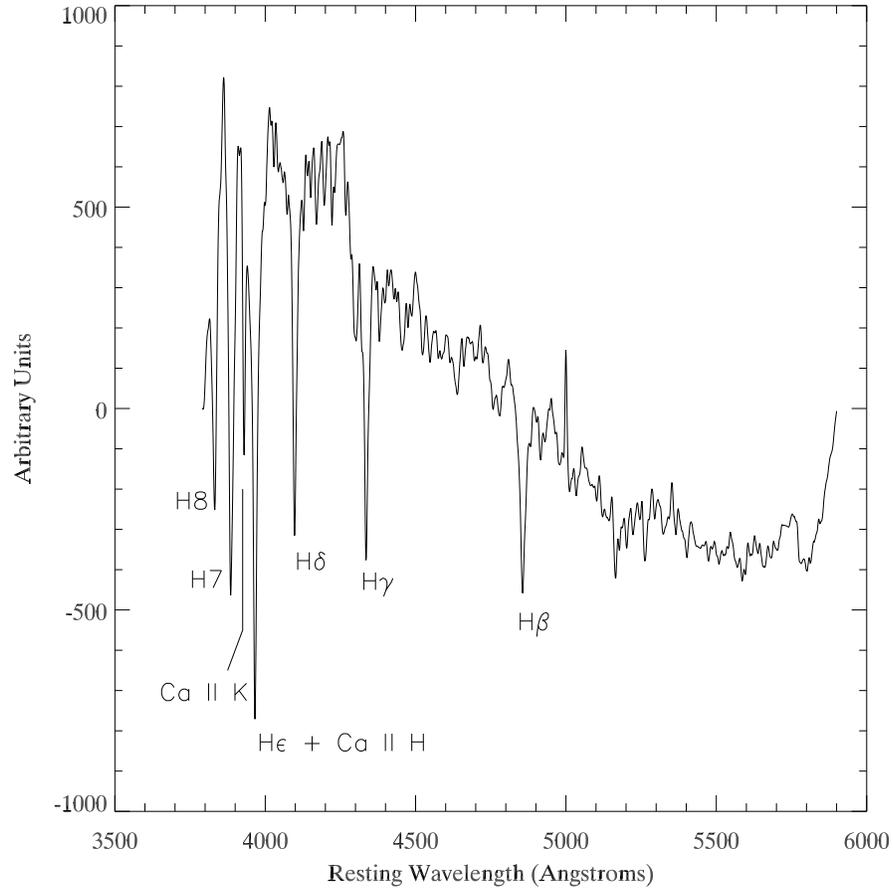,height=5in}}
\caption{The co-added spectrum for 6 strong post--starburst
galaxies visually identified in our SDSS Coma spectra.
This plot illustrates the strong Balmer absorption features
due to the young A stars. Also visible is the Calcium II
H \& K absorption features common in older galaxies.
We have labeled the common absorption lines.
\label{fig16}
}
\end{figure}

\begin{figure}
\centerline{\psfig{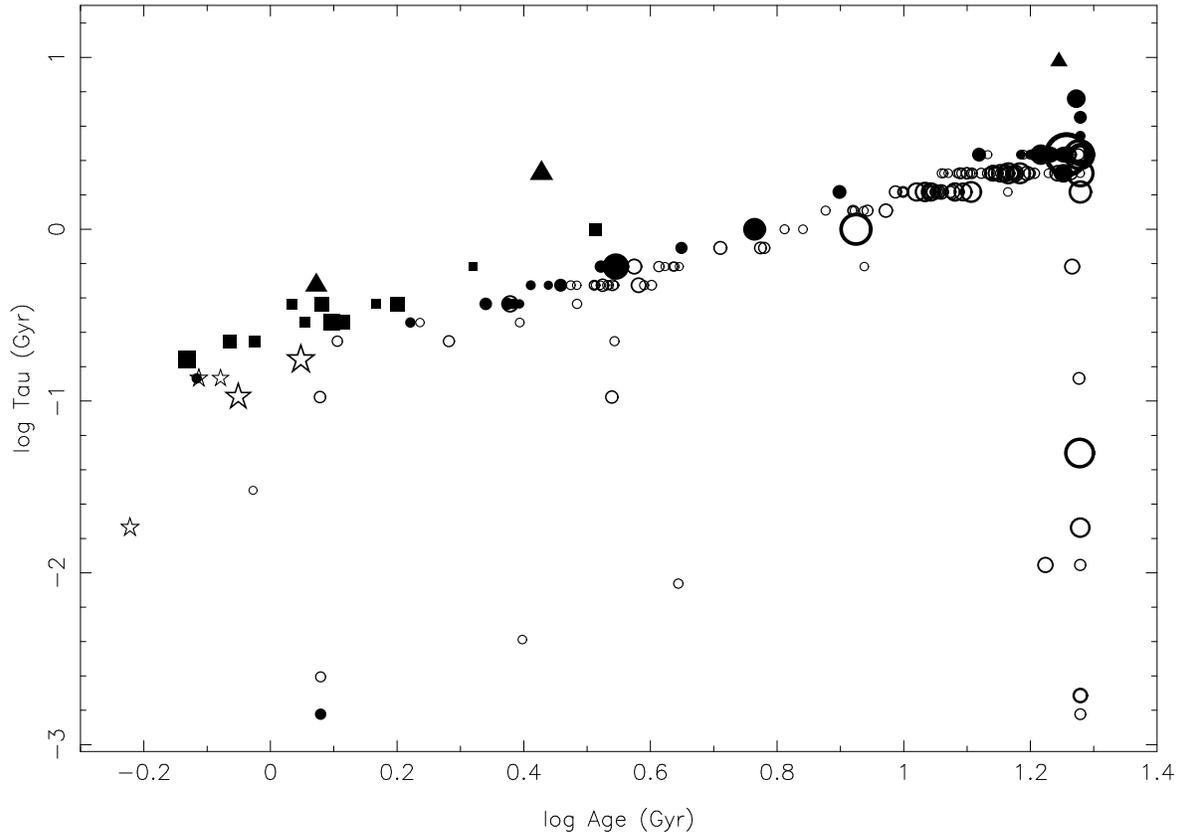}}
\caption{The star formation prescription versus age for all Coma galaxies
(see text). The maximum allowed age in the minimization is 19 Gyr.
\label{fig17}
} 
\end{figure}

\begin{figure}
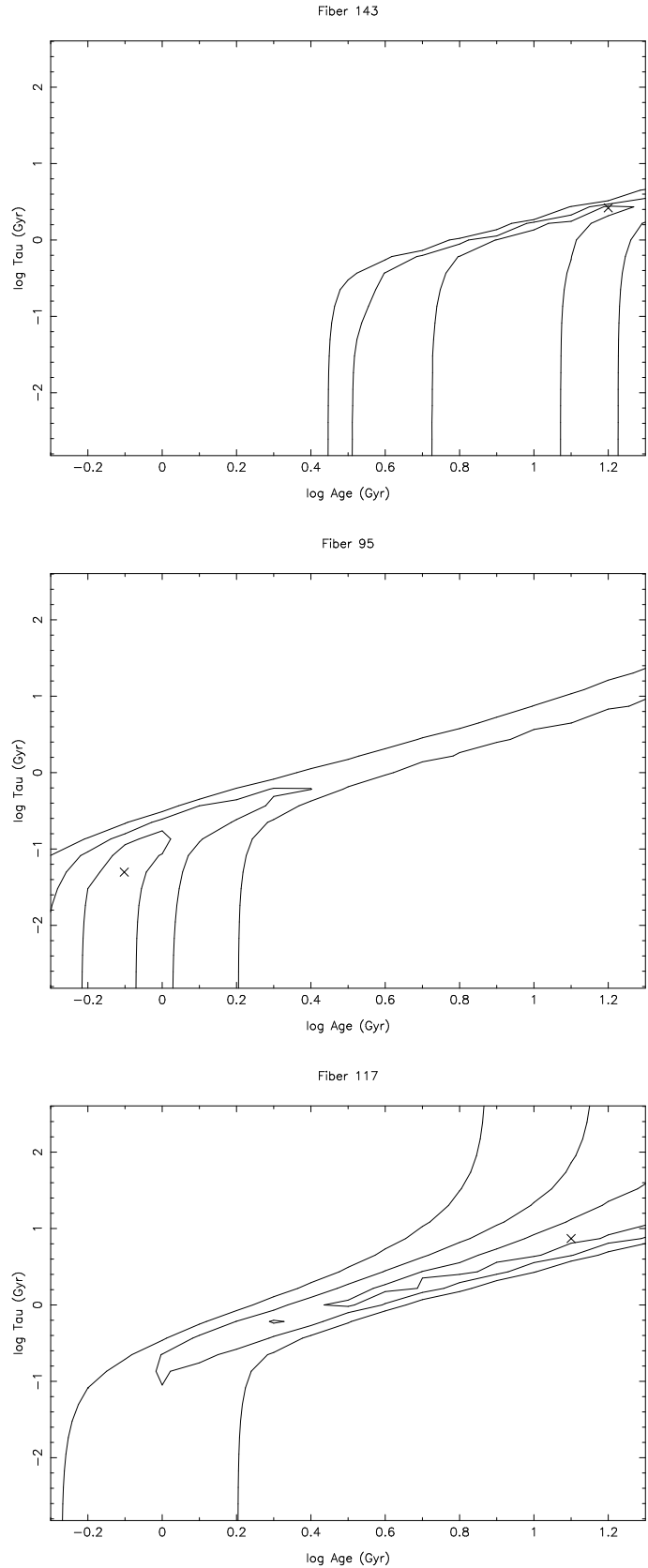

\vspace{-0.5cm}
\centerline{\psfig{file=castander.fig18a.ps,width=9.cm,angle=270}}
\vspace{0.5cm}
\centerline{\psfig{file=castander.fig18b.ps,width=9.cm,angle=270}}
\vspace{0.5cm}
\centerline{\psfig{file=castander.fig18c.ps,width=9.cm,angle=270}}
\caption{One, two and three sigma confidence levels contours for the
spectral fit of three galaxies of different type. The cross signals the
best fit age and $\tau$ parameter. Top: fiber 143 classified as {\bf AB};
Middle: fiber 95, {\bf PS}; Bottom: fiber 117, {\bf EM+AB}.
\label{fig18}
} 
\end{figure}

\begin{figure}
\centerline{\psfig{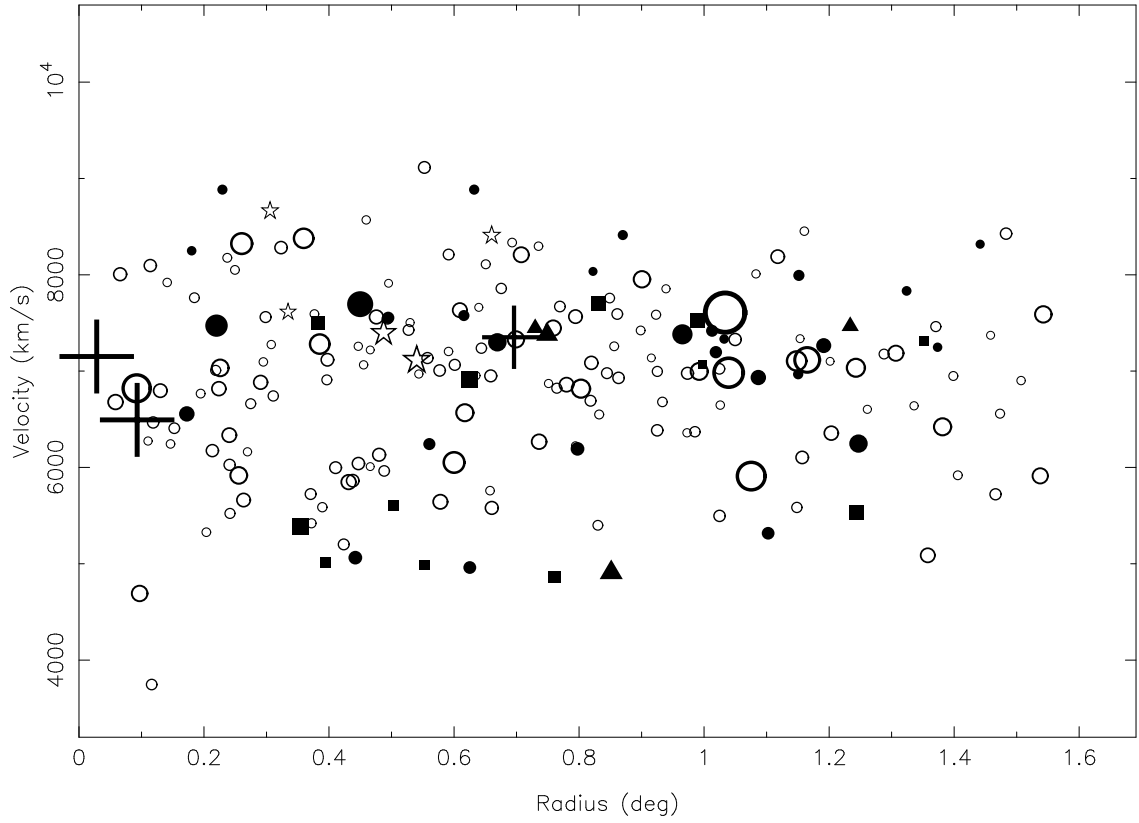}}
\caption{Distribution of radial velocities as a function of radius for the
observed Coma galaxies. The symbols are the same as in
Figure~\ref{fig3}.
\label{fig19}
}
\end{figure}

We explore here possible mechanisms for triggering star--formation in clusters
of galaxies. Several authors (Burns et al. 1994; Caldwell \& Rose 1997;
Metevier, Romer \& Ulmer 1999; Wang, Connolly \& Brunner 1997) have proposed
that shocks caused by cluster merger events can trigger star--formation in
cluster galaxies. However, Figures~\ref{fig3} and \ref{fig19} show that our
Coma galaxies with recent star--formation ({\bf EM+AB},{\bf AB+EM}, {\bf EM}
and {\bf PS}) are random distributed throughout the cluster with no apparent
correlation with the NGC 4839 group. This implies that this merger (or
post--merger) has had little effect on the recent star--formation rates of
Coma contrary to the scenario outlined in Burns et al. (1994) and Caldwell \&
Rose (1997). This difference may be the combination of several effects: {\it
1)} We have a complete, homogeneous, spectroscopically confirmed, sample of
Coma galaxies out to large cluster radii {\it i.e.} we have targeted all
bright galaxies out to the virial radius of Coma regardless of their location,
color or morphological type. Caldwell \& Rose (1997) were forced to
selectively sample specific areas in Coma, because of the smaller
field--of--view of their instrument, as well as preferentially targeting
early--type galaxies; {\it 2)} We have used different schemes, and thresholds,
for defining post--starburst and star--forming galaxies. For example, if we
lowered our threshold for the definition of a {\bf PS} galaxy (to older and
weaker lines) we may find a correlations with the NGC 4839 group.  We note
however that the extra wavelet {\bf PS} galaxies (which are older {\bf PS}
galaxies) show no sign of excess clustering with the NGC 4839 group; {\it 3)}
The NGC 4839 is approximately 5--10\% (CD96) of the mass of the core of Coma
and thus the merger of Coma and this group may not be severe enough to trigger
the levels of star--formation seen in more extreme head--on cluster--cluster
and group--group mergers (Metevier et al. 1999); {\it 4)} If the NGC 4839
group is still ``spiralling'' into Coma then we could simply be seeing the
combination of ram--pressure induced star--formation as the field galaxies
first infall into the cluster (see Dressler \& Gunn 1983) as well as the
expected recent star--formation in the NGC 4839 group before it fell in (see
Zabludoff et al. 1996). This would also explain the large velocity dispersion
measured for the emission--line galaxies (Carlberg et al. 1997). We must await
further data to fully discriminate between these possible models or others. It
is important to note that we only sample galaxies brighter than $b=18$ and
therefore our observations only apply to the brightest galaxy members and not
to the dwarf population (see Rakos et al. 2000). Moreover, our observations do
not distinguish between different morphological types (see Doi et
al. 1995a,b); this will be addressed using the main SDSS survey data.

\begin{figure}
\centerline{\psfig{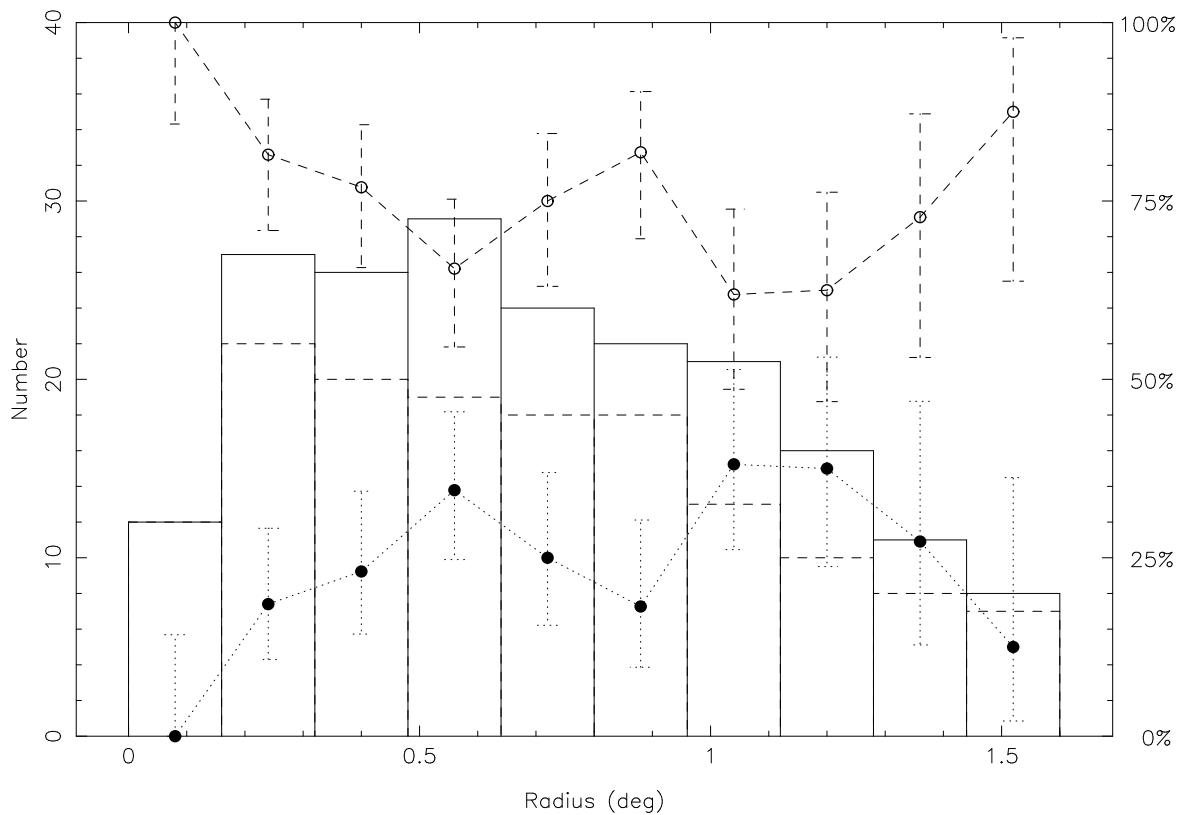}}
\caption{Number of observed Coma galaxies (solid histogram) and {\bf AB}
type galaxies (dashed histogram) as a function of projected radius. The
left vertical axis provides the scale. The fraction of old stellar
population galaxies ({\bf AB}; open circles and dashed line) and active
galaxies ({\bf PS, AB+EM, EM+AB, EM}; solid circles and dotted line) are
also plotted. The right vertical axis gives the scale. The errors in the
percentages have been computed using the prescription of Gehrels (1986).
\label{fig20}
} 
\end{figure}

The characteristic radius of the Coma cluster is $r_{200}\sim 1.5$ $h^{-1}$
Mpc (Geller et al 1999), or 1.25 degrees, and thus, we sample the galaxy
population out to the virial radius and slightly beyond. This provides a
unique opportunity to study the large--scale star--formation histories of
galaxies in this nearby cluster.  In Figure~\ref{fig20}, we present the
fraction of passive (absorption) and active (emission plus post--starburst)
galaxies as a function of projected distance from the center of Coma.  The
azimuthally averaged fraction of active galaxies ({\bf PS, AB+EM, EM+AB,
EM} types) appears to increase slightly with radius from 0.15 to 1.2
degrees$\footnote{At the distance of Coma, one degree on the sky
corresponds to 1.2 $h^{-1}$ Mpc.}$, but the observed percentages are also
consistent with a constant radial fraction of active galaxies in this
region. This trend is different from that observed in the CNOC sample at
higher redshift (Ellingson et al 2000), {\i.e.}, they observe a steeper
radial profile for the fraction of active galaxies in their high redshift
clusters.  However, the results are consistent given the errors and the
different analyses performed which render a detailed comparison
difficult. Moreover, the target selection used different criteria, although
similar, which need to be taken into account to provide a fair comparison.

We find no active galaxies in the central (projected) $\sim 200$ $h^{-1}$ Kpc
of the Coma cluster. This result is similar to that observed by Ellington et
al. (2000) who found a deficit of star--forming galaxies within 0.5$r_{200}$
of the cores of CNOC1 clusters (see also Rakos et al. 2000 who found a similar
trend in intermediate redshift clusters). Ellingson et al. (2000) propose that
the dense intracluster medium in the cores of clusters inhibits
star--formation producing a deficit of star--formation compared to the
outskirts of the cluster and the general field population.  Our data certainly
agrees with this scenario except we do see a reduction of active galaxies
beyond $r_{200}$, although this reduction is not statistically significant.
We defer a detailed discussion of Coma's star--forming galaxies, and their
relation to the X--ray gas, to a future paper.

\acknowledgements The Sloan Digital Sky Survey (SDSS) is a joint project of
The University of Chicago, Fermilab, the Institute for Advanced Study, the
Japan Participation Group, The Johns Hopkins University, the
Max-Planck-Institute for Astronomy, New Mexico State University, Princeton
University, the United States Naval Observatory, and the University of
Washington.  Apache Point Observatory, site of the SDSS telescopes, is
operated by the Astrophysical Research Consortium (ARC). Funding for the
project has been provided by the Alfred P. Sloan Foundation, the SDSS
member institutions, the National Aeronautics and Space Administration, the
National Science Foundation, the U.S. Department of Energy, Monbusho, and
the Max Planck Society.  The SDSS Web site is {\tt www.sdss.org}. We
acknowledge Michael Strauss for his insightful comments and thank an
anonymous referee for his/her helpful comments which have made the paper
better.

\clearpage

\begin{deluxetable}{ccccccrcc}
\tablenum{1}
\tablecolumns{9}
\tablecaption{Galaxy Redshifts.
\label{tbl1}}
\tablehead
{
\colhead{Fiber} & \colhead{GMP83} & 
\colhead{RA\tablenotemark{a}} & 
\colhead{DEC\tablenotemark{a}} & 
\colhead{$cz$} & \colhead{$\delta\,cz$} & 
\colhead{R} & \colhead{b} & \colhead{r}\\
number & number & (J2000) & (J2000) & km/s & km/s &  &  & 	
}
\startdata
  3 & 1411 & 13 02 32.8 & 27 17 43 &  7585 &  39 &   6.9 &  17.36 &  15.47 \nl
  4 & 1382 & 13 02 35.6 & 26 39 43 &  7372 &  72 &   3.0 &  17.97 &  16.04 \nl
 10 & 1060 & 13 03 13.7 & 27 22 08 &  6371 &  32 &  15.7 &  16.84 &  14.88 \nl
 11 & 1368 & 13 02 37.4 & 27 10 34 &  5497 &  33 &  15.4 &  16.52 &  14.65 \nl
 12 &  455 & 13 04 26.6 & 27 18 16 &  5534 &  31 &  16.9 &  15.77 &  14.52 \nl
 13 & 1392 & 13 02 35.2 & 27 26 20 &  5400 &  41 &   7.4 &  17.07 &  15.43 \nl
 14 & 1635 & 13 02  5.5 & 27 17 50 &  7258 &  35 &  10.5 &  17.50 &  15.57 \nl
 21 & 1245 & 13 02 52.7 & 27 51 59 &  8207 &  32 &  19.9 &  15.62 &  13.81 \nl
 27 & 1134 & 13 03  5.2 & 27 47 03 &  7671 &  32 &  18.9 &  16.74 &  14.88 \nl
 28 &  177 & 13 05  8.5 & 27 30 46 &  7177 &  35 &  10.9 &  17.05 &  15.25 \nl
 29 &  631 & 13 04  4.8 & 27 51 01 &  6360 &  34 &  11.4 &  17.77 &  15.96 \nl
 30 &  228 & 13 05  3.3 & 27 32 13 &  6603 &  61 &   4.3 &  17.88 &  16.21 \nl
 31 & 1616 & 13 02  7.9 & 27 38 54 &  6915 &  33 &  12.4 &  15.45 &  13.74 \nl
 32 & 1673 & 13 02  1.1 & 27 39 10 &  7065 &  32 &  17.6 &  16.61 &  14.70 \nl
 34 &  595 & 13 04 11.2 & 27 29 25 &  5317 &  34 &  10.3 &  16.50 &  14.67 \nl
 40 & 2235 & 13 01 10.7 & 27 14 47 &  6226 &  46 &   5.7 &  18.01 &  16.18 \nl
 43 & 2278 & 13 01  6.2 & 27 23 52 &  8109 &  33 &  14.4 &  17.42 &  15.60 \nl
 45 & 2639 & 13 00 29.2 & 27 19 59 &  8408 &  67 &   5.6 &  17.67 &  16.37 \nl
 50 & 2643 & 13 00 29.2 & 26 40 31 &  7185 &  31 &  23.3 &  15.60 &  13.71 \nl
 55 & 2431 & 13 00 49.9 & 27 24 20 &  6569 &  32 &  16.0 &  15.32 &  13.42 \nl
 58 & 1681 & 13 02  0.1 & 27 46 57 &  7111 &  36 &  12.2 &  16.08 &  14.43 \nl
 61 & 1961 & 13 01 36.5 & 27 42 28 &  7912 &  37 &   8.4 &  18.03 &  16.24 \nl
 65 & 1832 & 13 01 48.4 & 27 36 14 &  8210 &  32 &  17.0 &  16.70 &  14.82 \nl
 66 & 2219 & 13 01 12.3 & 27 36 16 &  7543 &  32 &  17.9 &  16.79 &  14.87 \nl
 68 & 2385 & 13 00 54.8 & 27 50 31 &  7096 &  33 &  13.2 &  17.68 &  15.89 \nl
 69 & 2157 & 13 01 17.6 & 27 48 33 &  7281 &  32 &  20.1 &  15.04 &  13.12 \nl
 70 & 2347 & 13 00 59.3 & 27 53 59 &  6883 &  32 &  19.7 &  15.88 &  13.98 \nl
 74 & 2355 & 13 00 58.4 & 27 39 07 &  5202 &  38 &   6.6 &  16.67 &  14.80 \nl
 75 & 1807 & 13 01 50.2 & 27 53 36 &  7562 &  36 &  10.9 &  15.89 &  14.00 \nl
 76 & 2185 & 13 01 15.2 & 27 40 09 &  7065 &  43 &   7.4 &  17.76 &  15.91 \nl
 84 & 3143 & 12 59 49.3 & 26 58 27 &  7072 &  32 &  13.6 &  17.88 &  16.53 \nl
 85 & 3618 & 12 59 16.7 & 27 06 21 &  8412 &  33 &  13.7 &  17.38 &  15.48 \nl
 86 & 3622 & 12 59 16.4 & 27 09 29 &  6692 &  32 &  17.6 &  16.67 &  14.78 \nl
 87 & 3310 & 12 59 37.4 & 27 20 09 &  6953 &  34 &  11.4 &  17.84 &  15.94 \nl
 88 & 2987 & 13 00  3.5 & 26 53 53 &  5911 &  32 &  16.1 &  14.43 &  12.52 \nl
 89 & 2948 & 13 00  6.3 & 27 18 02 &  7858 &  32 &  18.5 &  16.87 &  14.99 \nl
 91 & 3100 & 12 59 54.4 & 26 49 11 &  7994 &  36 &  12.1 &  16.98 &  15.34 \nl
 94 & 3661 & 12 59 13.7 & 27 24 09 &  5644 &  33 &  15.5 &  15.77 &  13.84 \nl
 95 & 2640 & 13 00 29.2 & 27 30 52 &  7395 &  38 &  14.1 &  16.33 &  15.02 \nl
 96 & 2945 & 13 00  6.3 & 27 46 32 &  6175 &  32 &  19.7 &  16.27 &  14.33 \nl
 97 & 2615 & 13 00 32.5 & 27 45 58 &  6662 &  32 &  19.3 &  16.99 &  15.10 \nl
 98 & 2599 & 13 00 33.7 & 27 38 15 &  7498 &  33 &  12.2 &  15.97 &  14.72 \nl
 99 & 2776 & 13 00 19.1 & 27 33 13 &  5862 &  32 &  17.3 &  16.19 &  14.27 \nl
100 & 2721 & 13 00 22.4 & 27 37 24 &  7592 &  33 &  14.1 &  17.57 &  15.69 \nl
101 & 2393 & 13 00 54.7 & 27 47 08 &  8282 &  69 &   3.1 &  16.22 &  14.31 \nl
102 & 2603 & 13 00 33.4 & 27 49 27 &  8176 &  33 &  13.5 &  17.48 &  15.69 \nl
103 & 2942 & 13 00  6.3 & 27 41 06 &  7561 &  32 &  17.8 &  16.64 &  14.82 \nl
105 & 2783 & 13 00 18.6 & 27 48 56 &  5328 &  36 &   8.4 &  17.59 &  15.66 \nl
106 & 2582 & 13 00 35.7 & 27 34 27 &  5064 &  36 &   8.9 &  16.24 &  14.39 \nl
107 & 2894 & 13 00 10.4 & 27 35 42 &  5587 &  32 &  14.4 &  17.26 &  15.36 \nl
108 & 2866 & 13 00 12.6 & 27 46 54 &  7008 &  33 &  14.2 &  16.97 &  15.22 \nl
110 & 2601 & 13 00 33.6 & 27 30 14 &  5602 &  32 &  15.2 &  16.88 &  15.60 \nl
111 & 2688 & 13 00 25.2 & 27 33 08 &  7258 &  33 &  13.6 &  17.80 &  15.96 \nl
115 & 4106 & 12 58 39.9 & 26 45 33 &  7467 &  31 &  19.3 &  17.01 &  15.76 \nl
117 & 4135 & 12 58 37.2 & 27 10 35 &  7705 &  32 &  13.8 &  15.81 &  14.33 \nl
120 & 3837 & 12 59  1.8 & 26 48 56 &  7117 &  33 &  17.0 &  14.60 &  12.59 \nl
123 & 4206 & 12 58 32.1 & 27 27 22 &  7007 &  32 &  17.8 &  16.48 &  14.58 \nl
124 & 4192 & 12 58 33.1 & 27 21 51 &  6950 &  32 &  18.1 &  16.34 &  14.41 \nl
126 & 4159 & 12 58 35.4 & 27 15 52 &  7380 &  32 &  15.9 &  16.15 &  14.77 \nl
127 & 4351 & 12 58 18.7 & 27 18 38 &  7447 &  31 &  18.4 &  17.04 &  15.39 \nl
130 & 4147 & 12 58 36.3 & 27 06 15 &  7954 &  31 &  22.0 &  15.41 &  13.49 \nl
131 & 4294 & 12 58 25.3 & 27 11 59 &  8035 &  38 &   7.2 &  17.90 &  16.04 \nl
132 & 4124 & 12 58 37.9 & 27 27 50 &  6243 &  43 &   5.4 &  16.67 &  14.72 \nl
133 & 3238 & 12 59 41.3 & 27 39 35 &  6742 &  32 &  18.8 &  16.88 &  14.95 \nl
134 & 3298 & 12 59 37.8 & 27 46 36 &  6766 &  34 &  11.2 &  17.50 &  15.75 \nl
135 & 3296 & 12 59 37.9 & 27 54 26 &  8005 &  33 &  17.4 &  16.08 &  14.14 \nl
136 & 3165 & 12 59 47.1 & 27 42 38 &  8324 &  31 &  21.5 &  14.91 &  13.02 \nl
137 & 3400 & 12 59 30.8 & 27 53 03 &  4692 &  35 &  12.2 &  15.55 &  13.63 \nl
138 & 3271 & 12 59 39.8 & 27 34 36 &  5012 &  32 &  13.4 &  16.63 &  15.23 \nl
139 & 3178 & 12 59 46.1 & 27 51 26 &  8096 &  31 &  23.8 &  16.27 &  14.45 \nl
140 & 3071 & 12 59 56.2 & 27 44 47 &  8883 &  50 &   8.2 &  17.39 &  16.07 \nl
141 & 3092 & 12 59 54.9 & 27 47 45 &  8250 &  32 &  16.6 &  17.66 &  15.81 \nl
142 & 3423 & 12 59 29.4 & 27 51 00 &  6797 &  34 &  16.4 &  15.95 &  14.07 \nl
143 & 3201 & 12 59 44.4 & 27 54 45 &  6679 &  32 &  19.6 &  15.70 &  13.79 \nl
144 & 3510 & 12 59 23.2 & 27 54 45 &  6821 &  36 &  11.5 &  14.46 &  12.59 \nl
145 & 3557 & 12 59 20.2 & 27 53 09 &  6468 &  32 &  17.9 &  16.53 &  14.70 \nl
146 & 3012 & 13 00  1.5 & 27 43 51 &  8051 &  41 &   7.4 &  17.52 &  15.74 \nl
147 & 3493 & 12 59 24.9 & 27 44 19 &  6027 &  32 &  17.5 &  16.55 &  14.68 \nl
148 & 3313 & 12 59 37.0 & 27 49 32 &  6244 &  32 &  15.7 &  17.64 &  15.83 \nl
149 & 3403 & 12 59 30.7 & 27 47 29 &  7762 &  31 &  20.3 &  17.06 &  15.23 \nl
150 & 3262 & 12 59 39.9 & 27 51 17 &  3747 &  40 &   6.9 &  16.77 &  14.94 \nl
151 & 3339 & 12 59 35.3 & 27 51 49 &  6274 &  32 &  14.9 &  17.81 &  15.91 \nl
152 & 3126 & 12 59 51.0 & 27 49 58 &  7921 &  33 &  14.1 &  17.69 &  15.74 \nl
153 & 3730 & 12 59  8.2 & 27 47 02 &  7033 &  33 &  18.0 &  15.37 &  13.40 \nl
155 & 3588 & 12 59 18.5 & 27 30 48 &  6009 &  37 &   8.2 &  18.01 &  16.31 \nl
156 & 3829 & 12 59  1.6 & 27 32 13 &  8570 &  42 &   6.9 &  17.73 &  15.80 \nl
158 & 3739 & 12 59  7.4 & 27 46 06 &  6336 &  33 &  14.9 &  15.80 &  13.94 \nl
160 & 3958 & 12 58 52.1 & 27 47 05 &  5662 &  33 &  15.8 &  15.96 &  14.14 \nl
161 & 3660 & 12 59 13.5 & 27 46 28 &  6818 &  31 &  20.8 &  15.85 &  13.91 \nl
162 & 3697 & 12 59 10.3 & 27 37 11 &  5725 &  32 &  17.2 &  16.67 &  14.80 \nl
163 & 3879 & 12 58 58.1 & 27 35 41 &  5998 &  32 &  18.7 &  16.39 &  14.52 \nl
164 & 3997 & 12 58 48.7 & 27 48 37 &  5918 &  33 &  15.9 &  15.37 &  13.48 \nl
165 & 4017 & 12 58 47.4 & 27 40 29 &  8378 &  32 &  18.8 &  15.04 &  13.11 \nl
166 & 3782 & 12 59  4.6 & 27 54 39 &  6408 &  32 &  19.2 &  16.75 &  14.84 \nl
167 & 3943 & 12 58 53.0 & 27 48 48 &  5523 &  33 &  13.9 &  16.96 &  15.11 \nl
168 & 3779 & 12 59  5.3 & 27 38 39 &  5387 &  33 &  11.9 &  15.36 &  13.85 \nl
169 & 3733 & 12 59  8.0 & 27 51 18 &  6556 &  32 &  19.4 &  15.87 &  13.99 \nl
172 & 3896 & 12 58 56.1 & 27 50 00 &  7473 &  33 &  13.8 &  14.99 &  13.27 \nl
173 & 4522 & 12 58  1.5 & 27 29 22 &  7633 &  31 &  20.3 &  15.78 &  13.92 \nl
174 & 4502 & 12 58  3.5 & 27 40 56 &  7220 &  32 &  16.7 &  18.02 &  16.10 \nl
175 & 4447 & 12 58  9.7 & 27 32 57 &  6970 &  33 &  15.0 &  17.83 &  15.89 \nl
178 & 4579 & 12 57 56.3 & 27 34 52 &  4984 &  31 &  21.6 &  16.94 &  15.81 \nl
179 & 4209 & 12 58 31.6 & 27 40 24 &  6910 &  32 &  18.3 &  17.01 &  15.13 \nl
180 & 4255 & 12 58 28.4 & 27 33 33 &  7554 &  36 &  11.6 &  16.63 &  14.94 \nl
181 & 4469 & 12 58  6.8 & 27 34 36 &  7504 &  36 &   9.0 &  17.85 &  15.91 \nl
182 & 4156 & 12 58 35.2 & 27 35 47 &  7694 &  35 &  13.8 &  14.71 &  12.96 \nl
183 & 4060 & 12 58 42.6 & 27 45 37 &  8662 &  66 &   5.7 &  17.79 &  16.48 \nl
184 & 4348 & 12 58 18.2 & 27 50 54 &  7612 &  45 &   8.8 &  18.04 &  16.69 \nl
185 & 4117 & 12 58 38.4 & 27 32 38 &  5964 &  32 &  17.1 &  16.77 &  14.90 \nl
187 & 4083 & 12 58 40.8 & 27 49 37 &  6163 &  37 &   8.2 &  17.81 &  15.98 \nl
190 & 4341 & 12 58 19.2 & 27 45 43 &  5420 &  32 &  15.1 &  17.32 &  15.52 \nl
191 & 4499 & 12 58  3.5 & 27 48 53 &  7118 &  32 &  20.4 &  16.18 &  14.33 \nl
192 & 4314 & 12 58 22.0 & 27 53 32 &  7276 &  32 &  16.0 &  17.76 &  15.89 \nl
193 & 4731 & 12 57 43.0 & 26 51 08 &  6356 &  31 &  20.3 &  15.85 &  13.99 \nl
194 & 4793 & 12 57 36.5 & 27 01 52 &  7328 &  32 &  19.6 &  16.27 &  14.28 \nl
196 & 4582 & 12 57 56.7 & 27 02 14 &  7419 &  33 &  13.3 &  16.70 &  14.83 \nl
197 & 4518 & 12 58  3.2 & 26 54 57 &  8188 &  32 &  18.0 &  15.98 &  14.11 \nl
201 & 4463 & 12 58  9.2 & 26 39 51 &  7314 &  31 &  17.0 &  17.34 &  15.78 \nl
202 & 4912 & 12 57 26.7 & 26 41 37 &  7248 &  34 &  11.0 &  17.70 &  15.77 \nl
204 & 4692 & 12 57 45.7 & 27 25 45 &  8334 &  34 &  13.0 &  17.56 &  15.73 \nl
205 & 4479 & 12 58  6.1 & 27 25 08 &  5759 &  34 &  11.3 &  17.62 &  15.76 \nl
209 & 4918 & 12 57 25.3 & 27 24 16 &  4863 &  31 &  21.3 &  16.31 &  14.37 \nl
210 & 4535 & 12 58  0.8 & 27 27 14 &  7662 &  32 &  16.4 &  17.95 &  16.04 \nl
212 & 4961 & 12 57 21.5 & 27 01 22 &  8009 &  33 &  14.6 &  17.80 &  16.00 \nl
213 & 4928 & 12 57 23.9 & 27 29 46 &  7329 &  41 &   8.5 &  15.43 &  13.40 \nl
216 & 4714 & 12 57 43.2 & 27 34 39 &  7204 &  33 &  14.2 &  17.71 &  15.87 \nl
218 & 4829 & 12 57 32.8 & 27 36 37 &  6052 &  34 &  14.5 &  14.92 &  12.97 \nl
219 & 4794 & 12 57 35.8 & 27 29 35 &  7299 &  32 &  20.1 &  15.41 &  13.42 \nl
220 & 4907 & 12 57 25.8 & 27 32 46 &  5580 &  33 &  16.0 &  16.09 &  14.21 \nl
221 & 5051 & 12 57  9.4 & 27 27 59 &  7447 &  32 &  19.4 &  15.55 &  13.60 \nl
222 & 4597 & 12 57 54.4 & 27 29 26 &  4962 &  34 &  10.8 &  16.47 &  14.62 \nl
223 & 4933 & 12 57 23.5 & 27 45 58 &  9115 &  32 &  19.0 &  16.39 &  14.43 \nl
224 & 4653 & 12 57 48.1 & 27 52 58 &  5849 &  32 &  19.0 &  15.72 &  13.88 \nl
225 & 4664 & 12 57 47.3 & 27 49 59 &  6041 &  32 &  19.1 &  16.25 &  14.39 \nl
226 & 5100 & 12 57  4.2 & 27 43 48 &  8884 &  36 &   9.9 &  17.34 &  15.63 \nl
227 & 4945 & 12 57 21.7 & 27 52 49 &  7427 &  32 &  18.7 &  16.72 &  15.00 \nl
228 & 4679 & 12 57 46.2 & 27 45 25 &  6133 &  32 &  18.0 &  16.11 &  14.17 \nl
229 & 4987 & 12 57 16.8 & 27 37 06 &  7238 &  32 &  18.7 &  16.86 &  14.97 \nl
231 & 4974 & 12 57 17.8 & 27 48 39 &  7136 &  33 &  15.7 &  16.59 &  14.98 \nl
232 & 5096 & 12 57  4.6 & 27 46 22 &  7577 &  36 &   9.0 &  17.00 &  15.17 \nl
234 & 5290 & 12 56 43.2 & 26 44 30 &  6950 &  33 &  15.2 &  17.44 &  15.56 \nl
236 & 5365 & 12 56 34.6 & 27 13 38 &  7197 &  39 &   8.5 &  16.60 &  14.87 \nl
237 & 5136 & 12 57  1.7 & 27 22 19 &  6979 &  32 &  19.9 &  16.66 &  14.70 \nl
238 & 5514 & 12 56 20.5 & 26 42 14 &  6558 &  33 &  15.0 &  17.39 &  15.54 \nl
240 & 5395 & 12 56 32.0 & 27 03 19 &  6105 &  32 &  18.5 &  16.19 &  14.33 \nl
241 & 5226 & 12 56 51.2 & 26 53 56 &  6247 &  32 &  19.9 &  15.42 &  13.68 \nl
242 & 5424 & 12 56 29.1 & 26 57 25 &  7035 &  32 &  19.8 &  15.20 &  13.32 \nl
243 & 5526 & 12 56 16.7 & 27 26 45 &  6386 &  31 &  20.6 &  16.50 &  14.56 \nl
244 & 5422 & 12 56 28.5 & 27 17 29 &  7526 &  33 &  10.7 &  15.74 &  14.18 \nl
246 & 5234 & 12 56 49.7 & 27 05 37 &  6934 &  33 &  14.3 &  15.89 &  13.87 \nl
247 & 5250 & 12 56 47.8 & 27 25 15 &  7758 &  32 &  17.2 &  17.12 &  15.29 \nl
248 & 5279 & 12 56 43.5 & 27 10 43 &  7607 &  33 &  18.4 &  14.00 &  11.98 \nl
249 & 5283 & 12 56 43.5 & 27 02 04 &  5586 &  32 &  18.1 &  16.88 &  15.03 \nl
250 & 5038 & 12 57 10.8 & 27 24 17 &  6193 &  33 &  13.0 &  16.25 &  14.33 \nl
251 & 5254 & 12 56 47.4 & 27 17 32 &  7854 &  36 &   9.4 &  17.86 &  16.15 \nl
252 & 5032 & 12 57 12.0 & 27 06 11 &  7333 &  34 &  11.2 &  17.66 &  15.78 \nl
253 & 5102 & 12 57  4.3 & 27 31 33 &  8296 &  33 &  13.0 &  17.66 &  15.70 \nl
254 & 5495 & 12 56 19.8 & 27 45 03 &  6859 &  32 &  19.7 &  15.80 &  13.87 \nl
255 & 5912 & 12 55 35.7 & 27 46 01 &  6680 &  33 &  14.4 &  17.17 &  15.29 \nl
256 & 5546 & 12 56 14.6 & 27 30 22 &  7422 &  32 &  18.3 &  17.51 &  15.60 \nl
257 & 5850 & 12 55 44.3 & 27 42 59 &  7138 &  32 &  14.7 &  18.03 &  16.06 \nl
258 & 5599 & 12 56  9.9 & 27 50 39 &  7566 &  31 &  21.9 &  15.96 &  13.95 \nl
259 & 5643 & 12 56  6.1 & 27 40 40 &  4909 &  31 &  21.3 &  16.00 &  14.51 \nl
260 & 5362 & 12 56 34.2 & 27 41 14 &  6872 &  32 &  17.4 &  17.83 &  15.93 \nl
261 & 5428 & 12 56 26.6 & 27 49 50 &  6268 &  32 &  20.4 &  15.69 &  13.80 \nl
262 & 5434 & 12 56 26.3 & 27 43 38 &  6823 &  31 &  21.2 &  16.96 &  15.03 \nl
264 & 5926 & 12 55 34.1 & 27 50 30 &  6997 &  31 &  20.6 &  16.94 &  15.00 \nl
266 & 5568 & 12 56 13.4 & 27 45 00 &  6817 &  91 &   2.9 &  15.19 &  13.66 \nl
267 & 5721 & 12 55 57.5 & 27 54 17 &  6551 &  32 &  14.9 &  17.44 &  15.48 \nl
268 & 5364 & 12 56 34.2 & 27 32 20 &  7086 &  32 &  20.1 &  16.03 &  14.07 \nl
271 & 5641 & 12 56  6.5 & 27 38 52 &  7593 &  32 &  19.9 &  16.74 &  14.88 \nl
272 & 5799 & 12 55 49.0 & 27 54 21 &  6930 &  31 &  21.4 &  16.49 &  14.57 \nl
273 & 6545 & 12 54 16.0 & 27 18 13 &  6422 &  32 &  20.1 &  15.32 &  13.32 \nl
275 & 6617 & 12 54  5.5 & 27 04 07 &  7589 &  32 &  18.9 &  15.41 &  13.46 \nl
279 & 6409 & 12 54 36.8 & 26 56 05 &  5913 &  32 &  16.5 &  15.59 &  13.75 \nl
280 & 5704 & 12 56  1.7 & 26 45 23 &  5722 &  33 &  15.3 &  16.57 &  14.67 \nl
282 & 6703 & 12 53 45.5 & 27 14 58 &  6902 &  35 &  11.4 &  17.73 &  15.88 \nl
285 & 6690 & 12 53 46.6 & 27 23 09 &  8317 &  43 &   5.0 &  17.84 &  15.92 \nl
287 & 6503 & 12 54 22.2 & 27 05 02 &  8427 &  32 &  21.1 &  16.51 &  14.52 \nl
289 & 5886 & 12 55 41.3 & 27 15 02 &  7105 &  33 &  17.8 &  15.05 &  13.06 \nl
291 & 5676 & 12 56  4.0 & 27 09 01 &  7337 &  33 &  15.8 &  17.97 &  16.05 \nl
292 & 6474 & 12 54 24.7 & 27 21 50 &  7833 &  39 &   8.6 &  17.55 &  15.64 \nl
293 & 6219 & 12 55  0.8 & 27 29 00 &  6968 &  36 &   7.9 &  17.34 &  15.51 \nl
294 & 6421 & 12 54 33.2 & 27 37 58 &  7265 &  33 &  13.8 &  15.96 &  13.91 \nl
295 & 6043 & 12 55 20.0 & 27 51 59 &  6978 &  32 &  16.3 &  16.33 &  14.46 \nl
296 & 5975 & 12 55 29.1 & 27 31 17 &  6983 &  32 &  21.5 &  14.36 &  12.49 \nl
297 & 6390 & 12 54 37.4 & 27 41 31 &  8453 &  32 &  19.1 &  17.66 &  15.79 \nl
302 &    - & 12 53 37.9 & 27 47 03 &  5088 &  32 &  18.2 &  15.80 &  13.89 \nl
303 & 5999 & 12 55 25.0 & 27 47 53 &  7382 &  32 &  19.7 &  15.20 &  12.93 \nl
304 & 6109 & 12 55 13.9 & 27 42 32 &  6647 &  32 &  15.1 &  17.70 &  15.76 \nl
306 &    - & 12 53 30.3 & 27 40 30 &  5919 &  31 &  20.8 &  17.50 &  15.59 \nl
307 & 5960 & 12 55 30.6 & 27 32 39 &  7024 &  32 &  14.8 &  16.76 &  14.80 \nl
308 & 6701 & 12 53 44.1 & 27 46 51 &  6640 &  32 &  16.5 &  17.81 &  15.96 \nl
309 & 6479 & 12 54 22.7 & 27 44 40 &  7102 &  36 &   9.5 &  18.04 &  16.27 \nl
311 &    - & 12 53 35.5 & 27 45 31 &  7463 &  33 &  14.4 &  16.95 &  15.08 \nl
312 & 5978 & 12 55 27.8 & 27 39 22 &  6997 &  32 &  18.8 &  15.33 &  13.36 \nl

\tablenotetext{a}{Position of the fiber. For some objects this is a bit off
of the galaxy.}
\enddata
\end{deluxetable}

\begin{deluxetable}{cccccccccccccccc}
{\tiny
\tablenum{2}
\tablecolumns{16}
\tablecaption{Galaxy Classifications.
\label{tbl2}}
\tablehead
{
Fiber   & Line & PCA & Wavelet & 
\multicolumn{3}{c}{Spectral\_1} & \multicolumn{3}{c}{Spectral\_2} & 
\multicolumn{6}{c}{Principal Component Analysis coefficients}\\
number  & Strengths &     &         & 
age (Gyr) & SF & $\chi^2/\nu$ & age (Gyr) & $\tau$ (Gyr) & $\chi^2/\nu$ &
$c_1$ & $c_2$ & $c_3$ & $\theta$ & $\phi$ & radius$^2$
}

\startdata

  3 & AB    & AB    & AB &  5.00 &  1 &  0.78 &   4.41 &   0.009 &  0.78 &  0.781 & -0.069 &  0.128 &   9.30 &  -5.03 & 0.632 \nl
  4 & AB    & AB    & AB &  6.14 &  2 &  0.51 &   4.20 &   0.607 &  0.51 &  0.846 & -0.071 &  0.109 &   7.33 &  -4.80 & 0.733 \nl
 10 & AB    & AB    & AB & 17.00 &  3 &  0.55 &  12.81 &   2.117 &  0.58 &  0.970 & -0.112 & -0.023 &  -1.35 &  -6.58 & 0.954 \nl
 11 & AB    & AB    & AB & 13.79 &  1 &  2.68 &  18.86 &   2.718 &  2.45 &  0.948 & -0.181 & -0.178 & -10.48 & -10.79 & 0.962 \nl
 12 & EM+AB & EM    & EM &  2.26 &  3 &  0.79 &   1.21 &   0.368 &  0.93 &  0.335 &  0.843 & -0.219 & -13.53 &  68.31 & 0.871 \nl
 13 & AB    & AB+EM & AB &  1.98 &  1 &  0.91 &   1.20 &   0.002 &  0.90 &  0.534 &  0.243 &  0.239 &  22.16 &  24.52 & 0.401 \nl
 14 & AB    & AB    & AB & 18.76 &  5 &  0.61 &   7.53 &   1.284 &  0.71 &  0.974 & -0.044 &  0.046 &   2.72 &  -2.60 & 0.953 \nl
 21 & AB    & AB    & AB & 17.04 &  2 &  2.52 &  16.63 &   2.718 &  2.62 &  0.957 & -0.127 & -0.040 &  -2.39 &  -7.55 & 0.935 \nl
 27 & AB    & AB    & AB & 19.00 &  4 &  0.96 &  12.27 &   2.117 &  1.06 &  0.979 & -0.022 &  0.062 &   3.64 &  -1.26 & 0.962 \nl
 28 & AB    & AB    & AB & 19.00 &  4 &  0.47 &  12.35 &   2.117 &  0.52 &  0.946 &  0.079 &  0.175 &  10.47 &   4.77 & 0.931 \nl
 29 & AB    & AB    & AB & 17.79 &  4 &  0.44 &  11.76 &   2.117 &  0.50 &  0.915 &  0.008 &  0.072 &   4.51 &   0.52 & 0.843 \nl
 30 & AB    & AB    & AB & 16.26 &  5 &  0.17 &   3.43 &   0.472 &  0.18 &  0.662 &  0.200 &  0.286 &  22.49 &  16.82 & 0.561 \nl
 31 & EM+AB & EM    & EM &  3.04 &  3 &  7.99 &   1.25 &   0.287 &  8.81 &  0.478 &  0.796 & -0.097 &  -5.97 &  59.01 & 0.872 \nl
 32 & AB    & AB    & AB & 19.00 &  4 &  0.74 &  12.58 &   2.117 &  0.82 &  0.988 & -0.041 &  0.025 &   1.44 &  -2.40 & 0.979 \nl
 34 & AB+EM & AB+EM & EM & 17.88 &  5 &  2.28 &  19.00 &   4.482 &  2.69 &  0.786 &  0.288 & -0.259 & -17.16 &  20.12 & 0.768 \nl
 40 & AB    & AB    & AB & 16.98 &  5 &  0.42 &   3.50 &   0.472 &  0.48 &  0.758 & -0.023 &  0.091 &   6.87 &  -1.75 & 0.584 \nl
 43 & AB    & AB    & AB & 17.16 &  4 &  0.62 &  11.48 &   2.117 &  0.72 &  0.958 &  0.023 &  0.145 &   8.60 &   1.36 & 0.939 \nl
 45 & PS    & PS    & PS &  1.26 &  1 &  0.37 &   0.60 &   0.018 &  0.39 &  0.483 &  0.484 &  0.587 &  40.61 &  45.06 & 0.813 \nl
 50 & AB    & AB    & AB & 19.00 &  2 &  2.47 &  18.18 &   2.718 &  2.36 &  0.987 & -0.116 & -0.008 &  -0.47 &  -6.73 & 0.988 \nl
 55 & AB    & AB    & AB & 13.46 &  2 &  0.89 &  11.02 &   1.649 &  0.94 &  0.984 & -0.067 &  0.005 &   0.31 &  -3.91 & 0.973 \nl
 58 & PS    & PS    & PS &  1.46 &  2 &  4.11 &   1.12 &   0.174 &  3.69 &  0.696 &  0.518 &  0.374 &  23.32 &  36.69 & 0.892 \nl
 61 & AB    & AB    & AB & 19.00 &  5 &  0.65 &  13.58 &   2.718 &  0.78 &  0.929 &  0.087 &  0.195 &  11.78 &   5.37 & 0.908 \nl
 65 & AB    & AB    & AB & 19.00 &  4 &  1.10 &  12.59 &   2.117 &  1.21 &  0.978 & -0.036 &  0.068 &   3.96 &  -2.11 & 0.962 \nl
 66 & AB    & AB    & AB & 18.61 &  4 &  0.85 &   9.97 &   1.649 &  0.96 &  0.984 & -0.046 &  0.062 &   3.59 &  -2.65 & 0.974 \nl
 68 & AB    & AB    & AB &  9.82 &  2 &  0.60 &   6.93 &   1.000 &  0.62 &  0.976 &  0.024 &  0.087 &   5.10 &   1.41 & 0.960 \nl
 69 & AB    & AB    & AB & 18.27 &  2 &  2.63 &  14.62 &   2.117 &  2.66 &  0.971 & -0.134 & -0.083 &  -4.83 &  -7.86 & 0.968 \nl
 70 & AB    & AB    & AB & 14.57 &  2 &  1.67 &  11.46 &   1.649 &  1.70 &  0.978 & -0.117 & -0.017 &  -1.00 &  -6.82 & 0.971 \nl
 74 & AB    & AB    & AB & 19.00 &  1 &  3.75 &  19.00 &   0.011 &  3.73 &  0.841 & -0.100 &  0.015 &   1.01 &  -6.76 & 0.717 \nl
 75 & AB    & AB    & AB & 18.00 &  1 & 15.71 &  19.00 &   0.002 & 15.19 &  0.954 & -0.122 & -0.056 &  -3.34 &  -7.29 & 0.928 \nl
 76 & AB    & AB    & AB & 13.00 &  1 &  3.19 &  14.60 &   1.649 &  3.18 &  0.844 & -0.008 &  0.177 &  11.87 &  -0.56 & 0.743 \nl
 84 & EM+AB & EM    & AB &  3.98 &  4 &  0.29 &   2.09 &   0.607 &  0.36 &  0.388 &  0.788 & -0.240 & -15.26 &  63.79 & 0.829 \nl
 85 & AB+EM & AB+EM & AB & 18.41 &  2 &  2.65 &  19.00 &   3.490 &  2.67 &  0.805 &  0.409 & -0.060 &  -3.78 &  26.92 & 0.819 \nl
 86 & AB    & AB    & AB & 19.00 &  3 &  0.98 &  16.14 &   2.718 &  1.05 &  0.981 & -0.066 &  0.035 &   2.06 &  -3.84 & 0.967 \nl
 87 & AB    & AB    & AB & 17.07 &  1 &  0.48 &  19.00 &   2.117 &  0.47 &  0.934 & -0.058 &  0.049 &   3.02 &  -3.57 & 0.878 \nl
 88 & AB    & AB    & AB & 13.83 &  1 &  2.25 &  18.94 &   2.718 &  2.10 &  0.971 & -0.143 & -0.089 &  -5.20 &  -8.40 & 0.972 \nl
 89 & AB    & AB    & AB & 17.58 &  2 &  0.70 &  14.46 &   2.117 &  0.73 &  0.967 &  0.004 &  0.128 &   7.54 &   0.22 & 0.952 \nl
 91 & AB+EM & PS    & PS &  1.98 &  1 &  1.23 &   1.20 &   0.002 &  1.20 &  0.755 &  0.463 &  0.420 &  25.39 &  31.52 & 0.961 \nl
 94 & AB    & AB    & AB & 17.53 &  1 &  2.41 &  16.74 &   0.011 &  2.33 &  0.950 & -0.204 & -0.165 &  -9.64 & -12.14 & 0.970 \nl
 95 & PS    & PS    & PS &  1.39 &  1 &  1.79 &   0.89 &   0.105 &  1.92 &  0.544 &  0.528 &  0.564 &  36.65 &  44.12 & 0.893 \nl
 96 & AB    & AB    & AB & 16.00 &  1 &  1.12 &  18.37 &   2.117 &  1.08 &  0.969 & -0.166 & -0.121 &  -7.02 &  -9.73 & 0.981 \nl
 97 & AB    & AB    & AB & 19.00 &  2 &  0.83 &  17.57 &   2.718 &  0.86 &  0.983 & -0.109 & -0.053 &  -3.05 &  -6.35 & 0.982 \nl
 98 & EM+AB & EM    & EM &  1.96 &  3 &  4.15 &   0.86 &   0.223 &  4.62 &  0.328 &  0.851 & -0.178 & -11.06 &  68.94 & 0.864 \nl
 99 & AB    & AB    & AB & 15.49 &  1 &  1.50 &  17.97 &   2.117 &  1.44 &  0.962 & -0.187 & -0.131 &  -7.59 & -10.99 & 0.977 \nl
100 & AB    & AB    & AB & 18.22 &  2 &  0.52 &  17.21 &   2.718 &  0.53 &  0.962 &  0.029 &  0.071 &   4.20 &   1.70 & 0.931 \nl
101 & AB    & AB+EM & AB & 19.00 &  1 &  0.20 &  19.00 &   0.002 &  0.19 &  0.435 &  0.164 &  0.192 &  22.41 &  20.69 & 0.253 \nl
102 & AB    & AB    & PS & 15.67 &  4 &  0.63 &   3.89 &   0.472 &  0.66 &  0.909 &  0.169 &  0.292 &  17.52 &  10.56 & 0.941 \nl
103 & AB    & AB    & AB & 16.17 &  2 &  1.03 &  13.88 &   2.117 &  1.08 &  0.982 & -0.078 & -0.032 &  -1.86 &  -4.56 & 0.971 \nl
105 & AB    & AB    & AB & 17.35 &  3 &  0.69 &  15.44 &   2.718 &  0.73 &  0.902 & -0.107 & -0.068 &  -4.30 &  -6.79 & 0.830 \nl
106 & AB+EM & AB+EM & AB & 18.09 &  5 &  2.06 &  13.15 &   2.718 &  2.52 &  0.791 &  0.342 &  0.009 &   0.62 &  23.40 & 0.743 \nl
107 & AB    & AB    & AB & 16.50 &  2 &  0.87 &  16.51 &   2.718 &  0.89 &  0.937 & -0.071 & -0.043 &  -2.61 &  -4.31 & 0.886 \nl
108 & AB    & AB    & AB & 14.27 &  4 &  1.12 &   4.11 &   0.607 &  1.15 &  0.964 & -0.029 & -0.004 &  -0.24 &  -1.73 & 0.930 \nl
110 & EM+AB & EM    & EM &  2.56 &  3 &  2.54 &   1.13 &   0.287 &  2.83 &  0.260 &  0.751 & -0.465 & -30.31 &  70.92 & 0.848 \nl
111 & AB    & AB    & AB & 17.33 &  2 &  0.50 &  16.82 &   2.718 &  0.51 &  0.934 & -0.002 &  0.034 &   2.07 &  -0.14 & 0.874 \nl
115 & EM    & EM    & EM &  0.02 &  5 &  6.54 &   0.06 &  90.017 &  6.04 & -0.086 &  0.534 & -0.529 & -44.34 &  80.80 & 0.572 \nl
117 & EM+AB & EM    & EM &  4.62 &  4 &  3.05 &   1.58 &   0.368 &  3.60 &  0.449 &  0.849 & -0.093 &  -5.54 &  62.11 & 0.932 \nl
120 & AB    & AB    & AB & 17.58 &  1 &  4.64 &  19.00 &   2.117 &  4.47 &  0.947 & -0.160 & -0.145 &  -8.57 &  -9.59 & 0.942 \nl
123 & AB    & AB    & AB & 10.33 &  2 &  1.47 &   5.94 &   0.779 &  1.56 &  0.977 & -0.143 & -0.062 &  -3.59 &  -8.33 & 0.978 \nl
124 & AB    & AB    & AB &  4.01 &  1 &  0.96 &   3.46 &   0.105 &  0.95 &  0.979 & -0.069 &  0.022 &   1.31 &  -4.01 & 0.964 \nl
126 & EM    & EM    & EM &  2.57 &  6 &  5.02 &   1.18 &   0.472 &  6.09 &  0.146 &  0.810 & -0.467 & -29.58 &  79.78 & 0.895 \nl
127 & EM    & EM    & EM & 12.66 &  7 &  2.13 &  17.58 &   9.488 &  1.95 &  0.142 &  0.778 & -0.552 & -34.91 &  79.63 & 0.930 \nl
130 & AB    & AB    & AB & 13.61 &  2 &  1.45 &  11.06 &   1.649 &  1.45 &  0.985 & -0.080 &  0.019 &   1.09 &  -4.62 & 0.976 \nl
131 & AB+EM & AB+EM & PS & 11.54 &  5 &  0.54 &   2.75 &   0.472 &  0.60 &  0.739 &  0.459 &  0.144 &   9.36 &  31.83 & 0.778 \nl
132 & AB+EM & AB+EM & AB & 15.82 &  6 &  0.67 &   2.37 &   0.368 &  0.74 &  0.645 &  0.268 & -0.080 &  -6.52 &  22.56 & 0.494 \nl
133 & AB    & AB    & AB & 19.00 &  3 &  0.54 &  16.03 &   2.718 &  0.56 &  0.965 & -0.067 & -0.031 &  -1.80 &  -3.97 & 0.938 \nl
134 & AB    & AB    & AB & 13.68 &  2 &  0.33 &  11.21 &   1.649 &  0.33 &  0.915 & -0.015 &  0.025 &   1.55 &  -0.92 & 0.837 \nl
135 & AB    & AB    & AB & 18.21 &  2 &  2.06 &  14.62 &   2.117 &  2.12 &  0.969 & -0.110 & -0.061 &  -3.58 &  -6.48 & 0.955 \nl
136 & AB    & AB    & AB & 14.25 &  1 &  1.87 &  19.00 &   2.718 &  1.77 &  0.970 & -0.101 &  0.015 &   0.87 &  -5.95 & 0.951 \nl
137 & AB    & AB    & AB & 17.83 &  2 &  3.63 &  14.50 &   2.117 &  3.68 &  0.935 & -0.173 & -0.163 &  -9.75 & -10.46 & 0.932 \nl
138 & EM+AB & EM    & AB &  0.97 &  2 &  2.10 &   0.94 &   0.223 &  2.30 &  0.437 &  0.782 & -0.146 &  -9.26 &  60.78 & 0.824 \nl
139 & AB    & AB    & AB & 18.83 &  2 &  1.75 &  14.95 &   2.117 &  1.71 &  0.978 & -0.111 &  0.013 &   0.76 &  -6.45 & 0.970 \nl
140 & AB+EM & PS    & PS &  1.22 &  1 &  1.22 &   0.76 &   0.135 &  1.24 &  0.511 &  0.570 &  0.492 &  32.75 &  48.10 & 0.828 \nl
141 & AB+EM & AB    & AB & 18.80 &  3 &  0.73 &  15.82 &   2.718 &  0.75 &  0.964 &  0.019 &  0.059 &   3.51 &   1.13 & 0.932 \nl
142 & AB    & AB    & AB & 15.10 &  1 &  2.88 &  17.74 &   2.117 &  2.75 &  0.945 & -0.179 & -0.180 & -10.59 & -10.72 & 0.958 \nl
143 & AB    & AB    & AB & 16.00 &  2 &  2.19 &  13.82 &   2.117 &  2.27 &  0.986 & -0.108 & -0.072 &  -4.14 &  -6.24 & 0.989 \nl
144 & AB    & AB    & AB & 19.00 &  1 &  0.46 &  18.95 &   0.050 &  0.45 &  0.932 & -0.121 & -0.105 &  -6.35 &  -7.39 & 0.893 \nl
145 & AB    & AB    & AB & 13.52 &  2 &  0.98 &  11.10 &   1.649 &  0.99 &  0.979 & -0.094 & -0.005 &  -0.29 &  -5.51 & 0.967 \nl
146 & AB    & AB    & AB & 19.00 &  5 &  0.37 &   6.48 &   1.000 &  0.40 &  0.820 &  0.157 &  0.209 &  14.07 &  10.82 & 0.740 \nl
147 & AB    & AB    & AB & 12.31 &  1 &  1.71 &  15.80 &   2.117 &  1.64 &  0.960 & -0.187 & -0.160 &  -9.29 & -11.01 & 0.983 \nl
148 & AB    & AB    & AB & 14.14 &  2 &  1.03 &  11.34 &   1.649 &  1.05 &  0.965 & -0.111 & -0.054 &  -3.17 &  -6.58 & 0.947 \nl
149 & AB    & AB    & AB & 14.14 &  2 &  1.12 &  11.31 &   1.649 &  1.12 &  0.984 & -0.026 &  0.032 &   1.84 &  -1.52 & 0.971 \nl
150 & AB    & AB    & AB & 19.00 &  1 &  0.86 &  19.00 &   0.002 &  0.82 &  0.695 & -0.125 & -0.042 &  -3.44 & -10.21 & 0.500 \nl
151 & AB    & AB    & AB &  8.80 &  1 &  0.52 &   8.66 &   0.607 &  0.52 &  0.951 & -0.095 & -0.022 &  -1.32 &  -5.69 & 0.914 \nl
152 & AB    & AB    & AB & 14.02 &  1 &  0.59 &  16.92 &   2.117 &  0.58 &  0.962 &  0.019 &  0.061 &   3.62 &   1.15 & 0.930 \nl
153 & AB    & AB    & AB & 14.09 &  1 &  3.07 &  19.00 &   2.718 &  2.87 &  0.957 & -0.159 & -0.143 &  -8.37 &  -9.45 & 0.962 \nl
155 & AB    & AB    & AB & 15.35 &  4 &  0.29 &   4.42 &   0.607 &  0.31 &  0.808 & -0.021 &  0.091 &   6.46 &  -1.46 & 0.661 \nl
156 & AB    & AB    & AB & 17.50 &  5 &  0.57 &   5.90 &   1.000 &  0.62 &  0.851 &  0.054 &  0.161 &  10.67 &   3.66 & 0.753 \nl
158 & AB    & AB    & AB & 19.00 &  2 &  0.51 &  18.16 &   2.718 &  0.50 &  0.960 & -0.132 & -0.113 &  -6.63 &  -7.80 & 0.951 \nl
160 & AB    & AB    & AB & 16.05 &  2 &  1.99 &  11.95 &   1.649 &  1.98 &  0.952 & -0.184 & -0.150 &  -8.79 & -10.96 & 0.964 \nl
161 & AB    & AB    & AB & 16.49 &  2 &  1.54 &  13.97 &   2.117 &  1.59 &  0.984 & -0.118 & -0.070 &  -4.05 &  -6.82 & 0.987 \nl
162 & AB    & AB    & AB & 14.30 &  2 &  1.44 &  11.38 &   1.649 &  1.44 &  0.957 & -0.173 & -0.155 &  -9.05 & -10.26 & 0.970 \nl
163 & AB    & AB    & AB & 17.89 &  2 &  1.71 &  14.53 &   2.117 &  1.69 &  0.963 & -0.173 & -0.159 &  -9.25 & -10.19 & 0.982 \nl
164 & AB    & AB    & AB & 17.18 &  2 &  2.90 &  12.41 &   1.649 &  2.83 &  0.961 & -0.161 & -0.160 &  -9.34 &  -9.51 & 0.976 \nl
165 & AB    & AB    & AB & 19.00 &  2 &  2.01 &  15.25 &   2.117 &  1.98 &  0.943 & -0.156 & -0.101 &  -6.06 &  -9.41 & 0.923 \nl
166 & AB    & AB    & AB & 13.89 &  2 &  1.02 &  11.22 &   1.649 &  1.02 &  0.982 & -0.134 & -0.089 &  -5.11 &  -7.78 & 0.990 \nl
167 & AB    & AB    & AB & 12.87 &  1 &  0.70 &  16.08 &   2.117 &  0.68 &  0.938 & -0.147 & -0.105 &  -6.28 &  -8.90 & 0.913 \nl
168 & EM+AB & EM    & EM &  1.15 &  1 & 10.68 &   0.74 &   0.174 & 12.44 &  0.205 &  0.632 & -0.356 & -28.16 &  72.04 & 0.567 \nl
169 & AB+EM & AB    & AB & 18.31 &  2 &  2.44 &  17.08 &   2.718 &  2.54 &  0.945 & -0.173 & -0.175 & -10.33 & -10.38 & 0.954 \nl
172 & AB+EM & AB+EM & AB & 17.24 &  5 &  1.78 &   5.81 &   1.000 &  2.34 &  0.965 &  0.199 &  0.123 &   7.10 &  11.65 & 0.986 \nl
173 & AB    & AB    & AB & 19.00 &  2 &  1.59 &  17.96 &   2.718 &  1.55 &  0.983 & -0.092 & -0.019 &  -1.10 &  -5.34 & 0.974 \nl
174 & AB    & AB    & AB & 17.48 &  2 &  0.68 &  14.49 &   2.117 &  0.68 &  0.961 & -0.039 & -0.004 &  -0.25 &  -2.33 & 0.926 \nl
175 & AB    & AB    & AB & 16.49 &  2 &  0.92 &  16.54 &   2.718 &  0.92 &  0.949 & -0.012 & -0.023 &  -1.41 &  -0.71 & 0.902 \nl
178 & EM+AB & EM    & EM &  2.00 &  4 &  1.64 &   1.08 &   0.368 &  2.02 &  0.194 &  0.791 & -0.472 & -30.10 &  76.19 & 0.887 \nl
179 & AB    & AB    & AB & 14.41 &  2 &  1.60 &  13.25 &   2.117 &  1.63 &  0.980 & -0.084 & -0.071 &  -4.12 &  -4.89 & 0.972 \nl
180 & AB+EM & AB+EM & PS & 12.64 &  5 &  2.61 &   3.33 &   0.607 &  3.02 &  0.801 &  0.465 &  0.208 &  12.64 &  30.14 & 0.902 \nl
181 & AB    & AB    & AB & 17.33 &  4 &  0.55 &  11.57 &   2.117 &  0.61 &  0.872 &  0.119 &  0.066 &   4.28 &   7.77 & 0.779 \nl
182 & AB+EM & AB+EM & PS & 14.38 &  5 & 12.00 &   3.51 &   0.607 & 14.82 &  0.921 &  0.283 &  0.173 &  10.16 &  17.06 & 0.957 \nl
183 & PS    & PS    & PS &  1.01 &  2 &  0.67 &   0.77 &   0.135 &  0.69 &  0.378 &  0.462 &  0.536 &  41.90 &  50.70 & 0.644 \nl
184 & PS    & PS    & PS &  1.29 &  1 &  0.68 &   0.83 &   0.135 &  0.70 &  0.482 &  0.592 &  0.527 &  34.60 &  50.84 & 0.861 \nl
185 & AB    & AB    & AB & 15.20 &  2 &  1.57 &  13.58 &   2.117 &  1.56 &  0.962 & -0.133 & -0.107 &  -6.28 &  -7.89 & 0.956 \nl
187 & AB    & AB    & AB &  4.55 &  2 &  0.51 &   3.32 &   0.472 &  0.51 &  0.772 & -0.024 & -0.067 &  -4.98 &  -1.78 & 0.602 \nl
190 & AB    & AB    & AB &  5.06 &  2 &  0.79 &   3.40 &   0.472 &  0.77 &  0.938 & -0.015 & -0.019 &  -1.16 &  -0.93 & 0.880 \nl
191 & AB    & AB    & AB & 19.00 &  2 &  2.23 &  14.97 &   2.117 &  2.17 &  0.973 & -0.127 & -0.097 &  -5.63 &  -7.44 & 0.971 \nl
192 & AB    & AB    & AB & 13.14 &  2 &  0.69 &  12.86 &   2.117 &  0.71 &  0.946 &  0.036 &  0.013 &   0.81 &   2.21 & 0.897 \nl
193 & AB    & AB    & AB &  3.53 &  1 &  1.57 &   3.81 &   0.472 &  1.53 &  0.981 & -0.070 & -0.001 &  -0.08 &  -4.07 & 0.967 \nl
194 & AB    & AB    & AB & 18.06 &  2 &  1.90 &  16.94 &   2.718 &  1.96 &  0.975 & -0.103 & -0.093 &  -5.44 &  -6.02 & 0.969 \nl
196 & AB+EM & AB+EM & AB & 15.53 &  5 &  1.92 &   4.46 &   0.779 &  2.56 &  0.949 &  0.161 &  0.061 &   3.63 &   9.60 & 0.930 \nl
197 & AB    & AB    & AB & 14.26 &  2 &  3.11 &  11.34 &   1.649 &  3.11 &  0.960 & -0.084 &  0.031 &   1.82 &  -5.02 & 0.929 \nl
201 & EM+AB & EM    & AB &  3.05 &  3 &  0.83 &   1.47 &   0.368 &  0.94 &  0.522 &  0.776 & -0.128 &  -7.81 &  56.06 & 0.892 \nl
202 & AB+EM & AB+EM & AB & 17.17 &  3 &  0.81 &  15.31 &   2.718 &  0.83 &  0.867 &  0.222 &  0.168 &  10.63 &  14.38 & 0.829 \nl
204 & AB    & AB    & AB &  3.41 &  2 &  0.67 &   2.98 &   0.472 &  0.68 &  0.867 &  0.181 &  0.300 &  18.71 &  11.77 & 0.875 \nl
205 & AB    & AB    & AB & 18.35 &  4 &  0.68 &   9.96 &   1.649 &  0.71 &  0.874 & -0.063 & -0.029 &  -1.89 &  -4.13 & 0.769 \nl
209 & EM+AB & EM    & EM & 13.15 &  8 &  2.82 &   3.26 &   1.000 &  3.08 &  0.542 &  0.516 & -0.569 & -37.26 &  43.58 & 0.884 \nl
210 & AB    & AB    & AB & 16.90 &  3 &  0.81 &  12.76 &   2.117 &  0.84 &  0.969 &  0.028 &  0.037 &   2.21 &   1.65 & 0.941 \nl
212 & AB    & AB    & AB &  3.59 &  2 &  0.30 &   3.05 &   0.472 &  0.30 &  0.907 &  0.114 &  0.233 &  14.31 &   7.16 & 0.890 \nl
213 & AB    & AB    & AB & 14.85 &  1 &  0.43 &  19.00 &   2.718 &  0.41 &  0.918 & -0.043 & -0.106 &  -6.57 &  -2.66 & 0.855 \nl
216 & AB    & AB    & AB & 10.86 &  2 &  0.71 &   8.65 &   1.284 &  0.74 &  0.980 & -0.014 &  0.002 &   0.09 &  -0.84 & 0.961 \nl
218 & AB    & AB    & AB & 18.14 &  1 &  5.03 &  19.00 &   1.649 &  4.92 &  0.923 & -0.222 & -0.226 & -13.38 & -13.53 & 0.951 \nl
219 & AB+EM & AB    & AB & 15.40 &  1 &  3.41 &  17.87 &   2.117 &  3.24 &  0.955 & -0.153 & -0.148 &  -8.72 &  -9.10 & 0.957 \nl
220 & AB    & AB    & AB & 18.74 &  2 &  2.28 &  14.86 &   2.117 &  2.17 &  0.955 & -0.184 & -0.168 &  -9.80 & -10.89 & 0.973 \nl
221 & AB    & AB    & AB & 19.00 &  2 &  2.75 &  15.00 &   2.117 &  2.67 &  0.965 & -0.146 & -0.105 &  -6.16 &  -8.63 & 0.963 \nl
222 & AB+EM & AB+EM & AB &  9.99 &  4 &  0.72 &   2.87 &   0.472 &  0.76 &  0.882 &  0.221 & -0.071 &  -4.47 &  14.07 & 0.832 \nl
223 & AB    & AB    & AB & 13.21 &  3 &  0.74 &   9.70 &   1.649 &  0.80 &  0.957 & -0.069 &  0.038 &   2.29 &  -4.12 & 0.922 \nl
224 & AB    & AB    & AB & 16.00 &  2 &  1.64 &  13.82 &   2.117 &  1.62 &  0.966 & -0.159 & -0.143 &  -8.32 &  -9.33 & 0.980 \nl
225 & AB    & AB    & AB & 13.43 &  2 &  1.97 &  10.98 &   1.649 &  1.91 &  0.978 & -0.128 & -0.099 &  -5.75 &  -7.47 & 0.982 \nl
226 & AB+EM & AB+EM & AB &  2.87 &  2 &  0.34 &   2.42 &   0.368 &  0.34 &  0.764 &  0.256 &  0.266 &  18.25 &  18.50 & 0.719 \nl
227 & AB    & AB    & PS &  2.23 &  1 &  1.46 &   1.91 &   0.223 &  1.45 &  0.917 &  0.214 &  0.311 &  18.27 &  13.12 & 0.984 \nl
228 & AB    & AB    & AB &  7.30 &  1 &  0.44 &   9.37 &   1.284 &  0.42 &  0.976 & -0.063 & -0.010 &  -0.56 &  -3.69 & 0.956 \nl
229 & AB    & AB    & AB & 15.60 &  2 &  1.42 &  13.69 &   2.117 &  1.45 &  0.986 & -0.090 & -0.050 &  -2.90 &  -5.21 & 0.983 \nl
231 & AB    & AB+EM & PS &  1.82 &  1 &  0.85 &   1.20 &   0.105 &  0.82 &  0.845 &  0.318 &  0.378 &  22.73 &  20.63 & 0.959 \nl
232 & AB+EM & AB+EM & PS & 18.77 &  6 &  0.81 &   3.42 &   0.607 &  1.08 &  0.854 &  0.403 &  0.065 &   3.97 &  25.25 & 0.895 \nl
234 & AB    & AB    & AB &  7.19 &  2 &  0.55 &   4.34 &   0.607 &  0.54 &  0.975 & -0.040 &  0.078 &   4.56 &  -2.34 & 0.957 \nl
236 & AB+EM & AB+EM & PS &  2.40 &  2 &  0.97 &   2.19 &   0.368 &  1.01 &  0.871 &  0.375 &  0.125 &   7.54 &  23.31 & 0.915 \nl
237 & AB    & AB    & AB & 10.29 &  2 &  0.53 &   6.02 &   0.779 &  0.51 &  0.987 & -0.104 & -0.029 &  -1.70 &  -6.00 & 0.986 \nl
238 & AB    & AB    & AB &  3.13 &  1 &  0.47 &   3.05 &   0.368 &  0.46 &  0.985 & -0.098 & -0.021 &  -1.20 &  -5.67 & 0.980 \nl
240 & AB    & AB    & AB &  7.49 &  2 &  1.74 &   5.13 &   0.779 &  1.79 &  0.975 & -0.154 & -0.104 &  -6.00 &  -8.98 & 0.985 \nl
241 & AB+EM & EM    & EM & 17.71 &  6 &  2.13 &  18.72 &   5.755 &  1.80 &  0.564 &  0.459 & -0.304 & -22.66 &  39.15 & 0.621 \nl
242 & AB    & AB    & AB & 12.87 &  2 &  1.74 &  10.80 &   1.649 &  1.78 &  0.989 & -0.107 & -0.047 &  -2.68 &  -6.18 & 0.992 \nl
243 & AB    & AB    & AB & 15.69 &  2 &  0.88 &  11.83 &   1.649 &  0.87 &  0.974 & -0.168 & -0.095 &  -5.52 &  -9.79 & 0.985 \nl
244 & EM+AB & EM    & AB &  1.27 &  2 &  0.99 &   1.30 &   0.287 &  1.05 &  0.620 &  0.676 & -0.005 &  -0.29 &  47.49 & 0.841 \nl
246 & AB+EM & AB    & AB & 19.00 &  2 &  1.67 &  17.83 &   2.718 &  1.74 &  0.981 & -0.006 & -0.071 &  -4.14 &  -0.33 & 0.967 \nl
247 & AB    & AB    & AB &  2.93 &  1 &  0.52 &   3.24 &   0.472 &  0.50 &  0.968 &  0.084 &  0.188 &  10.97 &   4.98 & 0.980 \nl
248 & AB    & AB    & AB & 19.00 &  2 &  4.70 &  18.07 &   2.718 &  4.48 &  0.937 & -0.174 & -0.147 &  -8.78 & -10.50 & 0.929 \nl
249 & AB    & AB    & AB &  9.88 &  2 &  0.93 &   8.31 &   1.284 &  0.95 &  0.961 & -0.099 & -0.037 &  -2.18 &  -5.88 & 0.935 \nl
250 & AB+EM & AB+EM & AB & 15.31 &  5 &  1.02 &   7.92 &   1.649 &  1.36 &  0.933 &  0.054 & -0.224 & -13.48 &   3.29 & 0.924 \nl
251 & AB    & AB    & AB &  2.86 &  2 &  0.30 &   2.43 &   0.368 &  0.31 &  0.821 &  0.096 &  0.208 &  14.10 &   6.69 & 0.727 \nl
252 & AB+EM & AB+EM & AB & 14.62 &  5 &  0.49 &   3.62 &   0.607 &  0.59 &  0.902 &  0.185 &  0.092 &   5.68 &  11.59 & 0.855 \nl
253 & AB    & AB    & AB & 19.00 &  4 &  0.49 &  12.14 &   2.117 &  0.53 &  0.940 &  0.014 &  0.125 &   7.56 &   0.85 & 0.899 \nl
254 & AB    & AB    & AB & 18.63 &  1 &  2.70 &  18.45 &   0.607 &  2.62 &  0.965 & -0.164 & -0.121 &  -7.03 &  -9.66 & 0.974 \nl
255 & AB    & AB    & AB & 16.10 &  4 &  0.50 &   4.00 &   0.472 &  0.51 &  0.959 & -0.031 &  0.034 &   2.06 &  -1.85 & 0.922 \nl
256 & AB    & AB    & AB & 18.18 &  4 &  0.58 &   8.37 &   1.284 &  0.66 &  0.980 & -0.034 &  0.036 &   2.10 &  -1.96 & 0.963 \nl
257 & AB    & AB    & AB & 18.68 &  4 &  0.50 &  10.00 &   1.649 &  0.56 &  0.969 & -0.054 &  0.024 &   1.39 &  -3.18 & 0.943 \nl
258 & AB    & AB    & AB & 19.00 &  2 &  1.55 &  15.64 &   2.117 &  1.52 &  0.977 & -0.131 & -0.058 &  -3.34 &  -7.62 & 0.976 \nl
259 & EM    & EM    & EM &  2.76 &  8 & 11.34 &   2.68 &   2.117 &  9.45 &  0.016 &  0.680 & -0.696 & -45.65 &  88.64 & 0.948 \nl
260 & AB    & AB    & AB & 17.99 &  4 &  0.55 &   8.30 &   1.284 &  0.61 &  0.964 & -0.022 &  0.044 &   2.60 &  -1.32 & 0.932 \nl
261 & AB    & AB    & AB & 14.72 &  1 &  1.96 &  17.50 &   2.117 &  1.87 &  0.960 & -0.187 & -0.162 &  -9.39 & -11.03 & 0.984 \nl
262 & AB    & AB    & AB & 19.00 &  2 &  0.87 &  18.04 &   2.718 &  0.84 &  0.979 & -0.141 & -0.082 &  -4.75 &  -8.19 & 0.986 \nl
264 & AB    & AB    & AB & 17.01 &  2 &  0.84 &  14.23 &   2.117 &  0.87 &  0.985 & -0.119 & -0.040 &  -2.32 &  -6.87 & 0.985 \nl
266 & AB    & AB    & AB & 19.00 &  1 &  0.53 &  18.99 &   0.018 &  0.53 &  0.486 & -0.032 & -0.131 & -15.01 &  -3.82 & 0.255 \nl
267 & AB    & AB    & AB & 16.87 &  4 &  0.61 &   3.49 &   0.223 &  0.68 &  0.957 & -0.083 &  0.016 &   0.97 &  -4.98 & 0.922 \nl
268 & AB    & AB    & AB & 15.11 &  1 &  2.19 &  17.74 &   2.117 &  2.08 &  0.969 & -0.162 & -0.114 &  -6.63 &  -9.47 & 0.978 \nl
271 & AB    & AB    & AB & 19.00 &  4 &  0.78 &   8.77 &   1.284 &  0.82 &  0.969 &  0.023 &  0.160 &   9.39 &   1.34 & 0.964 \nl
272 & AB    & AB    & AB & 19.00 &  1 &  1.35 &  18.91 &   0.135 &  1.30 &  0.981 & -0.139 & -0.065 &  -3.76 &  -8.05 & 0.987 \nl
273 & AB    & AB    & AB & 16.35 &  2 &  1.38 &  12.04 &   1.649 &  1.40 &  0.978 & -0.126 & -0.075 &  -4.37 &  -7.33 & 0.977 \nl
275 & AB    & AB    & AB & 17.12 &  2 &  0.98 &  14.21 &   2.117 &  1.01 &  0.984 & -0.087 & -0.010 &  -0.55 &  -5.04 & 0.975 \nl
279 & AB    & AB    & AB &  2.79 &  2 &  0.93 &   2.39 &   0.368 &  0.98 &  0.962 & -0.122 & -0.006 &  -0.38 &  -7.24 & 0.940 \nl
280 & AB    & AB    & AB & 13.57 &  2 &  0.77 &  11.04 &   1.649 &  0.80 &  0.960 & -0.183 & -0.154 &  -8.97 & -10.77 & 0.979 \nl
282 & AB    & AB    & AB &  3.05 &  1 &  0.22 &   2.50 &   0.004 &  0.22 &  0.839 & -0.066 &  0.125 &   8.44 &  -4.47 & 0.723 \nl
285 & AB+EM & AB+EM & AB &  9.79 &  4 &  0.25 &   2.47 &   0.368 &  0.26 &  0.508 &  0.234 &  0.147 &  14.68 &  24.72 & 0.334 \nl
287 & AB    & AB    & AB & 18.61 &  2 &  0.95 &  14.82 &   2.117 &  0.95 &  0.962 & -0.146 & -0.043 &  -2.51 &  -8.65 & 0.949 \nl
289 & AB    & AB    & AB & 18.29 &  2 &  3.51 &  12.78 &   1.649 &  3.43 &  0.946 & -0.173 & -0.143 &  -8.45 & -10.39 & 0.944 \nl
291 & AB    & AB    & AB & 12.25 &  4 &  0.41 &   3.24 &   0.472 &  0.44 &  0.951 &  0.035 &  0.121 &   7.24 &   2.09 & 0.921 \nl
292 & AB+EM & AB+EM & AB &  8.44 &  4 &  0.61 &   2.58 &   0.472 &  0.67 &  0.798 &  0.344 &  0.090 &   5.95 &  23.32 & 0.762 \nl
293 & AB+EM & AB+EM & AB &  1.90 &  2 &  0.80 &   1.66 &   0.287 &  0.86 &  0.818 &  0.368 &  0.086 &   5.46 &  24.19 & 0.812 \nl
294 & AB+EM & AB    & AB & 17.72 &  5 &  0.94 &   5.90 &   1.000 &  1.27 &  0.962 &  0.138 &  0.067 &   3.93 &   8.15 & 0.949 \nl
295 & AB    & AB    & AB &  4.77 &  2 &  0.55 &   3.34 &   0.472 &  0.54 &  0.971 & -0.047 &  0.094 &   5.55 &  -2.80 & 0.954 \nl
296 & AB    & AB    & AB &  7.43 &  1 &  3.04 &   8.40 &   1.000 &  2.99 &  0.962 & -0.184 & -0.108 &  -6.30 & -10.82 & 0.971 \nl
297 & AB    & AB    & AB &  2.67 &  1 &  0.36 &   2.48 &   0.287 &  0.36 &  0.948 & -0.102 &  0.080 &   4.78 &  -6.13 & 0.916 \nl
302 & AB    & AB    & AB &  4.42 &  2 &  1.55 &   3.75 &   0.607 &  1.59 &  0.948 & -0.181 & -0.208 & -12.17 & -10.79 & 0.975 \nl
303 & AB+EM & AB    & AB & 16.67 &  2 &  1.25 &  16.45 &   2.718 &  1.35 &  0.956 & -0.147 & -0.126 &  -7.40 &  -8.75 & 0.952 \nl
304 & AB    & AB    & AB & 12.39 &  4 &  0.47 &   3.26 &   0.472 &  0.51 &  0.928 & -0.109 &  0.020 &   1.20 &  -6.68 & 0.873 \nl
306 & AB    & AB    & AB &  7.14 &  2 &  0.41 &   4.32 &   0.607 &  0.41 &  0.972 & -0.139 & -0.054 &  -3.15 &  -8.11 & 0.967 \nl
307 & AB    & AB    & AB &  5.65 &  2 &  0.63 &   3.47 &   0.472 &  0.64 &  0.958 & -0.009 &  0.106 &   6.34 &  -0.51 & 0.929 \nl
308 & AB    & AB    & AB &  2.00 &  2 &  0.41 &   1.72 &   0.287 &  0.43 &  0.924 & -0.071 &  0.095 &   5.87 &  -4.39 & 0.867 \nl
309 & AB    & AB    & AB &  1.58 &  1 &  2.09 &   0.94 &   0.030 &  2.05 &  0.646 & -0.246 &  0.030 &   2.48 & -20.81 & 0.479 \nl
311 & AB    & AB    & AB &  1.55 &  2 &  0.86 &   1.28 &   0.223 &  0.93 &  0.883 & -0.020 &  0.198 &  12.63 &  -1.27 & 0.820 \nl
312 & AB    & AB    & AB & 11.77 &  2 &  1.33 &  10.48 &   1.649 &  1.42 &  0.976 & -0.140 & -0.060 &  -3.49 &  -8.17 & 0.976 \nl

\enddata
}
\end{deluxetable}

\begin{deluxetable}{ccccccc}
\tablenum{3}
\tablecolumns{7}
\tablecaption{Line Strengths versus PCA classification
\label{tbl3}}
\tablehead{PCA $\backslash$ Line Strengths & AB & PS & AB+EM & EM+AB & EM & Total}
\startdata
AB    & 144 & 0 &  6 &  0 & 0 & 150 \nl
PS    &   0 & 5 &  2 &  0 & 0 &   7 \nl
AB+EM &   3 & 0 & 19 &  0 & 0 &  22 \nl
EM    &   0 & 0 &  1 & 12 & 4 &  17 \nl \hline
Total & 147 & 5 & 28 & 12 & 4 & 196 \nl

\enddata
\end{deluxetable}

\begin{deluxetable}{ccccccc}
\tablenum{4}
\tablecolumns{7}
\tablecaption{Line Strengths versus Wavelet classification
\label{tbl4}}
\tablehead{Wavelet $\backslash$ Line Strengths & AB & PS & AB+EM & EM+AB & EM & Total}
\startdata
AB    & 144 & 0 & 19 &  4 & 0 & 167 \nl
PS    &   3 & 5 &  7 &  0 & 0 &  15 \nl
EM    &   0 & 0 &  2 &  8 & 4 &  14 \nl \hline
Total & 147 & 5 & 28 & 12 & 4 & 196 \nl 
\enddata
\end{deluxetable}

\begin{deluxetable}{cccccc}
\tablenum{5}
\tablecolumns{6}
\tablecaption{PCA versus Wavelet classification schemes
\label{tbl5}}
\tablehead{Wavelet $\backslash$ PCA & AB & PS & AB+EM & EM & Total}
\startdata
AB    & 148 & 0 & 15 &  4 &  167 \nl
PS    &   2 & 7 &  6 &  0 &   15 \nl
EM    &   0 & 0 &  1 & 13 &   14 \nl \hline
Total & 150 & 7 & 22 & 17 &  196 \nl 
\enddata
\end{deluxetable}

\newpage

\clearpage

\clearpage

\clearpage

\clearpage

\clearpage

\clearpage

\end{document}